\documentclass[11pt,paper,twoside]{JHEP3}
\usepackage{graphicx}                     %%% for pictures and figures
\usepackage{epsfig}                     %%% for pictures and figures
\usepackage{psfig}                     %%% for pictures and figures
\usepackage{amssymb}
\usepackage{amsmath}
\usepackage{float}
\usepackage{dcolumn}
\usepackage{fancyhdr}
\usepackage{amsfonts}
\newcommand{\ve}{\varepsilon}
\newcommand{\beq}{\begin{equation}}
\newcommand{\eeq}{\end{equation}}
\newcommand{\beqa}{\begin{eqnarray}}
\newcommand{\eeqa}{\end{eqnarray}}

\title{Chiral extrapolations of baryon masses for \\[0.3em]
 unquenched three-flavor
lattice simulations\thanks{Work supported in part by funds provided from
the EU to the project ``Study of Strongly Interacting Matter'' under contract no.
RII3-CT-2004-506078.}}
 
 \author{Matthias Frink\\
        Institut f\"ur Kernphysik (Theorie), Forschungszentrum J\"ulich, 
        D-52425 J\"ulich, Germany\\and\\
           Helmholtz-Institut f\"ur Strahlen- und Kernphysik (Theorie), 
           Universit\"at  Bonn\\
           Nu{\ss}allee 14-16, D-53115 Bonn, Germany \\
           E-mail: \email{m.frink@fz-juelich.de}}
 \author{Ulf-G. Mei{\ss}ner\\
         Helmholtz-Institut f\"ur Strahlen- und Kernphysik (Theorie), 
         Universit\"at  Bonn\\
         Nu{\ss}allee 14-16, D-53115 Bonn, Germany \\
         and\\
         Institut f\"ur Kernphysik (Theorie), Forschungszentrum J\"ulich, 
        D-52425 J\"ulich, Germany\\ E-mail: \email{meissner@itkp.uni-bonn.de}}

\abstract{
We have analyzed the chiral expansion of the baryon masses to second order
in the quark masses, including strong isospin violation. The calculations
are performed in Lorentz-invariant baryon chiral perturbation theory. 
Chiral extrapolation functions for three-flavor unquenched lattice 
simulations are derived.  The matching to the two-flavor case and 
the sigma terms are also considered.
}
\preprint{HISKP-TH-04/06}

\begin{document}

%%%%%%%%%%%%%%%%%%%%%%%%%%%%%%%%%%%
\section{Introduction}
\setcounter{footnote}{0}
%%%%%%%%%%%%%%%%%%%%%%%%%%%%%%%%%%%%

Lattice QCD supplied with chiral perturbation theory is a powerful non-perturbative
tool to analyze the structure of QCD in the confinement regime. It has been realized that
lattice simulations performed at unphysically large quark masses can be extrapolated
systematically to the physical limit based on functions derived in chiral
perturbation theory or extensions
thereof, see e.g. \cite{panel,Karl}. Many observables like the pion mass, nucleon form factors
and so on have been discussed in the literature, we focus here on the simplest baryonic two-point
functions, namely the masses of the ground state octet. While there is still on-going debate about the
precise implementation of such schemes in  terms of regulator prescriptions and the
corresponding resummation of classes of higher order terms (see e.g. the discussions
in Refs.\cite{Adelaide,Adelaide2,BHMcutoff,PHW}), there is little controversy about the
inherent usefulness of such studies. This is documented by the excellent fits one
obtains for the quark mass expansion of the nucleon mass applied to two-flavor simulations,
which typically have pion masses larger than 400 MeV, see e.g. \cite{CPPACS,JLQCD,QCDSF}.
It should also be pointed out that such two-flavor studies can be constrained by chiral
perturbation theory analyses of pion-nucleon scattering, since these allow one to pin down
some  combinations of the low-energy constants (LECs) appearing in the quark mass expansion of the
nucleon mass.  Recently, detailed finite size studies for the nucleon mass ($N_F =2$)  
have also become available \cite{AliKhan:2003cu} \footnote{Note that in this paper we will
only be concerned with unquenched lattice QCD, for a study of nucleon properties in partially
quenched QCD see \cite{BS} and references therein.}. Still, simulations at lower quark masses
are needed to further sharpen these analyses.

\medskip\noindent
The situation is entirely different in the three--flavor sector ($N_F=3$). Only recently
a large lattice effort was initiated to study tree--flavor dynamical QCD \cite{Kanaya}, and
it will take some time before detailed results are available. In the meson sector, one
has little doubt that chiral perturbation theory can be applied, although there are
speculations of a strong flavor dependence of certain order parameters \cite{orsay}. In
the baryon sector, the issue of the convergence of the chiral expansion is not settled,
but it is fair to say that more higher order calculations are needed for a
final  clarification\footnote{For a recent discussion about the onset of the 
chiral regime, see e.g. \cite{Silas}.}.
Calculations of the nucleon (baryon) mass(es)
to third and fourth order in various variants of chiral perturbation theory
have been performed, see e.g. \cite{Juerg}-\cite{Mainz}. It is obvious that for reliable
extrapolation functions for the baryon masses one has to go beyond the leading non-analytic terms
$\sim m_{\rm quark}^{3/2}$, because these do not even provide a reliable extrapolation function
for two flavors. It is thus mandatory to consider at least the terms quadratic in the quark masses.
This will be the topic of the present paper. In contrast to the earlier work \cite{BM}, we perform
the calculation in a Lorentz-invariant framework based on the so-called infrared regularization (IR) 
\cite{BL}. Furthermore, we also include all strong isospin breaking
effects $\sim m_u-m_d$ up-to-and-including 
fourth order in the chiral expansion. Concerning the inclusion of the spin-3/2 decuplet, we
follow exactly the lines of \cite{BHMcutoff}, were it is discussed how such effects can
be absorbed in the values of certain LECs.
We are well aware that present day lattice studies are far from
reaching the required accuracy to resolve these fine effects, 
but we hope that the representations given here will become useful in the future.
 
\medskip\noindent
The manuscript is organized as follows. The effective Lagrangian underlying our calculation
is briefly presented in Section~\ref{sec:Lagr}. The calculation of the baryon masses is outlined in
Section~\ref{sec:calc} and the resulting quark mass expansions of the octet ground state masses
are collected in Section~\ref{sec:extra}. To constrain combinations of the appearing low-energy
constants, we perform  matching to the SU(2) case in Section~\ref{sec:match}. Sigma terms are
discussed in  Section ~\ref{sec:sigma}. We briefly summarize in Section ~\ref{sec:sum}. The 
appendices contain  the representation of the baryon masses in terms of the Goldstone
boson masses and one-loop integrals, a general discussion of the quark mass
dependence of tadpole diagrams, and the $\beta$--functions related to the
renormalization of the one-loop graphs.
% and some further technicalities.
%\newpage

%%%%%%%%%%%%%%%%%%%%%%%%%%%%%%%%%%%%%%%%
\section{Effective Lagrangian}
\label{sec:Lagr} 
%%%%%%%%%%%%%%%%%%%%%%%%%%%%%%%%%%%%%%%%

In this section, we collect the terms of the effective chiral Lagrangian
underlying our calculation. We only display the terms relevant to the topic
under investigation and keep the discussion as brief as possible. We employ
the standard power counting in which external momenta count as order $p$
(more precisely meant are meson four- and baryon-three momenta),
whereas a quark mass insertion is booked as ${\cal O}(p^2)$.  Here, $p$ denotes
a genuine small parameter. Our calculation
includes all terms up-to-and-including order $p^4$ from the effective
Lagrangian (one-loop approximation). The chiral limit is defined by
$m_u = m_d = m_s = 0$, where $m_u \,(m_d,m_s)$ denotes the up (down, strange)
quark mass. The basic fields in our effective Lagrangian are the eight pseudoscalar
(Pseudo-)Goldstone bosons $(\pi^\pm, \pi^0, K^\pm, K^0, \bar{K}^0, \eta)$ chirally
coupled to the octet of the spin-1/2 baryons. The effects of the spin-3/2 decuplet
are included in the low-energy constants appearing at the various  orders.

\subsection{Meson Lagrangian}
\label{sec:MLagr} 

First, we consider the mesonic Lagrangian \cite{GL85}. The current quark mass matrix
${\cal M} = {\rm diag}(m_u,m_d,m_s)$ is contained in the external source $\chi
= 2B(s(x) + ip(x))$, $s(x) = {\cal M}+\ldots$ and $B = |\langle 0|\bar q
q|0\rangle|/F_\pi^2$ measures the strength of 
the quark condensate in the chiral limit. To one-loop accuracy, the
pertinent effective Lagrangian includes terms of chiral dimension two and
four,
\beq
{\cal L}_{\rm eff} =  {\cal L}^{(2)}  + {\cal L}^{(4)} ~.
\eeq
The relevant terms are given by
\beq
{\cal L}^{(2)} = \frac{F_\pi^2}{4}{\rm Tr} \left\{ u_\mu u^\mu \right\} +
                 \frac{F_\pi^2}{4}{\rm Tr} \left\{ \chi_+\right\}~,
\eeq
\beqa
{\cal L}^{(4)} &=& L_1 \, \left[ {\rm Tr} \left\{ u_\mu u^\mu
  \right\}\right]^2 + L_2\, {\rm Tr} \left\{ u_\mu u_\nu \right\}
  \, {\rm Tr} \left\{ u^\mu u^\nu \right\}
  + L_3 \,  {\rm Tr} \left\{ u_\mu u^\mu u_\nu u^\nu  \right\}
\nonumber\\
 &+& L_4 \, {\rm Tr} \left\{ u_\mu u^\mu \right\}\,{\rm Tr} \left\{\chi_+\right\}
  + L_5 \, {\rm Tr}  \left\{ u_\mu u^\mu \chi_+\right\} 
  + L_6\, \left[ {\rm Tr} \left\{ \chi_+\right\}\right]^2
\nonumber\\
 &+& L_7\, \left[ {\rm Tr} \left\{ \chi_-\right\}\right]^2
  + L_8\, \frac{1}{2}\,{\rm Tr} \left\{\chi_+ \chi_+ +  \chi_-\chi_-\right\}~,
\eeqa
with
\beq
u_\mu = i u^\dagger \partial_\mu U u^\dagger \sim {\cal O}(p)~, \quad
\chi_\pm =  u^\dagger \chi u^\dagger \pm  u \chi^\dagger u \sim {\cal O}(p^2) ~.
\eeq
Here, $F_\pi$ is the pseudoscalar decay constant (in the chiral limit) and the
trace ``Tr'' refers to flavor space.
The $L_i$ are low-energy constants, they are in general scale--dependent
and absorb the infinities generated by the one-loop graphs.
The pseudoscalar Goldstone fields ($\phi = \pi, K, \eta$) are collected in
the  $3 \times 3$ unimodular, unitary matrix $U$,
\begin{equation}
 U(\phi) = u^2 (\phi) = \exp \lbrace i \phi / F_\pi \rbrace~,
\end{equation}
with 
\beq
 \phi =  \sqrt{2}  \left( 
\begin{matrix}  \frac{1}{\sqrt 2} \pi^0 + \frac{1}{\sqrt 6} \eta
&\pi^+ &K^+  \\
\pi^-
        & -\frac{1}{\sqrt 2} \pi^0 + \frac{1}{\sqrt 6} \eta & K^0
         \\ \left.\right.
K^-
        &  \bar{K^0}&- \frac{2}{\sqrt 6} \eta  
\end{matrix}\right) 
\eeq
in terms of the physical fields without $\pi^0-\eta$ mixing.  Under SU(3)$_L \times$SU(3)$_R$, $U$ 
transforms as $U \to U'=LUR^\dagger$, with $L,R \in$ SU(3)$_{L,R}$.
The neutral pion and the eta mix, the
physical fields are related to the pure SU(3) components via
\beq
\left( \begin{matrix} 
  \pi^0 \\  \eta 
\end{matrix}\right) = \left(  \begin{matrix} \cos \ve &\sin\ve\\
                                            -\sin \ve &\cos\ve
                                             \end{matrix}\right) \,
\left( \begin{matrix} 
  \phi_3 \\  \phi_8 
\end{matrix}\right)~.
\eeq
The (lowest order) mixing angle can be fixed from the condition 
that the  second order  expression for the meson self-energy is diagonal,
\beq
\Sigma^{(2)}_{\pi^0 \eta} = \Sigma^{(2)}_{\eta \pi^0} = 0~,
\eeq
resulting in
\beq
\tan 2\ve = \sqrt 3 \frac{m_d - m_u}{2m_s - m_u -m_d} = \frac{ \sqrt 3}{2} 
\frac{m_d - m_u}{m_s -\hat  m}~,\label{mixang}
\eeq
where $\hat m = (m_u + m_d)/2$ is the average light quark mass. Numerically, the
mixing angle is quite small. Note that the
effect of further mixing with the $\eta '$  is contained in the numerical values
of some of the low-energy constants. 
From this, one reads off the leading terms in the quark mass expansion of the 
meson masses (as denoted by the bar),
\beqa\label{mesmass2}
\bar M_{\pi^+}^2 &=& B \, (m_u+m_d) = 2B \, \hat m~, \nonumber\\
\bar M_{\pi^0}^2 &=& 2B \, \left[ \frac{m_u+m_d+m_s}{3} +\frac{m_u+m_d-2m_s}{6}\cos (2\ve)
                    + \frac{\sqrt{3}}{6}(m_u-m_d) \sin (2\ve) \right] \nonumber\\
                 &=& 2B\, \left[ \frac{2\hat m+m_s}{3} +\frac{\hat m-m_s}{3}
                                       \frac{1}{\cos (2\ve)}\right]~,\nonumber\\
\bar M_{K^+}^2 &=& B \,  (m_u+m_s) = B\, \left[ \hat m - \frac{1}{\sqrt{3}} \tan(2\ve )
                                     (m_s - \hat m) + m_s \right]~, \nonumber\\
\bar M_{K^0}^2 &=& B \,  (m_d+m_s) = B\, \left[ \hat m + \frac{1}{\sqrt{3}} \tan(2\ve )
                                     (m_s - \hat m) + m_s \right]~, \nonumber\\
\bar M_{\eta}^2 &=&  2B \, \left[ \frac{m_u+m_d+m_s}{3} - \frac{m_u+m_d-2m_s}{6}\cos (2\ve)
                    - \frac{\sqrt{3}}{6}(m_u-m_d) \sin (2\ve) \right] \nonumber\\
                 &=& 2B\, \left[ \frac{2\hat m+m_s}{3} -\frac{\hat m-m_s}{3}
                                       \frac{1}{\cos (2\ve)}\right]~,
\eeqa
and by CPT $M_{\pi^-} = M_{\pi^+}$, $M_{\bar K^0} = M_{K^0}$. The complete
one-loop expressions for the $\pi^+$ and $K^+$ are collected in Appendix~\ref{sec:meson}. Note that
in the following we use the charged pion and kaon masses as our reference
values. If we use Eqs.~(\ref{mesmass2}) together with the leading isospin violating
corrections $\sim m_u - m_d$ and $\sim e^2$, where the latter are  the electromagnetic
effects, to relate the Goldstone boson to the the quark masses, we get $ \ve = 1\cdot 
10^{-2}$. This justifies expanding all functions of $\ve$ in powers of this small parameter
to the appropriate order, as will be done in what follows.

\subsection{Meson-Baryon Lagrangian}
\label{sec:MBLagr} 

%~\footnote{The extension of the framework 
%utilized to include explicit $\eta'$ degrees of freedom is given in
%\cite{BB}.}

The ground state baryons with $J^P = \frac{1}{2}^+$ are collected in the 
SU(3) matrix
\beq
 B  =   \left( 
\begin{matrix}  \frac{1}{\sqrt 2} \Sigma^0 + \frac{1}{\sqrt 6} \Lambda
&\Sigma^+ &p  \\
\Sigma^-
        & -\frac{1}{\sqrt 2} \Sigma^0 + \frac{1}{\sqrt 6} \Lambda & n
         \\ \left.\right.
\Xi^-
        &  \Xi^0 &- \frac{2}{\sqrt 6} \Lambda  
\end{matrix}\right) 
\eeq
in terms of the physical fields without $\Lambda - \Sigma^0$ mixing. It turns out convenient to diagonalize the second order baryon self--energy contribution with the help of the lowest order mixing angle $\ve$ in Eq.~(\ref{mixang}). Note that to higher order in the chiral expansion the concept of a unique $\Lambda - \Sigma^0$ mixing angle breaks down, which calls for the introduction of two different angles.
Under chiral SU(3)$\times$SU(3), $B$ transforms as any
matter field, $B \to B' = KBK^\dagger$, with $K(U,L,R)$ the compensator field
representing an element of the conserved subgroup SU(3)$_V$. The corresponding
effective meson-baryon Lagrangian contains terms of even and odd dimension,
\beq\label{leffsum}
{\cal L}_{\rm eff} = {\cal L}^{(1)} + {\cal L}^{(2)} + {\cal L}^{(3)} + {\cal L}^{(4)} ~,
\eeq
to the accuracy required here. The form of the lowest order Lagrangian is standard,
\beq\label{leff1}
{\cal L}^{(1)} =  {\rm Tr}\left\{\overline B \left(i \gamma^{\mu}D_{\mu}
-m_0\right)B\right\}+\frac{D}{2} {\rm Tr}\left\{\overline B \gamma^{\mu}
\gamma^5 \left\{u_{\mu},B\right\}\right\}+
\frac{F}{2} {\rm Tr}\left\{\overline B \gamma^{\mu}\gamma^5 
\left[u_{\mu},B\right]\right\}~,
\eeq
with $D \simeq 0.81$ and $F \simeq 0.46$ \cite{ratcliffe} 
the two axial-vector couplings. $D_\mu$
is the chiral covariant derivative. For our analysis it suffices to use
$D_\mu = \partial_\mu$. Furthermore, $m_0$ is the average octet mass in the chiral
limit. This large mass scale, which is of the order of the scale of chiral
symmetry breaking, $\Lambda_\chi \simeq 4\pi F_\pi$, requires special
treatment as detailed below. At dimension two, we have considerably more
terms (for the construction principles see \cite{Krause,FMMS,FMlarge})
\begin{eqnarray}\label{leff2}
{\cal L}^{(2)}&=&b_D  {\rm Tr}\left\{\overline B\left\{\chi_{+},B\right\}
\right\}+b_F {\rm Tr}\left\{\overline B\left[\chi_{+},B\right]\right\}
+b_0 {\rm Tr}\left\{\overline B B\right\}{\rm Tr}\left\{\chi_{+}\right\}
\nonumber \\ &+& b_1  {\rm Tr}\left\{\overline B \left[u_{\mu},
\left[u^{\mu},B\right]\right]\right\}+
b_2  {\rm Tr}\left\{\overline B\left\{u_{\mu},\left[u^{\mu},B\right]\right\}
\right\}\nonumber\\
&+&b_3  {\rm Tr}\left\{\overline B\left\{u_{\mu},\left
\{u^{\mu},B\right\}\right\}\right\}
+ b_4 {\rm Tr}\left\{\overline B B\right\} 
{\rm Tr}\left\{u_{\mu}u^{\mu}\right\}\nonumber\\
&+&\frac{b_5}{2} \Big( {\rm Tr}\left\{\overline B i \gamma^{\mu}
\left[u_{\mu},\left[u_{\nu},\left[D^{\nu},B\right]\right]\right]\right\}
+{\rm Tr}\left\{\overline B i \gamma^{\mu}
\left[D_{\nu},\left[u^{\nu},\left[u_{\mu},B\right]\right]\right]\right\}\Big)
\nonumber\\
& +&\frac{b_6}{2} \Big( {\rm Tr}\left
\{\overline B i \gamma^{\mu}\left[u_{\mu},\left\{u_{\nu},
\left[D^{\nu},B\right]\right\}\right]\right\}
+{\rm Tr}\left\{\overline B i \gamma^{\mu}
\left[D_{\nu},\left\{u^{\nu},\left[u_{\mu},B\right]\right\}\right]\right\}\Big)
\nonumber\\ 
&+&\frac{b_7}{2}\Big( {\rm Tr}\left\{\overline B i \gamma^{\mu}\left\{u_{\mu},
\left\{u_{\nu},\left[D^{\nu},B\right]\right\}\right\}\right\}+
{\rm Tr}\left\{\overline B i \gamma^{\mu}\left[D_{\nu},
\left\{u^{\nu},\left\{u_{\mu},B\right\}\right\}\right]\right\}\Big)
\nonumber\\
&+& 
b_8 {\rm Tr}\left\{\overline B i \gamma^{\mu}\left[D_{\nu},B\right]
\right\}{\rm Tr}\left\{u_{\mu}u^{\nu}\right\} \nonumber\\
&+&\frac{b_{9}}{2}\Big({\rm Tr}\left
\{\overline B u_{\mu}\right\}i\gamma^{\mu}{\rm Tr}\left\{u_{\nu}\left[D^{\nu},
B\right]\right\}+{\rm Tr}\left\{\overline B u_{\nu}\right\}
i\gamma^{\mu}{\rm Tr}\left\{u_{\mu}\left[D^{\nu},B\right]\right\}\nonumber\\ 
& & +{\rm Tr}\left\{\overline B\left[D_{\nu},u^{\nu}\right]\right\} 
i\gamma^{\mu}{\rm Tr}\left\{u_{\mu}B\right\}+{\rm Tr}\left\{\overline 
B u_{\nu}\right\}i\gamma^{\mu}{\rm Tr}\left\{\left[D^{\nu},u_{\mu}\right]
B\right\}\Big)~.
\end{eqnarray}
The LECs $b_i$ ($i=0,D,F,1,\ldots,4$) have dimension mass$^{-1}$, whereas the $b_j$
($j=5,\ldots,9$) have dimension  mass$^{-2}$. 
The first three of these terms $\sim b_{0,D,F}$ stick out since they parameterize
the explicit symmetry breaking and contribute already at tree level to the masses and the
corresponding sigma terms. All other terms only play a role in the fourth
order loop graphs. Note that we have more terms than were given in \cite{BM}.
There, all terms with one or two covariant derivatives were absorbed in the
structures $\sim b_{1,2,3,4}$. This can be done as long as one works for a
set of fixed quark masses. For our purpose, we need to retain all terms that
lead to structures of different quark mass dependences. A more detailed discussion
of this topic is relegated to App.~\ref{app:tadpole}.
We also note that the complete minimal  meson-baryon Lagrangian
can be found in Ref.\cite{FMlarge}.
We have no contribution from the dimension three terms, thus  
\beq
{\cal L}^{(3)} = 0~.
\eeq
Finally, there are tree contributions from the dimension four Lagrangian (we have
performed a relabeling of the last two terms as compared to \cite{BM})
\beqa
{\cal L}^{(4)}
& = & d_1 \, {\rm Tr} \left\{\overline{B} [\chi_+ , [ \chi_+ , B]] \right\}   
 + d_2 \, {\rm Tr} \left\{\overline{B} [\chi_+ , \{ \chi_+ , B\} ] \right\} \nonumber \\   
& + & d_3 \, {\rm Tr} \left\{\overline{B} \{ \chi_+ , \{ \chi_+ , B\} \} \right\}   
 + d_4 \, {\rm Tr}\left\{\overline{B} \chi_+ \right\} \,
          {\rm Tr}\left\{\chi_+ B \right\} \nonumber \\
& + & d_5 \, {\rm Tr} \left\{ \overline{B}  [ \chi_+ , B] \right\} 
          \, {\rm Tr}\left\{\chi_+\right\}   
% + d_6 \, {\rm Tr} (\bar{B}  \{ \chi_+ , B\} ) {\rm Tr}(\chi_+)
% \nonumber \\   
 +  d_6 \, {\rm Tr} \left\{\overline{B}B\right\}\, 
           {\rm Tr}\left\{\chi_+\right\} {\rm Tr}\left\{\chi_+\right\} \nonumber \\     
& + & d_7 \, {\rm Tr} \left\{\overline{B}B\right\} {\rm Tr}\left\{\chi_+^2\right\} \, \, \, .
\label{leff4}
\eeqa
The LECs $d_i$ are of dimension mass$^{-3}$. They are the sum of an infinite part to
absorb the infinities from the boson loops and a finite part, that is of
relevance here. Whenever we write down any of these LECs, it refers to its
finite (renormalized) value at the renormalization scale $\lambda=m_0$ 
(as discussed in more detail below). Note that some
of the terms in Eq.~(\ref{leff4}) are quark mass renormalizations of some of
the second order terms listed in Eq.~(\ref{leff2}), but for our purpose we have
to keep these contributions separated since they lead to different quark mass
dependences.

%%%%%%%%%%%%%%%%%%%%%%%%%%%%%%%%%%%%%%%%%%%%%%%%%%%%%%%%%%%%%%%%%%%%%%%%%%%%%%%%%%%%%%%%
\section{Calculation of baryon masses}
\label{sec:calc}
%%%%%%%%%%%%%%%%%%%%%%%%%%%%%%%%%%%%%%%%%%%%%%%%%%%%%%%%%%%%%%%%%%%%%%%%%%%%%%%%%%%%%%%%
To calculate the baryon masses in chiral perturbation theory,
we consider the baryonic two-point functions
\beq\label{twopint}
\int d^4 x \, e^{ip\cdot x} \, \langle 0|T\{ B (x) \bar B' (0)\} | 0\rangle 
 = \left(\frac{i}{p \!\!\!/ - m_0 - \Sigma}\right)_{BB'}~,
\eeq
with $\Sigma ( p \!\!\!/)$ the self-energy. Furthermore,
$B\,(B')$ stands for any member of the ground state octet, $B = p,n,\Lambda, \Sigma^0, \Sigma^+,
\Sigma^-, \Xi^0, \Xi^-$. The matrix notation is necessary because of the $\Lambda-\Sigma^0$ mixing.
The baryon mass at a given order in the chiral expansion can be expressed in terms of the 
corresponding self-energy at the baryon pole for the diagonal states as
\beqa\label{Bmass}
m_B^{(0)} &=& m_0~, \nonumber \\
m_B^{(2)} &=& \Sigma_{BB}^{(2)} ( p \!\!\!/ = m_0)~,\nonumber \\
m_B^{(3)} &=& \Sigma_{BB}^{(3)}( p \!\!\!/ = m_0)~,\nonumber \\
m_B^{(4)} &=&  \Sigma_{BB}^{(4)}( p \!\!\!/ = m_0) + \frac{\partial}{\partial  p \!\!\!/}
\left(\Sigma_{BB}^{(3)}( p \!\!\!/ = m_0)\right) \,  \Sigma_{BB}^{(2)} ( p \!\!\!/ = m_0)~,
\eeqa
where the superscript $(n)$ denotes the chiral dimension. We remark that due to parity there 
are no first order contributions and $m_0$ denotes the octet mass in the chiral limit of
vanishing up, down, and strange quark masses. For the $\Lambda$, one has
\beqa\label{Lmass}
m_\Lambda^{(0)} &=& m_0~, \nonumber \\
m_\Lambda^{(2)} &=& \Sigma_{\Lambda\Lambda}^{(2)} ( p \!\!\!/ = m_0)~,\nonumber \\
m_\Lambda^{(3)} &=& \Sigma_{\Lambda\Lambda}^{(3)}( p \!\!\!/ = m_0)~,\nonumber \\
m_\Lambda^{(4)} &=&  \Sigma_{\Lambda\Lambda}^{(4)}( p \!\!\!/ = m_0) 
                 +    \frac{\partial}{\partial  p \!\!\!/}
\left(\Sigma_{\Lambda\Lambda}^{(3)}( p \!\!\!/ = m_0)\right) \,
 \Sigma_{\Lambda\Lambda}^{(2)} ( p \!\!\!/ = m_0)
\nonumber\\ &&\quad\qquad + \frac{\Sigma^{(3)}_{\Sigma^0 \Lambda} ( p \!\!\!/ = m_0) \, 
\Sigma^{(3)}_{\Lambda\Sigma^0}( p \!\!\!/ = m_0) }{
 \Sigma_{\Lambda\Lambda}^{(2)} ( p \!\!\!/ = m_0)  
                          -  \Sigma_{\Sigma^0\Sigma^0}^{(2)} ( p \!\!\!/ = m_0)}~,\label{nondiag}
\eeqa
and a similar expression holds for the $\Sigma^0$.
It can be obtained from Eq.~(\ref{Lmass}) by the interchange
$\Lambda \leftrightarrow \Sigma^0$. The chiral expansion is equivalent to the quark mass expansion.
The latter can symbolically be written as
\beq\label{massq}
m_B = m_0 + \sum_{q=u,d,s} A_q^B m_q + \sum_{q=u,d,s} B_q^B m_q^{3/2} +  
 \sum_{q=u,d,s} C_q^B m_q^2 + \ldots~,
\eeq
and the state-dependent coefficients $A_q^B, B_q^B$ and $C_q^B$ can be read off from 
Eqs.~(\ref{Bmass},~\ref{Lmass}) if one utilizes the relations between the Goldstone boson and 
the quark masses.
%\FIGURE[p]{
\begin{figure}[tb]
  \begin{center}
    \psfig{file=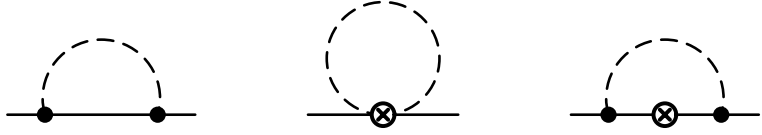,width=4.in}
    \caption{Loop contributions to the baryon masses at third and fourth order.
     Solid (dashed) lines refer to baryons (Goldstone bosons). The solid dot 
     (circlecross) denotes an insertion from the dimension one (two) meson-baryon
     Lagrangian.}
\label{fig:dia}
  \end{center}
\vspace{-0.4cm}
\end{figure}
%}

\medskip\noindent
We now turn to the explicit calculation of the baryon masses. At first and second order, one
only has contributions from tree graphs, whereas at third order one-loop  and at 
fourth order tree and one-loop diagrams contribute. The relevant one-loop graphs are displayed in 
Fig.~\ref{fig:dia}. We briefly discuss the structure of the related loop integrals. The
finite baryon mass in the chiral limit spoils the one-to-one correspondence between the
chiral and the loop expansion. There exist various methods to overcome this problem, 
we use here the so-called infrared regularization to separate the soft (chiral) loop 
contributions from the hard ones. The latter can be absorbed in local terms given by
the effective chiral Lagrangian (for a more detailed discussion we refer to \cite{BL}).
The essential trick in this method is to write a Feynman parameter representation for
any loop integral and split the integration range [0,1] as [0,$\infty$]-[1,$\infty$].
The first term defines the irregular part which contains all the interesting chiral physics
and obeys the power counting. The regular part given by the second integral (which
is generated by momenta of the size of the hard scale $m_0$) is a polynomial
in external momenta and quark masses.
This splitting is symmetry-preserving and can be applied to any
loop graph. In what follows, the symbol $\int_I$ stands for the irregular part obtained
by applying this procedure. As done in \cite{BL}, we use dimensional regularization for
separating the divergent parts and set $\lambda = m_0$, where $\lambda$ is the scale of
dimensional regularization. In principle, the IR method generates some higher
divergences, we follow the same strategy as in \cite{BL} and \cite{KM} and
simply ignore these (a different approach is followed in \cite{PHW}). 

\medskip\noindent
We now turn to the explicit representation of the relevant
loop diagrams. The tadpole graph (one meson propagator) is standard
\beq\label{D2}
\bar\Delta_P = \int_I \frac{d^dk}{(2\pi)^d} \frac{i}{k^2 - \bar M_P^2} 
\stackrel{d \to 4}{\longrightarrow} 
2ö\bar M_P^2 \left( \bar L + \frac{1}{2} \bar \mu_P \right)~,
\eeq
with
\beqa\label{mubar}
\bar \mu_P &=& \frac{1}{16\pi^2} \ln \left( \frac{\bar M_P^2}{m_0^2} \right)~, \nonumber\\
\bar {L} &=& \frac{m_0^{d-4}}{16\pi^2} \left[ \frac{1}{d-4} + \frac{1}{2} \left(
\gamma_E - 1 - \ln 4\pi \right)\right]~,
\eeqa
where $P$ stands for any Goldstone boson, $d$ denotes the number of space-time
dimensions and $\gamma_E$ is the Euler-Mascheroni constant. This type of graph 
contributes only at fourth order since the tadpole with the lowest order 
$\overline BBPP$-vertex
vanishes.
The self-energy graph with lowest order insertions is given by (one meson and one nucleon
propagator)
\beq\label{Ibar}
I_P (q^2) = -\int_I \frac{d^dk}{(2\pi)^d} \frac{i}{k^2 - \bar M_P^2}\frac{1}{(q-k)^2 -m_0^2}
  \stackrel{d \to 4}{\longrightarrow} 
- \frac{q^2 - m_0^2 + \bar M_P^2}{q^2} \bar L + \bar I_P (q^2)~.
\eeq
This diagram gives the complete third order contribution.
From  the finite part, we only need the quark (meson) mass expansion at $q^2 = m_0^2$,
\beq
\bar I_P (m_0^2) = -\frac{1}{16\pi}\frac{\bar M_P}{m_0} 
-  \frac{1}{16\pi^2}\frac{\bar M_P^2}{2m_0^2}
- \frac{\bar M_P^2}{2m_0^2}\, \bar \mu_P + {\cal O} (p^3)~.
\eeq
The first term $\sim \bar M_P$ in this integral generates the leading non-analytic 
terms in the quark mass expansion of the baryon masses.
Similarly, the self-energy with one dimension two insertion (one meson and two nucleon
propagators) can be expressed in terms
of the loop function (note that we only need a special case of the corresponding 
function defined in \cite{BL})
\beq\label{I12bar}
\tilde{I}_P^{12} (q^2) = \int_I \frac{d^dk}{(2\pi)^d} 
\frac{i}{k^2 - \bar M_P^2}\frac{1}{[(q-k)^2 -m_0^2]^2}
  \stackrel{d \to 4}{\longrightarrow} 
- \frac{1}{q^2} \bar L + \bar{I}_P^{12} (q^2)~,
\eeq
and at $q^2 = m_0^2$ the finite piece can be expanded as
\beq
\bar I_P^{12} (m_0^2) =-\frac{1}{16\pi^2}\frac{1}{2m_0^2}
- \frac{1}{2m_0^2}\,\bar \mu_P + {\cal O} (p)~.
\eeq
The third order self-energy contributions and the fourth order tadpoles with 
insertions $\sim b_{0,D,F}$ have already been evaluated in \cite{ET} in the
isospin limit. In App.~\ref{app:renorm}, we discuss briefly the renormalization
of the various loop contributions and collect the pertinent $\beta$-functions. 

%%%%%%%% 4th order and chiral extrapolation %%%%%%%%%%%%%%%%%%%%%%%%%%%%%%%%%%%%%%%%%
\section{Chiral extrapolation functions}
\label{sec:extra}
%%%%%%%%%%%%%%%%%%%%%%%%%%%%%%%%%%%%%%%%%%%%%%%%%%%%%%%%%%%%%%%%%%%%%%%%%%%%%%%%%%%%%

This section contains the main part of this paper, namely the explicit expressions
for the baryon masses in terms of the quark masses. We express all masses in terms
of $\hat m$, $\ve$ and $m_s$ and utilize the chiral limit value of $B$ throughout. 
To arrive at these results,
we express the formulas for the baryon masses as functions of the meson masses
given in App.~\ref{sec:meson} in terms of the quark masses. We only retain the
linear terms in the parameter $\ve$ that parameterizes isospin violation. Note that
in the strict chiral counting, the second order terms from
the effective meson-baryon Lagrangian also generate terms $\sim (m_d-m_u)^2$ to 
the accuracy we are working here. Since these terms are numerically irrelevant, 
we do not display them here (although we have calculated them). We now give the
various terms proportional to $m_q$, $m_q^{3/2}$ and $m_q^2$ for the different 
baryons. The second order contributions $\sim m_q$  are
\beq
m_B^{(2)} = \left(\gamma_{1,B} + \tilde{\gamma}_{1,B} \frac{\ve}{\sqrt{3}} \right) \, B\, \hat m 
          + \left(\gamma_{2,B} + \tilde{\gamma}_{2,B} \frac{\ve}{\sqrt{3}} \right) \, B\, m_s~, 
\eeq
where the $\gamma_{i,B}$ are the isospin--symmetric coefficients defined via
\beq
\gamma_{i,\Sigma} \equiv \gamma_{i,\Sigma^+} =  \gamma_{i,\Sigma^-} = \gamma_{i,\Sigma^0}~,~~
\gamma_{i,N} \equiv \gamma_{i,p} =  \gamma_{i,n}~,~~
\gamma_{i,\Xi} \equiv \gamma_{i,\Xi^-} =  \gamma_{i,\Xi^0}~~(i=1,2)~.
\eeq
These  second order coefficients $\gamma_B$ are 
\beq
\begin{tabular}{ll}
$\gamma_{1,\Sigma}=-8\left(b_0+b_D\right)$~,\qquad\qquad\qquad& 
$\gamma_{2,\Sigma}=-4b_0 $~,\qquad\qquad\qquad\\
$\gamma_{1,N}=-8b_0-4\left(b_D+b_F\right)$~,&
$\gamma_{2,N}=4\left(-b_0-b_D+b_F\right)$ ~,\\
$\gamma_{1,\Xi}=-8b_0+4\left(-b_D+b_F\right)$~,&
$\gamma_{2,\Xi}=-4\left(b_0+b_D+b_F\right)$ ~,\\
$\gamma_{1,\Lambda}=-8b_0-\frac{8}{3}b_D$~,&
$\gamma_{2,\Lambda}=-4b_0-\frac{16}{3}b_D $~,\\
\end{tabular}
\eeq
in agreement with earlier calculations.
The corresponding coefficients of the isospin breaking corrections $\sim \ve$ are 
\beq
\begin{tabular}{ll}
$\tilde{\gamma}_{1,\Sigma^+}=-16b_F$~,&
$\tilde{\gamma}_{2,\Sigma^+}=16b_F$~,\\
$\tilde{\gamma}_{1,\Sigma^-}=16b_F$~,&
$\tilde{\gamma}_{2,\Sigma^-}=-16b_F$~,\\
%\tilde{\gamma}_{1,\Sigma^0}&=&\\
%\tilde{\gamma}_{2,\Sigma^0}&=&\\
$\tilde{\gamma}_{1,p}=-8\left(b_D+b_F\right)$~,&
$\tilde{\gamma}_{2,p}=8\left(b_D+b_F\right)$~,\\
$\tilde{\gamma}_{1,\Xi^-}=8\left(-b_D+b_F\right)$~,&
$\tilde{\gamma}_{2,\Xi^-}=8\left(b_D-b_F\right)$~,\\
$\tilde{\gamma}_{1,n}=8\left(b_D+b_F\right)$~,&
$\tilde{\gamma}_{2,n}=-8\left(b_D+b_F\right)$~,\\
$\tilde{\gamma}_{1,\Xi^0}=8\left(b_D-b_F\right)$~,&
$\tilde{\gamma}_{2,\Xi^0}=8\left(-b_D+b_F\right)$~.\\
%\tilde{\gamma}_{1,\Lambda}&=&\\
%\tilde{\gamma}_{2,\Lambda}&=&\\
\end{tabular}
\eeq
\noindent Note that for the $\Lambda$ and $\Sigma^0$ the isospin breaking effects (and in particular the non-diagonal contributions in Eqs.~(\ref{nondiag}))
are of order $(m_u-m_d)^2$.
The third order contributions $\sim m_q^{3/2}$ can be most compactly written as
\beqa
m_B^{(3)} &=& \left(\delta_{1,B} + \tilde{\delta}_{1,B} \frac{\ve}{\sqrt{3}} \right) 
              \, \frac{\sqrt{2}\, B^{3/2}}{4\pi F_\pi^2} \, \hat m^{3/2}
           +  \delta_{2,B} \, \frac{B^{3/2}}{4\pi F_\pi^2} \, (\hat m+m_s)^{3/2} \\
          &+& \left(\delta_{3,B} + \tilde{\delta}_{3,B} \frac{\ve}{\sqrt{3}} \right) 
              \, \frac{\sqrt{2}\, B^{3/2}}{4 \sqrt{3}\pi F_\pi^2} \, (\hat m + 2m_s)^{3/2}
           +  \tilde{\delta}_{4,B}\, \frac{\ve}{\sqrt{3}}
              \, \frac{B^{3/2}}{4\pi F_\pi^2} \, (\hat m-m_s)(\hat m+m_s)^{1/2}~, \nonumber
\eeqa
where again we have isospin--symmetric coefficients,
\beq
\delta_{i,\Sigma} \equiv \delta_{i,\Sigma^+} =  \delta_{i,\Sigma^-} = \delta_{i,\Sigma^0}~,~~
\delta_{i,N} \equiv \delta_{i,p} =  \delta_{i,n}~,~~
\delta_{i,\Xi} \equiv \delta_{i,\Xi^-} =  \delta_{i,\Xi^0}~~(i=1,2,3)~, 
\eeq
which are explicitly given by
%The third order coefficients $\delta_B$ are given by:
\beq
\begin{tabular}{ll}
$\delta_{1,\Sigma}=-\frac{1}{3}D^2-2F^2$  ~,&
$\delta_{2,\Sigma}=-\frac{1}{2}\left(D^2+F^2\right)$  ~,\\
$\delta_{3,\Sigma}=-\frac{1}{9}D^2$     ~,&
%\delta_{4,\Sigma}&=&     ~,\nonumber\\
%\delta_{5,\Sigma}&=&     ~,\nonumber\\
$\delta_{1,N}=-\frac{3}{4}D^2-\frac{3}{2}DF-\frac{3}{4}F^2$  ~,\\
$\delta_{2,N}=-\frac{5}{12}D^2+\frac{1}{2}DF-\frac{3}{4}F^2$  ~,&
$\delta_{3,N}=-\frac{1}{36}D^2+\frac{1}{6}DF-\frac{1}{4}F^2$  ~,\\
%\delta_{4,N}&=&  ~,\nonumber\\
%\delta_{5,N}&=&  ~,\nonumber\\
$\delta_{1,\Xi}=-\frac{3}{4}D^2+\frac{3}{2}DF-\frac{3}{4}F^2$  ~,&
$\delta_{2,\Xi}=-\frac{5}{12}D^2-\frac{1}{2}DF-\frac{3}{4}F^2$  ~,\\
$\delta_{3,\Xi}=-\frac{1}{36}D^2-\frac{1}{6}DF-\frac{1}{4}F^2$  ~,&
%\delta_{4,\Xi}&=&  ~,\nonumber\\
%\delta_{5,\Xi}&=&  ~,\nonumber\\
$\delta_{1,\Lambda}=-D^2$  ~,\\
$\delta_{2,\Lambda}=-\frac{1}{6}D^2-\frac{3}{2}F^2$  ~,&
$\delta_{3,\Lambda}=-\frac{1}{9}D^2 $ ~.\\
%\delta_{4,\Lambda}&=&  ~,\nonumber\\
%\delta_{5,\Lambda}&=&  ~,
\end{tabular}
\eeq
The coefficients of the third order isospin breaking $\ve$--corrections are
\beq
\begin{tabular}{ll}
$\tilde{\delta}_{1,\Sigma^+}=-2DF $  ~,&
%\tilde{\delta}_{2,\Sigma^+}&=&   ~,\nonumber\\
$\tilde{\delta}_{3,\Sigma^+}=\frac{2}{3}DF$   ~,\\
$\tilde{\delta}_{4,\Sigma^+}= -3DF$  ~,&
%\tilde{\delta}_{5,\Sigma^+}&=&   ~,\nonumber\\
$\tilde{\delta}_{1,\Sigma^-}=2DF$   ~,\\
%\tilde{\delta}_{2,\Sigma^-}&=&   ~,\nonumber\\
$\tilde{\delta}_{3,\Sigma^-}=-\frac{2}{3}DF$   ~,&
$\tilde{\delta}_{4,\Sigma^-}=3DF$   ~,\\
%\tilde{\delta}_{5,\Sigma^-}&=&   ~,\nonumber\\
%\tilde{\delta}_{1,\Sigma^0}&=&   ~,\nonumber\\
%\tilde{\delta}_{2,\Sigma^0}&=&   ~,\nonumber\\
%\tilde{\delta}_{3,\Sigma^0}&=&   ~,\nonumber\\
%\tilde{\delta}_{4,\Sigma^0}&=&   ~,\nonumber\\
%\tilde{\delta}_{5,\Sigma^0}&=&   ~,\nonumber\\
$\tilde{\delta}_{1,p}=\frac{1}{2}D^2-DF-\frac{3}{2}F^2$   ~,&
%\tilde{\delta}_{2,p}&=&   ~,\nonumber\\
$\tilde{\delta}_{3,p}=-\frac{1}{6}D^2+\frac{1}{3}DF+\frac{1}{2}F^2$   ~,\\
$\tilde{\delta}_{4,p}=\frac{1}{4}D^2-\frac{3}{2}DF-\frac{3}{4}F^2$   ~,&
%\tilde{\delta}_{5,p}&=&   ~,\nonumber\\
$\tilde{\delta}_{1,\Xi^-}=\frac{1}{2}D^2+DF-\frac{3}{2}F^2$   ~,\\
%\tilde{\delta}_{2,\Xi^-}&=&   ~,\nonumber\\
$\tilde{\delta}_{3,\Xi^-}=-\frac{1}{6}D^2-\frac{1}{3}DF+\frac{1}{2}F^2$   ~,&
$\tilde{\delta}_{4,\Xi^-}=\frac{1}{4}D^2+\frac{3}{2}DF-\frac{3}{4}F^2$   ~,\\
%\tilde{\delta}_{5,\Xi^-}&=&   ~,\nonumber\\
$\tilde{\delta}_{1,n}=-\frac{1}{2}D^2+DF+\frac{3}{2}F^2$   ~,&
%\tilde{\delta}_{2,n}&=&   ~,\nonumber\\
$\tilde{\delta}_{3,n}=\frac{1}{6}D^2-\frac{1}{3}DF-\frac{1}{2}F^2$   ~,\\
$\tilde{\delta}_{4,n}=-\frac{1}{4}D^2+\frac{3}{2}DF+\frac{3}{4}F^2$   ~,&
%\tilde{\delta}_{5,n}&=&   ~,\nonumber\\
$\tilde{\delta}_{1,\Xi^0}=-\frac{1}{2}D^2-DF+\frac{3}{2}F^2$   ~,\\
%\tilde{\delta}_{2,\Xi^0}&=&   ~,\nonumber\\
$\tilde{\delta}_{3,\Xi^0}=\frac{1}{6}D^2+\frac{1}{3}DF-\frac{1}{2}F^2$   ~,&
$\tilde{\delta}_{4,\Xi^0}=-\frac{1}{4}D^2-\frac{3}{2}DF+\frac{3}{4}F^2 $  ~.\\
%\tilde{\delta}_{5,\Xi^0}&=&   ~,\nonumber\\
%\tilde{\delta}_{1,\Lambda}&=&   ~,\nonumber\\
%\tilde{\delta}_{2,\Lambda}&=&   ~,\nonumber\\
%\tilde{\delta}_{3,\Lambda}&=&   ~,\nonumber\\
%\tilde{\delta}_{4,\Lambda}&=&   ~,\nonumber\\
%\tilde{\delta}_{5,\Lambda}&=&   ~,\nonumber\\
\end{tabular}
\eeq
Since these are the leading one-loop corrections, they only depend on the
lowest order axial coupling constants $D$ and $F$. The major result of this
investigation are the complete fourth order corrections $\sim m_q^2$. 
In terms of the quark masses, they can most conveniently be written as  
\beqa
m_B^{(4)} &=&\left(\epsilon_{1,B} + \tilde{\epsilon}_{1,B}\frac{\ve}{\sqrt{3}} \right) \,B^2\, \hat m^2 
           + \left(\epsilon_{2,B} + \tilde{\epsilon}_{2,B}\frac{\ve}{\sqrt{3}}\right) \, B^2\,  m_s^2
           + \left(\epsilon_{3,B} + \tilde{\epsilon}_{3,B}\frac{\ve}{\sqrt{3}} \right) \, B^2\, \hat m \, m_s     
           \nonumber\\
        &+&  \left(\epsilon_{4,B} + \tilde{\epsilon}_{4,B}\frac{\ve}{\sqrt{3}} \right) \, 
             \frac{B^2}{16\pi^2 F_\pi^2}\, \hat m^2 \, \ln \left(\frac{2B\,\hat m}{m_0^2}\right)
        \nonumber\\
        &+&  \left(\epsilon_{5,B} + \tilde{\epsilon}_{5,B}\frac{\ve}{\sqrt{3}} \right) \, 
             \frac{B^2}{16\pi^2 F_\pi^2}\, \hat m^2 \, \ln \left(\frac{B\,(\hat m+m_s)}{m_0^2}\right)
        \nonumber\\        
        &+&  \left(\epsilon_{6,B} + \tilde{\epsilon}_{6,B}\frac{\ve}{\sqrt{3}} \right) \, 
             \frac{B^2}{16\pi^2 F_\pi^2}\, \hat m^2 \, \ln \left(\frac{2B\,(\hat m+2m_s)}{3m_0^2}\right)
        \nonumber\\
%        &+&  \left(\epsilon_{7,B} + \tilde{\epsilon}_{7,B}\frac{\ve}{\sqrt{3}} \right) \, 
%             \frac{B^2}{16\pi^2 F_\pi^2}\,  m_s^2 \, \ln \left(\frac{2B\,\hat m}{m_0}\right)
%        \nonumber\\
        &+&  \left(\epsilon_{7,B} + \tilde{\epsilon}_{7,B}\frac{\ve}{\sqrt{3}} \right) \, 
             \frac{B^2}{16\pi^2 F_\pi^2}\,  m_s^2 \, \ln \left(\frac{B\,(\hat m+m_s)}{m_0^2}\right)
        \nonumber\\        
        &+&  \left(\epsilon_{8,B} + \tilde{\epsilon}_{8,B}\frac{\ve}{\sqrt{3}} \right) \, 
             \frac{B^2}{16\pi^2 F_\pi^2}\,  m_s^2 \, \ln \left(\frac{2B\,(\hat m+2m_s)}{3m_0^2}\right)
        \nonumber\\
        &+&  \left(\epsilon_{9,B} + \tilde{\epsilon}_{9,B}\frac{\ve}{\sqrt{3}} \right) \, 
             \frac{B^2}{16\pi^2 F_\pi^2}\, \hat m \, m_s \, \ln \left(\frac{2B\,\hat m}{m_0^2}\right)
        \nonumber\\
        &+&  \left(\epsilon_{10,B} + \tilde{\epsilon}_{10,B}\frac{\ve}{\sqrt{3}} \right) \, 
             \frac{B^2}{16\pi^2 F_\pi^2}\, \hat m \, m_s \, \ln \left(\frac{B\,(\hat m+m_s)}{m_0^2}\right)
        \nonumber\\        
        &+&  \left(\epsilon_{11,B} + \tilde{\epsilon}_{11,B}\frac{\ve}{\sqrt{3}} \right) \, 
             \frac{B^2}{16\pi^2 F_\pi^2}\, \hat m \, m_s \, \ln \left(\frac{2B\,(\hat m+2m_s)}{3m_0^2}\right)~,
\eeqa
where again we defined isospin-symmetric coefficients as before
\beq
\epsilon_{i,\Sigma} \equiv \epsilon_{i,\Sigma^+} =  \epsilon_{i,\Sigma^-} = \epsilon_{i,\Sigma^0}~,~~
\epsilon_{i,N} \equiv \epsilon_{i,p} =  \epsilon_{i,n}~,~~
\epsilon_{i,\Xi} \equiv \epsilon_{i,\Xi^-} =  \epsilon_{i,\Xi^0}~~(i=1,\ldots,11) ~.
\eeq
Note that to this order in $\ve$, there are no terms $\sim m_s^2 \ln \hat m$.
The largest contributions stem indeed from the terms of order $m_s^2$ and 
$m_s^2 \ln m_s$ at the physical quark masses, but in actual lattice simulations
the up and down quarks might have a similar mass as the strange quark. It is therefore
important to retain all these terms when connecting lattice results to the
physical world.

\medskip\noindent
The fourth order contributions $\epsilon_B$ read:
\beqa
\epsilon_{1,\Sigma}&=& \frac{1}{(4\pi)^2 F_\pi^2}\left(-\frac{67}{27}\frac{D^2}{m_0}
-9\frac{F^2}{m_0}+\left(\frac{136}{9}D^2+8F^2\right)b_D+16DFb_F
-\frac{9}{2}m_0b_5\right.\nonumber\\&-&\left.\frac{139}{54}m_0b_7
-\frac{37}{9}m_0b_8-m_0b_9\right)-64\left(d_3+d_6\right)-32d_7       ~,\nonumber\\
\epsilon_{2,\Sigma}&=& \frac{1}{(4\pi)^2 F_\pi^2}\left(-\frac{43}{27}\frac{D^2}{m_0}
-\frac{F^2}{m_0}-8\left(D^2+F^2\right)b_D-16DFb_F
-\frac{m_0}{2}b_5\right.\nonumber\\&-&\left.\frac{43}{54}m_0b_7-\frac{13}{9}m_0b_8\right)
-16\left(d_6+d_7\right)  ~,\nonumber\\
\epsilon_{3,\Sigma}&=&\frac{1}{(4\pi)^2 F_\pi^2}\left(-\frac{70}{27}\frac{D^2}{m_0}
-2\frac{F^2}{m_0}-\frac{64}{9}D^2b_D-m_0b_5-\frac{35}{27}m_0b_7-\frac{22}{9}m_0b_8\right)
-64d_6  ~,\nonumber\\
\epsilon_{4,\Sigma}&=& -\frac{4}{3}\frac{D^2}{m_0}-8\frac{F^2}{m_0}+24b_0
+\left(24+\frac{32}{3}D^2\right)b_D-32b_1-16b_3-24b_4
+8m_0b_5\nonumber\\&+&4m_0b_7+6m_0b_8+2m_0b_9~,\nonumber
\eeqa
\beqa
\epsilon_{5,\Sigma}&=& -\frac{D^2}{m_0}-\frac{F^2}{m_0}+8b_0
+\left(4+12\left(D^2+F^2\right)\right)b_D+24DFb_F-4\left(b_1+b_3\right)
-8b_4\nonumber\\&+&m_0\left(b_5+b_7\right)+2m_0b_8 ~,\nonumber\\
\epsilon_{6,\Sigma}&=& -\frac{4}{27}\frac{D^2}{m_0}+\frac{8}{9}\left(b_0+b_D\right)-\frac{16}{27}b_3-\frac{8}{9}b_4+\frac{4}{27}m_0b_7+\frac{2}{9}m_0b_8~, \nonumber\\
%\epsilon_{7,\Sigma}&=& ~,\nonumber\\
\epsilon_{7,\Sigma}&=& -\frac{D^2}{m_0}-\frac{F^2}{m_0}+8b_0+\left(4-12\left(D^2+F^2\right)\right)b_D-24DFb_F-4\left(b_1+b_3\right)-8b_4\nonumber\\&+&m_0\left(b_5+b_7\right)+2m_0b_8 ~,\nonumber\\
\epsilon_{8,\Sigma}&=&-\frac{16}{27}\frac{D^2}{m_0}+\frac{32}{9}b_0-\frac{64}{27}b_3-\frac{32}{9}b_4+\frac{16}{27}m_0b_7+\frac{8}{9}m_0b_8 ~,\nonumber
\eeqa
\beqa
\epsilon_{9,\Sigma}&=&-\frac{32}{3}D^2b_D ~,\nonumber\\
\epsilon_{10,\Sigma}&=& -2\left(\frac{D^2}{m_0}+\frac{F^2}{m_0}\right)+16b_0+8\left(b_D-b_1-b_3\right)-16b_4+2m_0\left(b_5+b_7\right)+4m_0b_8 ~,\nonumber\\
\epsilon_{11,\Sigma}&=&-\frac{16}{27}\frac{D^2}{m_0}+\frac{32}{9}b_0+\frac{16}{9}b_D-\frac{64}{27}b_3-\frac{32}{9}b_4+\frac{16}{27}m_0b_7+\frac{8}{9}m_0b_8 ~,
\eeqa
\beqa
\epsilon_{1,N}&=&   \frac{1}{(4\pi)^2 F_\pi^2}\left(-\frac{209}{54}\frac{D^2}{m_0}-\frac{43}{9}\frac{DF}{m_0}-\frac{29}{6}\frac{F^2}{m_0}+\left(-\frac{52}{9}D^2+\frac{40}{3}DF-4F^2\right)b_D\right.\nonumber\\&+&\left.\left(\frac{20}{3}D^2-8DF+12F^2\right)b_F-\frac{29}{12}m_0b_5-\frac{43}{36}m_0b_6-\frac{245}{108}m_0b_7-\frac{37}{9}m_0b_8-\frac{m_0}{4}b_9\right)\nonumber\\&-&16\left(d_1+d_2+d_3\right)-32d_5-64d_6-32d_7       ~,\nonumber\\
\epsilon_{2,N}&=&   \frac{1}{(4\pi)^2 F_\pi^2}\left(-\frac{53}{54}\frac{D^2}{m_0}+\frac{17}{9}\frac{DF}{m_0}-\frac{17}{6}\frac{F^2}{m_0}+\left(\frac{52}{9}D^2-\frac{40}{3}DF+4F^2\right)b_D\right.\nonumber\\&+&\left.\left(-\frac{20}{3}D^2+8DF-12F^2\right)b_F-\frac{17}{12}m_0b_5+\frac{17}{36}m_0b_6-\frac{89}{108}m_0b_7-\frac{13}{9}m_0b_8-\frac{m_0}{4}b_9\right)\nonumber\\&+&16\left(-d_1+d_2-d_3+d_5-d_6-d_7\right)  ~,\nonumber
\eeqa
\beqa
\epsilon_{3,N}&=&  \frac{1}{(4\pi)^2 F_\pi^2}\left(-\frac{49}{27}\frac{D^2}{m_0}+\frac{26}{9}\frac{DF}{m_0}-\frac{13}{3}\frac{F^2}{m_0}-\frac{13}{6}m_0b_5+\frac{13}{18}m_0b_6-\frac{85}{54}m_0b_7\right.\nonumber\\&-&\left.\frac{22}{9}m_0b_8-\frac{m_0}{2}b_9\right)+32\left(d_1-d_3\right)+16d_5-64d_6   ~,\nonumber\\
\epsilon_{4,N}&=&  -3\frac{D^2}{m_0}-6\frac{DF}{m_0}-3\frac{F^2}{m_0}+24b_0+12\left(b_D+b_F-b_1-b_2-b_3\right)-24b_4\nonumber\\&+&3m_0\left(b_5+b_6+b_7\right)+6m_0b_8   ~,\nonumber
\eeqa
\beqa
\epsilon_{5,N}&=&-\frac{5}{6}\frac{D^2}{m_0}+\frac{DF}{m_0}-\frac{3}{2}\frac{F^2}{m_0}+8b_0+\left(6-\frac{26}{3}D^2+20DF-6F^2\right)b_D\nonumber\\&+&\left(-2+10D^2-12DF+18F^2\right)b_F-6b_1+2b_2-6b_3-8b_4+\frac{3}{2}m_0b_5-\frac{m_0}{2}b_6\nonumber\\&+&\frac{3}{2}m_0b_7+2m_0b_8+\frac{m_0}{2}b_9 ~,\nonumber\\
\epsilon_{6,N}&=& -\frac{1}{27}\frac{D^2}{m_0}+\frac{2}{9}\frac{DF}{m_0}-\frac{1}{3}\frac{F^2}{m_0}+\frac{8}{9}b_0+\frac{4}{9}\left(b_D+b_F\right)-\frac{4}{3}b_1+\frac{4}{9}b_2-\frac{4}{27}b_3-\frac{8}{9}b_4\nonumber\\&+&\frac{m_0}{3}b_5-\frac{m_0}{9}b_6+\frac{m_0}{27}b_7+\frac{2}{9}m_0b_8 ~,\nonumber
%\epsilon_{7,N}&=& ~,\nonumber\\
\eeqa
\beqa
\epsilon_{7,N}&=&-\frac{5}{6}\frac{D^2}{m_0}+\frac{DF}{m_0}-\frac{3}{2}\frac{F^2}{m_0}+8b_0+\left(6+\frac{26}{3}D^2-20DF+6F^2\right)b_D\nonumber\\&+&\left(-2-10D^2+12DF-18F^2\right)b_F-6b_1+2b_2-6b_3-8b_4+\frac{3}{2}m_0b_5-\frac{m_0}{2}b_6\nonumber\\&+&\frac{3}{2}m_0b_7+2m_0b_8+\frac{m_0}{2}b_9 ~,\nonumber\\
\epsilon_{8,N}&=& -\frac{4}{27}\frac{D^2}{m_0}+\frac{8}{9}\frac{DF}{m_0}-\frac{4}{3}\frac{F^2}{m_0}+\frac{32}{9}\left(b_0+b_D-b_F\right)-\frac{16}{3}b_1+\frac{16}{9}b_2-\frac{16}{27}b_3-\frac{32}{9}b_4\nonumber\\&+&\frac{4}{3}m_0b_5-\frac{4}{9}m_0b_6+\frac{4}{27}m_0b_7+\frac{8}{9}m_0b_8 ~,\nonumber
%\epsilon_{9,N}&=& ~,\nonumber\\
\eeqa
\beqa
\epsilon_{10,N}&=& -\frac{5}{3}\frac{D^2}{m_0}+2\frac{DF}{m_0}-3\frac{F^2}{m_0}+16b_0+12b_D-4b_F-12b_1+4b_2-12b_3-16b_4\nonumber\\&+&3m_0b_5-m_0b_6+3m_0b_7+4m_0b_8 +m_0b_9 ~,\nonumber\\
\epsilon_{11,N}&=& -\frac{4}{27}\frac{D^2}{m_0}+\frac{8}{9}\frac{DF}{m_0}-\frac{4}{3}\frac{F^2}{m_0}+\frac{32}{9}b_0+\frac{8}{3}b_D-\frac{8}{9}b_F-\frac{16}{3}b_1+\frac{16}{9}b_2-\frac{16}{27}b_3-\frac{32}{9}b_4\nonumber\\&+&\frac{4}{3}m_0b_5-\frac{4}{9}m_0b_6+\frac{4}{27}m_0b_7+\frac{8}{9}m_0b_8   ~,
\eeqa
\beqa
\epsilon_{1,\Xi}&=&    \frac{1}{(4\pi)^2 F_\pi^2}\left(-\frac{209}{54}\frac{D^2}{m_0}+\frac{43}{9}\frac{DF}{m_0}-\frac{29}{6}\frac{F^2}{m_0}-\left(\frac{52}{9}D^2+\frac{40}{3}DF+4F^2\right)b_D\right.\nonumber\\&-&\left.\left(\frac{20}{3}D^2+8DF+12F^2\right)b_F-\frac{29}{12}m_0b_5+\frac{43}{36}m_0b_6-\frac{245}{108}m_0b_7-\frac{37}{9}m_0b_8-\frac{m_0}{4}b_9\right)\nonumber\\&+&16\left(-d_1+d_2-d_3\right)+32d_5-64d_6-32d_7       ~,\nonumber
\eeqa
\beqa
\epsilon_{2,\Xi}&=&  \frac{1}{(4\pi)^2 F_\pi^2}\left(-\frac{53}{54}\frac{D^2}{m_0}-\frac{17}{9}\frac{DF}{m_0}-\frac{17}{6}\frac{F^2}{m_0}+\left(\frac{52}{9}D^2+\frac{40}{3}DF+4F^2\right)b_D\right.\nonumber\\&+&\left.\left(\frac{20}{3}D^2+8DF+12F^2\right)b_F-\frac{17}{12}m_0b_5-\frac{17}{36}m_0b_6-\frac{89}{108}m_0b_7-\frac{13}{9}m_0b_8-\frac{m_0}{4}b_9\right)\nonumber\\&-&16\left(d_1+d_2+d_3+d_5+d_6+d_7\right) ~,\nonumber
\eeqa
\beqa
\epsilon_{3,\Xi}&=& \frac{1}{(4\pi)^2 F_\pi^2}\left(-\frac{49}{27}\frac{D^2}{m_0}-\frac{26}{9}\frac{DF}{m_0}-\frac{13}{3}\frac{F^2}{m_0}-\frac{13}{6}m_0b_5-\frac{13}{18}m_0b_6-\frac{85}{54}m_0b_7\right.\nonumber\\&-&\left.\frac{22}{9}m_0b_8-\frac{m_0}{2}b_9\right)+32\left(d_1-d_3\right)-16d_5-64d_6  ~,\nonumber\\
\epsilon_{4,\Xi}&=& -3\frac{D^2}{m_0}+6\frac{DF}{m_0}-3\frac{F^2}{m_0}+24b_0+12\left(b_D-b_F-b_1+b_2-b_3\right)-24b_4\nonumber\\&+&3m_0\left(b_5-b_6+b_7\right)+6m_0b_8  ~,\nonumber\\
\epsilon_{5,\Xi}&=&-\frac{5}{6}\frac{D^2}{m_0}-\frac{DF}{m_0}-\frac{3}{2}\frac{F^2}{m_0}+8b_0+\left(6-\frac{26}{3}D^2-20DF-6F^2\right)b_D\nonumber\\&+&\left(2-10D^2-12DF-18F^2\right)b_F-6b_1-2b_2-6b_3-8b_4+\frac{3}{2}m_0b_5+\frac{m_0}{2}b_6\nonumber\\&+&\frac{3}{2}m_0b_7+2m_0b_8+\frac{m_0}{2}b_9 ~,\nonumber
\eeqa
\beqa
\epsilon_{6,\Xi}&=& -\frac{1}{27}\frac{D^2}{m_0}-\frac{2}{9}\frac{DF}{m_0}-\frac{1}{3}\frac{F^2}{m_0}+\frac{8}{9}b_0+\frac{4}{9}\left(b_D-b_F\right)-\frac{4}{3}b_1-\frac{4}{9}b_2-\frac{4}{27}b_3-\frac{8}{9}b_4\nonumber\\&+&\frac{m_0}{3}b_5+\frac{m_0}{9}b_6+\frac{m_0}{27}b_7+\frac{2}{9}m_0b_8 ~,\nonumber\\
%\epsilon_{7,\Xi}&=& ~,\nonumber\\
\epsilon_{7,\Xi}&=&-\frac{5}{6}\frac{D^2}{m_0}-\frac{DF}{m_0}-\frac{3}{2}\frac{F^2}{m_0}+8b_0+\left(6+\frac{26}{3}D^2+20DF+6F^2\right)b_D\nonumber\\&+&\left(2+10D^2+12DF+18F^2\right)b_F-6b_1-2b_2-6b_3-8b_4+\frac{3}{2}m_0b_5+\frac{m_0}{2}b_6\nonumber\\&+&\frac{3}{2}m_0b_7+2m_0b_8+\frac{m_0}{2}b_9 ~,\nonumber
\eeqa
\beqa
\epsilon_{8,\Xi}&=&-\frac{4}{27}\frac{D^2}{m_0}-\frac{8}{9}\frac{DF}{m_0}-\frac{4}{3}\frac{F^2}{m_0}+\frac{32}{9}\left(b_0+b_D+b_F\right)-\frac{16}{3}b_1-\frac{16}{9}b_2-\frac{16}{27}b_3-\frac{32}{9}b_4\nonumber\\&+&\frac{4}{3}m_0b_5+\frac{4}{9}m_0b_6+\frac{4}{27}m_0b_7+\frac{8}{9}m_0b_8 ~,\nonumber\\
%\epsilon_{9,\Xi}&=& ~,\nonumber\\
\epsilon_{10,\Xi}&=& -\frac{5}{3}\frac{D^2}{m_0}-2\frac{DF}{m_0}-3\frac{F^2}{m_0}+16b_0+12b_D+4b_F-12b_1-4b_2-12b_3-16b_4\nonumber\\&+&3m_0b_5+m_0b_6+3m_0b_7+4m_0b_8 +m_0b_9 ~,\nonumber\\
\epsilon_{11,\Xi}&=& -\frac{4}{27}\frac{D^2}{m_0}-\frac{8}{9}\frac{DF}{m_0}-\frac{4}{3}\frac{F^2}{m_0}+\frac{32}{9}b_0+\frac{8}{3}b_D+\frac{8}{9}b_F-\frac{16}{3}b_1-\frac{16}{9}b_2-\frac{16}{27}b_3-\frac{32}{9}b_4\nonumber\\&+&\frac{4}{3}m_0b_5+\frac{4}{9}m_0b_6+\frac{4}{27}m_0b_7+\frac{8}{9}m_0b_8  ~,
\eeqa
\beqa
\epsilon_{1,\Lambda}&=&  \frac{1}{(4\pi)^2 F_\pi^2}\left(-\frac{121}{27}\frac{D^2}{m_0}-3\frac{F^2}{m_0}-\left(\frac{200}{9}D^2+8F^2\right)b_D-16DFb_F-\frac{3}{2}m_0b_5\right.\nonumber\\&-&\left.\frac{43}{18}m_0b_7-\frac{37}{9}m_0b_8-\frac{m_0}{9}b_9\right)-\frac{64}{3}d_3-\frac{32}{3}d_4-64d_6-32d_7        ~,\nonumber
\\
\epsilon_{2,\Lambda}&=&\frac{1}{(4\pi)^2 F_\pi^2}\left(-\frac{25}{27}\frac{D^2}{m_0}-3\frac{F^2}{m_0}+\left(\frac{8}{9}D^2+8F^2\right)b_D+16DFb_F-\frac{3}{2}m_0b_5\right.\nonumber\\&-&\left.\frac{19}{18}m_0b_7-\frac{13}{9}m_0b_8-\frac{4}{9}m_0b_9\right)-\frac{128}{3}d_3-\frac{32}{3}d_4-16\left(d_6+d_7\right)  ~,\nonumber\\
\epsilon_{3,\Lambda}&=&\frac{1}{(4\pi)^2 F_\pi^2}\left(-\frac{34}{27}\frac{D^2}{m_0}-6\frac{F^2}{m_0}+\frac{64}{3}D^2b_D-3m_0b_5-\frac{11}{9}m_0b_7-\frac{22}{9}m_0b_8-\frac{4}{9}m_0b_9\right)\nonumber\\&+&\frac{64}{3}d_4-64d_6  ~,\nonumber\\
\epsilon_{4,\Lambda}&=& -4\frac{D^2}{m_0}+24b_0+\left(8-32D^2\right)b_D-16b_3-24b_4+4m_0b_7+6m_0b_8 ~,\nonumber\\
\epsilon_{5,\Lambda}&=& -\frac{1}{3}\frac{D^2}{m_0}-3\frac{F^2}{m_0}+8b_0+\left(\frac{20}{3}-\frac{4}{3}D^2-12F^2\right)b_D-24DFb_F-12b_1-\frac{4}{3}b_3-8b_4\nonumber\\&+&3m_0b_5+\frac{m_0}{3}b_7+2m_0b_8  ~,\nonumber\\
\epsilon_{6,\Lambda}&=& -\frac{4}{27}\frac{D^2}{m_0}+\frac{8}{9}b_0+\frac{8}{27}b_D-\frac{16}{9}b_3-\frac{8}{9}b_4+\frac{4}{9}m_0b_7+\frac{2}{9}m_0\left(b_8+b_9\right)~, \nonumber\\
%\epsilon_{7,\Lambda}&=& ~,\nonumber\\
\epsilon_{7,\Lambda}&=& -\frac{1}{3}\frac{D^2}{m_0}-3\frac{F^2}{m_0}+8b_0+\left(\frac{20}{3}+\frac{4}{3}D^2+12F^2\right)b_D+24DFb_F-12b_1-\frac{4}{3}b_3-8b_4\nonumber\\&+&3m_0b_5+\frac{m_0}{3}b_7+2m_0b_8 ~,\nonumber\\
\epsilon_{8,\Lambda}&=&-\frac{16}{27}\frac{D^2}{m_0}+\frac{32}{9}b_0+\frac{128}{27}b_D-\frac{64}{9}b_3-\frac{32}{9}b_4+\frac{16}{9}m_0b_7+\frac{8}{9}m_0\left(b_8+b_9\right) ~,\nonumber\\
\epsilon_{9,\Lambda}&=&32D^2b_D ~,\nonumber\\
\epsilon_{10,\Lambda}&=& -\frac{2}{3}\frac{D^2}{m_0}-6\frac{F^2}{m_0}+16b_0+\frac{40}{3}b_D-24b_1-\frac{8}{3}b_3-16b_4+6m_0b_5+\frac{2}{3}m_0b_7+4m_0b_8  ~,\nonumber\\
\epsilon_{11,\Lambda}&=&-\frac{16}{27}\frac{D^2}{m_0}+\frac{32}{9}b_0+\frac{80}{27}b_D-\frac{64}{9}b_3-\frac{32}{9}b_4+\frac{16}{9}m_0b_7+\frac{8}{9}m_0\left(b_8+b_9\right) ~.
\eeqa

\noindent
The corresponding $\ve$-coefficients are given by
\beqa
\tilde{\epsilon}_{1,\Sigma^+}&=&  \frac{1}{(4\pi)^2 F_\pi^2}\left(-\frac{172}{9}\frac{DF}{m_0}+\frac{464}{3}DFb_D+\left(8+\frac{232}{3}D^2+120F^2\right)b_F\right.\nonumber\\&-&\left.8b_2-\frac{16}{9}m_0b_6\right)-128\left(d_2+d_5\right)        ~,\nonumber\\
\tilde{\epsilon}_{2,\Sigma^+}&=&  \frac{1}{(4\pi)^2 F_\pi^2}\left(\frac{140}{9}\frac{DF}{m_0}+48DFb_D+\left(-8+24\left(D^2+F^2\right)\right)b_F+8b_2\right.\nonumber\\&+&\left.\frac{8}{9}m_0b_6\right)+64d_5       ~,\nonumber\\
\tilde{\epsilon}_{3,\Sigma^+}&=&  \frac{1}{(4\pi)^2 F_\pi^2}\left(\frac{32}{9}\frac{DF}{m_0}-\frac{608}{3}DFb_D-\left(\frac{304}{3}D^2+144F^2\right)b_F+\frac{8}{9}m_0b_6\right)\nonumber\\&+&128d_2+64d_5        ~,\nonumber
\eeqa
\beqa
\tilde{\epsilon}_{4,\Sigma^+}&=&   -8\frac{DF}{m_0}+64DFb_D+\left(32\left(1+D^2\right)+96F^2\right)b_F-16b_2+4m_0b_6           ~,\nonumber\\
\tilde{\epsilon}_{5,\Sigma^+}&=&  -8\frac{DF}{m_0}+96DFb_D+\left(16+48\left(D^2+F^2\right)\right)b_F-16b_2+4m_0b_6           ~,\nonumber\\
\tilde{\epsilon}_{6,\Sigma^+}&=&   \frac{8}{9}\frac{DF}{m_0}-\frac{32}{9}b_F+\frac{16}{9}b_2-\frac{4}{9}m_0b_6            ~,\nonumber
%\tilde{\epsilon}_{7,\Sigma^+}&=&     ~,\nonumber\\
\eeqa
\beqa
\tilde{\epsilon}_{7,\Sigma^+}&=&   8\frac{DF}{m_0}+16\left(-b_F+b_2\right)-4m_0b_6~,             
 \phantom{8\frac{DF}{m_0}+16\left(-b_F+b_2\right)-4m_0b_6}
\nonumber\\
\tilde{\epsilon}_{8,\Sigma^+}&=&    \frac{32}{9}\left(\frac{DF}{m_0}-b_F\right)+\frac{64}{9}b_2-\frac{16}{9}m_0b_6   ~,\nonumber\\
\tilde{\epsilon}_{9,\Sigma^+}&=&-64DFb_D -\left(16+32D^2+96F^2\right)b_F            ~,\nonumber\\
\tilde{\epsilon}_{10,\Sigma^+}&=&  -96DFb_D -48\left(D^2+F^2\right)b_F    ~,\nonumber\\
\tilde{\epsilon}_{11,\Sigma^+}&=&   \frac{32}{9}\frac{DF}{m_0}-\frac{80}{9}b_F+\frac{64}{9}b_2-\frac{16}{9}m_0b_6    ~,
\eeqa
\beqa
\tilde{\epsilon}_{1,\Sigma^-}&=&  \frac{1}{(4\pi)^2 F_\pi^2}\left(\frac{172}{9}\frac{DF}{m_0}-\frac{464}{3}DFb_D-\left(8+\frac{232}{3}D^2+120F^2\right)b_F\right.\nonumber\\&+&\left.8b_2+\frac{16}{9}m_0b_6\right)+128\left(d_2+d_5\right)        ~,\nonumber\\
\tilde{\epsilon}_{2,\Sigma^-}&=&  \frac{1}{(4\pi)^2 F_\pi^2}\left(-\frac{140}{9}\frac{DF}{m_0}-48DFb_D+\left(8-24\left(D^2+F^2\right)\right)b_F-8b_2\right.\nonumber\\&-&\left.\frac{8}{9}m_0b_6\right)-64d_5       ~,\nonumber\\
\tilde{\epsilon}_{3,\Sigma^-}&=&  \frac{1}{(4\pi)^2 F_\pi^2}\left(-\frac{32}{9}\frac{DF}{m_0}+\frac{608}{3}DFb_D+\left(\frac{304}{3}D^2+144F^2\right)b_F-\frac{8}{9}m_0b_6\right)\nonumber\\&-&128d_2-64d_5        ~,\nonumber\\
\tilde{\epsilon}_{4,\Sigma^-}&=&   8\frac{DF}{m_0}-64DFb_D-\left(32\left(1+D^2\right)+96F^2\right)b_F+16b_2-4m_0b_6           ~,\nonumber%\\
\eeqa
\beqa
\tilde{\epsilon}_{5,\Sigma^-}&=&  8\frac{DF}{m_0}-96DFb_D-\left(16+48\left(D^2+F^2\right)\right)b_F+16b_2-4m_0b_6           ~,\nonumber\\
\tilde{\epsilon}_{6,\Sigma^-}&=&   -\frac{8}{9}\frac{DF}{m_0}+\frac{32}{9}b_F-\frac{16}{9}b_2+\frac{4}{9}m_0b_6            ~,\nonumber\\
%\tilde{\epsilon}_{7,\Sigma^-}&=&     ~,\nonumber\\
\tilde{\epsilon}_{7,\Sigma^-}&=&  - 8\frac{DF}{m_0}+16\left(b_F-b_2\right)+4m_0b_6             ~,\nonumber\\
\tilde{\epsilon}_{8,\Sigma^-}&=&    \frac{32}{9}\left(-\frac{DF}{m_0}+b_F\right)-\frac{64}{9}b_2+\frac{16}{9}m_0b_6   ~,\nonumber%\\
\eeqa
\beqa
\tilde{\epsilon}_{9,\Sigma^-}&=&64DFb_D +\left(16+32D^2+96F^2\right)b_F            ~,\nonumber\\
\tilde{\epsilon}_{10,\Sigma^-}&=&  96DFb_D +48\left(D^2+F^2\right)b_F    ~,\nonumber\\
\tilde{\epsilon}_{11,\Sigma^-}&=&  - \frac{32}{9}\frac{DF}{m_0}+\frac{80}{9}b_F-\frac{64}{9}b_2+\frac{16}{9}m_0b_6    ~,
\eeqa
\beqa
\tilde{\epsilon}_{1,p}&=&   \frac{1}{(4\pi)^2 F_\pi^2}\left(\frac{25}{9}\frac{D^2}{m_0}-\frac{86}{9}\frac{DF}{m_0}-\frac{25}{3}\frac{F^2}{m_0}+\left(4+\frac{556}{9}D^2+\frac{136}{3}DF+60F^2\right)b_D\right.\nonumber\\&+&\left.\left(4+\frac{68}{3}D^2+120DF+60F^2\right)b_F-4\left(b_1+b_2+b_3\right)-\frac{8}{3}m_0b_5\right.\nonumber\\&+&\left.\frac{8}{9}m_0\left(-b_6+b_7\right)\right)-64\left(d_1+d_2+d_3+d_5\right)     ~,\nonumber\\
\tilde{\epsilon}_{2,p}&=&   \frac{1}{(4\pi)^2 F_\pi^2}\left(-\frac{17}{9}\frac{D^2}{m_0}+\frac{70}{9}\frac{DF}{m_0}+\frac{17}{3}\frac{F^2}{m_0}+\left(-4-\frac{68}{9}D^2-8DF+12F^2\right)b_D\right.\nonumber\\&+&\left.\left(-4\left(1+D^2\right)+24DF+12F^2\right)b_F+4\left(b_1+b_2+b_3\right)+\frac{4}{3}m_0b_5\right.\nonumber\\&+&\left.\frac{4}{9}m_0\left(b_6-b_7\right)\right)+64\left(-d_1+d_3\right)+32d_5     ~,\nonumber%\\
\eeqa
\beqa
\tilde{\epsilon}_{3,p}&=&   \frac{1}{(4\pi)^2 F_\pi^2}\left(-\frac{8}{9}\frac{D^2}{m_0}+\frac{16}{9}\frac{DF}{m_0}+\frac{8}{3}\frac{F^2}{m_0}-\left(\frac{488}{9}D^2+\frac{112}{3}DF+72F^2\right)b_D\right.\nonumber\\&-&\left.\left(\frac{56}{3}D^2+144DF+72F^2\right)b_F+\frac{4}{3}m_0b_5+\frac{4}{9}m_0\left(b_6-b_7\right)\right)+128d_1\nonumber\\&+&64d_2+32d_5   ~,\nonumber\\
\tilde{\epsilon}_{4,p}&=&  2\frac{D^2}{m_0}-4\frac{DF}{m_0}-6\frac{F^2}{m_0}+\left(16+48D^2+96DF+48F^2\right)\left(b_D+b_F\right)-24b_1\nonumber\\&+&8\left(-b_2+b_3\right)+6m_0b_5+2m_0\left(b_6-b_7\right)   ~,\nonumber%\\
\eeqa
\beqa
\tilde{\epsilon}_{5,p}&=&    \frac{2}{3}\frac{D^2}{m_0}-4\frac{DF}{m_0}-2\frac{F^2}{m_0}+\left(8+\frac{104}{3}D^2-16DF+24F^2\right)b_D\nonumber\\&+&\left(8\left(1-D^2\right)+48DF+24F^2\right)b_F-8\left(b_1+b_2+b_3\right)+2m_0\left(b_5+b_6+b_7+b_9\right)   ~,\nonumber\\
\tilde{\epsilon}_{6,p}&=&   - \frac{2}{9}\frac{D^2}{m_0}+\frac{4}{9}\frac{DF}{m_0}+\frac{2}{3}\frac{F^2}{m_0}-\frac{16}{9}\left(b_D+b_F\right)+\frac{8}{3}b_1+\frac{8}{9}\left(b_2-b_3\right)-\frac{2}{3}m_0b_5\nonumber\\&+&\frac{2}{9}m_0\left(-b_6+b_7\right)   ~,\nonumber\\
%\tilde{\epsilon}_{7,p}&=&     ~,\nonumber\\
\tilde{\epsilon}_{7,p}&=&   - \frac{2}{3}\frac{D^2}{m_0}+4\frac{DF}{m_0}+2\frac{F^2}{m_0}-\left(8+\frac{64}{3}D^2\right)b_D+8\left(-b_F+b_1+b_2+b_3\right)\nonumber\\&-&2m_0\left(b_5+b_6+b_7+b_9\right)    ~,\nonumber%\\
\eeqa
\beqa
\tilde{\epsilon}_{8,p}&=&   - \frac{8}{9}\frac{D^2}{m_0}+\frac{16}{9}\frac{DF}{m_0}+\frac{8}{3}\frac{F^2}{m_0}-\frac{16}{9}\left(b_D+b_F\right)+\frac{32}{3}b_1+\frac{32}{9}\left(b_2-b_3\right)-\frac{8}{3}m_0b_5\nonumber\\&+&\frac{8}{9}m_0\left(-b_6+b_7\right)       ~,\nonumber\\
\tilde{\epsilon}_{9,p}&=& -\left(8+48D^2+96DF+48F^2\right)\left(b_D+b_F\right)    ~,\nonumber\\
\tilde{\epsilon}_{10,p}&=&  \left(-\frac{40}{3}D^2+16DF-24F^2\right)b_D+\left(8D^2-48DF-24F^2\right)b_F   ~,\nonumber\\
\tilde{\epsilon}_{11,p}&=&   - \frac{8}{9}\frac{D^2}{m_0}+\frac{16}{9}\frac{DF}{m_0}+\frac{8}{3}\frac{F^2}{m_0}-\frac{40}{9}\left(b_D+b_F\right)+\frac{32}{3}b_1+\frac{32}{9}\left(b_2-b_3\right)-\frac{8}{3}m_0b_5\nonumber\\&+&\frac{8}{9}m_0\left(-b_6+b_7\right)     ~,
\eeqa
\beqa
\tilde{\epsilon}_{1,\Xi^-}&=&   \frac{1}{(4\pi)^2 F_\pi^2}\left(\frac{25}{9}\frac{D^2}{m_0}+\frac{86}{9}\frac{DF}{m_0}-\frac{25}{3}\frac{F^2}{m_0}+\left(4+\frac{556}{9}D^2-\frac{136}{3}DF+60F^2\right)b_D\right.\nonumber\\&+&\left.\left(-4-\frac{68}{3}D^2+120DF-60F^2\right)b_F+4\left(-b_1+b_2-b_3\right)-\frac{8}{3}m_0b_5\right.\nonumber\\&+&\left.\frac{8}{9}m_0\left(b_6+b_7\right)\right)+64\left(-d_1+d_2-d_3+d_5\right)     ~,\nonumber\\
\tilde{\epsilon}_{2,\Xi^-}&=&   \frac{1}{(4\pi)^2 F_\pi^2}\left(-\frac{17}{9}\frac{D^2}{m_0}-\frac{70}{9}\frac{DF}{m_0}+\frac{17}{3}\frac{F^2}{m_0}+\left(-4-\frac{68}{9}D^2+8DF+12F^2\right)b_D\right.\nonumber\\&+&\left.\left(4\left(1+D^2\right)+24DF-12F^2\right)b_F+4\left(b_1-b_2+b_3\right)+\frac{4}{3}m_0b_5\right.\nonumber\\&-&\left.\frac{4}{9}m_0\left(b_6+b_7\right)\right)+64\left(-d_1+d_3\right)-32d_5     ~,\nonumber\\
\tilde{\epsilon}_{3,\Xi^-}&=&   \frac{1}{(4\pi)^2 F_\pi^2}\left(-\frac{8}{9}\frac{D^2}{m_0}-\frac{16}{9}\frac{DF}{m_0}+\frac{8}{3}\frac{F^2}{m_0}+\left(-\frac{488}{9}D^2+\frac{112}{3}DF-72F^2\right)b_D\right.\nonumber\\&+&\left.\left(\frac{56}{3}D^2-144DF+72F^2\right)b_F+\frac{4}{3}m_0b_5-\frac{4}{9}m_0\left(b_6+b_7\right)\right)+128d_1\nonumber\\&-&64d_2-32d_5   ~,\nonumber\\
\tilde{\epsilon}_{4,\Xi^-}&=&  2\frac{D^2}{m_0}+4\frac{DF}{m_0}-6\frac{F^2}{m_0}+\left(16+48D^2-96DF+48F^2\right)\left(b_D-b_F\right)-24b_1\nonumber\\&+&8\left(b_2+b_3\right)+6m_0b_5-2m_0\left(b_6+b_7\right)   ~,\nonumber%\\
\eeqa
\beqa
\tilde{\epsilon}_{5,\Xi^-}&=&    \frac{2}{3}\frac{D^2}{m_0}+4\frac{DF}{m_0}-2\frac{F^2}{m_0}+\left(8+\frac{104}{3}D^2+16DF+24F^2\right)b_D\nonumber\\&+&\left(8\left(-1+D^2\right)+48DF-24F^2\right)b_F+8\left(-b_1+b_2-b_3\right)+2m_0\left(b_5-b_6+b_7+b_9\right)   ~,\nonumber\\
\tilde{\epsilon}_{6,\Xi^-}&=&   - \frac{2}{9}\frac{D^2}{m_0}-\frac{4}{9}\frac{DF}{m_0}+\frac{2}{3}\frac{F^2}{m_0}+\frac{16}{9}\left(-b_D+b_F\right)+\frac{8}{3}b_1-\frac{8}{9}\left(b_2+b_3\right)-\frac{2}{3}m_0b_5\nonumber\\&+&\frac{2}{9}m_0\left(b_6+b_7\right)   ~,\nonumber\\
%\tilde{\epsilon}_{7,\Xi^-}&=&     ~,\nonumber\\
\tilde{\epsilon}_{7,\Xi^-}&=&   - \frac{2}{3}\frac{D^2}{m_0}-4\frac{DF}{m_0}+2\frac{F^2}{m_0}-\left(8+\frac{64}{3}D^2\right)b_D+8\left(b_F+b_1-b_2+b_3\right)\nonumber\\&+&2m_0\left(-b_5+b_6-b_7-b_9\right)    ~,\nonumber\\
\tilde{\epsilon}_{8,\Xi^-}&=&   - \frac{8}{9}\frac{D^2}{m_0}-\frac{16}{9}\frac{DF}{m_0}+\frac{8}{3}\frac{F^2}{m_0}+\frac{16}{9}\left(-b_D+b_F\right)+\frac{32}{3}b_1-\frac{32}{9}\left(b_2+b_3\right)-\frac{8}{3}m_0b_5\nonumber\\&+&\frac{8}{9}m_0\left(b_6+b_7\right)       ~,\nonumber%\\
\eeqa
\beqa
\tilde{\epsilon}_{9,\Xi^-}&=& \left(-8-48D^2+96DF-48F^2\right)\left(b_D-b_F\right)    ~,\nonumber\\
\tilde{\epsilon}_{10,\Xi^-}&=& - \left(\frac{40}{3}D^2+16DF+24F^2\right)b_D+\left(-8D^2-48DF+24F^2\right)b_F   ~,\nonumber\\
\tilde{\epsilon}_{11,\Xi^-}&=&   - \frac{8}{9}\frac{D^2}{m_0}-\frac{16}{9}\frac{DF}{m_0}+\frac{8}{3}\frac{F^2}{m_0}+\frac{40}{9}\left(-b_D+b_F\right)+\frac{32}{3}b_1-\frac{32}{9}\left(b_2+b_3\right)-\frac{8}{3}m_0b_5\nonumber\\&+&\frac{8}{9}m_0\left(b_6+b_7\right)     ~,
\eeqa
\beqa
\tilde{\epsilon}_{1,n}&=&   \frac{1}{(4\pi)^2 F_\pi^2}\left(-\frac{25}{9}\frac{D^2}{m_0}+\frac{86}{9}\frac{DF}{m_0}+\frac{25}{3}\frac{F^2}{m_0}-\left(4+\frac{556}{9}D^2+\frac{136}{3}DF+60F^2\right)b_D\right.\nonumber\\&-&\left.\left(4+\frac{68}{3}D^2+120DF+60F^2\right)b_F+4\left(b_1+b_2+b_3\right)+\frac{8}{3}m_0b_5\right.\nonumber\\&+&\left.\frac{8}{9}m_0\left(b_6-b_7\right)\right)+64\left(d_1+d_2+d_3+d_5\right)     ~,\nonumber\\
\tilde{\epsilon}_{2,n}&=&   \frac{1}{(4\pi)^2 F_\pi^2}\left(\frac{17}{9}\frac{D^2}{m_0}-\frac{70}{9}\frac{DF}{m_0}-\frac{17}{3}\frac{F^2}{m_0}+\left(4+\frac{68}{9}D^2+8DF-12F^2\right)b_D\right.\nonumber\\&+&\left.\left(4\left(1+D^2\right)-24DF-12F^2\right)b_F-4\left(b_1+b_2+b_3\right)-\frac{4}{3}m_0b_5\right.\nonumber\\&+&\left.\frac{4}{9}m_0\left(-b_6+b_7\right)\right)+64\left(d_1-d_3\right)-32d_5     ~,\nonumber%\\
\eeqa
\beqa
\tilde{\epsilon}_{3,n}&=&   \frac{1}{(4\pi)^2 F_\pi^2}\left(\frac{8}{9}\frac{D^2}{m_0}-\frac{16}{9}\frac{DF}{m_0}-\frac{8}{3}\frac{F^2}{m_0}+\left(\frac{488}{9}D^2+\frac{112}{3}DF+72F^2\right)b_D\right.\nonumber\\&+&\left.\left(\frac{56}{3}D^2+144DF+72F^2\right)b_F-\frac{4}{3}m_0b_5+\frac{4}{9}m_0\left(-b_6+b_7\right)\right)-128d_1\nonumber\\&-&64d_2-32d_5   ~,\nonumber\\
\tilde{\epsilon}_{4,n}&=& - 2\frac{D^2}{m_0}+4\frac{DF}{m_0}+6\frac{F^2}{m_0}-\left(16+48D^2+96DF+48F^2\right)\left(b_D+b_F\right)+24b_1\nonumber\\&+&8\left(b_2-b_3\right)-6m_0b_5+2m_0\left(-b_6+b_7\right)   ~,\nonumber%\\
\eeqa
\beqa
\tilde{\epsilon}_{5,n}&=&   - \frac{2}{3}\frac{D^2}{m_0}+4\frac{DF}{m_0}+2\frac{F^2}{m_0}+\left(-8-\frac{104}{3}D^2+16DF-24F^2\right)b_D\nonumber\\&+&\left(8\left(-1+D^2\right)-48DF-24F^2\right)b_F+8\left(b_1+b_2+b_3\right)-2m_0\left(b_5+b_6+b_7+b_9\right)   ~,\nonumber\\
\tilde{\epsilon}_{6,n}&=&    \frac{2}{9}\frac{D^2}{m_0}-\frac{4}{9}\frac{DF}{m_0}-\frac{2}{3}\frac{F^2}{m_0}+\frac{16}{9}\left(b_D+b_F\right)-\frac{8}{3}b_1+\frac{8}{9}\left(-b_2+b_3\right)+\frac{2}{3}m_0b_5\nonumber\\&+&\frac{2}{9}m_0\left(b_6-b_7\right)   ~,\nonumber\\
%\tilde{\epsilon}_{7,n}&=&     ~,\nonumber\\
\tilde{\epsilon}_{7,n}&=&    \frac{2}{3}\frac{D^2}{m_0}-4\frac{DF}{m_0}-2\frac{F^2}{m_0}+\left(8+\frac{64}{3}D^2\right)b_D+8\left(b_F-b_1-b_2-b_3\right)\nonumber\\&+&2m_0\left(b_5+b_6+b_7+b_9\right)    ~,\nonumber%\\
\eeqa
\beqa
\tilde{\epsilon}_{8,n}&=&    \frac{8}{9}\frac{D^2}{m_0}-\frac{16}{9}\frac{DF}{m_0}
-\frac{8}{3}\frac{F^2}{m_0}+\frac{16}{9}\left(b_D+b_F\right)-\frac{32}{3}b_1
+\frac{32}{9}\left(-b_2+b_3\right)+\frac{8}{3}m_0b_5\nonumber\\&+&
\frac{8}{9}m_0\left(b_6-b_7\right)       ~,\nonumber\\
\tilde{\epsilon}_{9,n}&=& \left(8+48D^2+96DF+48F^2\right)\left(b_D+b_F\right)    ~,\nonumber\\
\tilde{\epsilon}_{10,n}&=&  \left(\frac{40}{3}D^2-16DF+24F^2\right)b_D
+\left(-8D^2+48DF+24F^2\right)b_F   ~,\nonumber\\
\tilde{\epsilon}_{11,n}&=&    \frac{8}{9}\frac{D^2}{m_0}-\frac{16}{9}\frac{DF}{m_0}
-\frac{8}{3}\frac{F^2}{m_0}+\frac{40}{9}\left(b_D+b_F\right)-\frac{32}{3}b_1
+\frac{32}{9}\left(-b_2+b_3\right)+\frac{8}{3}m_0b_5\nonumber\\&+&\frac{8}{9}m_0\left(b_6-b_7\right)     ~,
\eeqa
\beqa
\tilde{\epsilon}_{1,\Xi^0}&=&   \frac{1}{(4\pi)^2 F_\pi^2}\left(-\frac{25}{9}\frac{D^2}{m_0}
-\frac{86}{9}\frac{DF}{m_0}+\frac{25}{3}\frac{F^2}{m_0}+\left(-4-\frac{556}{9}D^2
+\frac{136}{3}DF-60F^2\right)b_D\right.\nonumber\\&+&\left.\left(4+\frac{68}{3}D^2
-120DF+60F^2\right)b_F+4\left(b_1-b_2+b_3\right)+\frac{8}{3}m_0b_5\right.\nonumber\\&-&\left.\frac{8}{9}m_0\left(b_6+b_7\right)\right)+64\left(d_1-d_2+d_3-d_5\right)     ~,\nonumber\\
\tilde{\epsilon}_{2,\Xi^0}&=&   \frac{1}{(4\pi)^2 F_\pi^2}\left(\frac{17}{9}\frac{D^2}{m_0}
+\frac{70}{9}\frac{DF}{m_0}-\frac{17}{3}\frac{F^2}{m_0}+\left(4+\frac{68}{9}D^2-8DF
-12F^2\right)b_D\right.\nonumber\\&+&\left.\left(-4\left(1+D^2\right)-24DF+12F^2\right)b_F
+4\left(-b_1+b_2-b_3\right)-\frac{4}{3}m_0b_5\right.\nonumber\\&+&\left.\frac{4}{9}m_0\left(b_6+b_7\right)\right)+64\left(d_1-d_3\right)+32d_5     ~,\nonumber%\\
\eeqa
\beqa
\tilde{\epsilon}_{3,\Xi^0}&=&   \frac{1}{(4\pi)^2 F_\pi^2}\left(\frac{8}{9}\frac{D^2}{m_0}
+\frac{16}{9}\frac{DF}{m_0}-\frac{8}{3}\frac{F^2}{m_0}+\left(\frac{488}{9}D^2-\frac{112}{3}DF
+72F^2\right)b_D\right.\nonumber\\&+&\left.\left(-\frac{56}{3}D^2+144DF-72F^2\right)b_F
-\frac{4}{3}m_0b_5+\frac{4}{9}m_0\left(b_6+b_7\right)\right)-128d_1\nonumber\\&+&64d_2+32d_5   ~,\nonumber\\
\tilde{\epsilon}_{4,\Xi^0}&=& - 2\frac{D^2}{m_0}-4\frac{DF}{m_0}+6\frac{F^2}{m_0}
+\left(-16-48D^2+96DF-48F^2\right)\left(b_D-b_F\right)+24b_1\nonumber\\&-&8\left(b_2+b_3\right)
-6m_0b_5+2m_0\left(b_6+b_7\right)   ~,\nonumber%\\
\eeqa
\beqa
\tilde{\epsilon}_{5,\Xi^0}&=&   - \frac{2}{3}\frac{D^2}{m_0}-4\frac{DF}{m_0}+2\frac{F^2}{m_0}-\left(8+\frac{104}{3}D^2+16DF+24F^2\right)b_D\nonumber\\&+&\left(8\left(1-D^2\right)-48DF+24F^2\right)b_F+8\left(b_1-b_2+b_3\right)+2m_0\left(-b_5+b_6-b_7-b_9\right)   ~,\nonumber\\
\tilde{\epsilon}_{6,\Xi^0}&=&    \frac{2}{9}\frac{D^2}{m_0}+\frac{4}{9}\frac{DF}{m_0}-\frac{2}{3}\frac{F^2}{m_0}+\frac{16}{9}\left(b_D-b_F\right)-\frac{8}{3}b_1+\frac{8}{9}\left(b_2+b_3\right)+\frac{2}{3}m_0b_5\nonumber\\&-&\frac{2}{9}m_0\left(b_6+b_7\right)   ~,\nonumber\\
%\tilde{\epsilon}_{7,\Xi^0}&=&     ~,\nonumber\\
\tilde{\epsilon}_{7,\Xi^0}&=&    \frac{2}{3}\frac{D^2}{m_0}+4\frac{DF}{m_0}
-2\frac{F^2}{m_0}+\left(8+\frac{64}{3}D^2\right)b_D+8\left(-b_F-b_1+b_2-b_3\right)
\nonumber\\&+&2m_0\left(b_5-b_6+b_7+b_9\right)    ~,\nonumber%\\
\eeqa
\beqa
\tilde{\epsilon}_{8,\Xi^0}&=&    \frac{8}{9}\frac{D^2}{m_0}+\frac{16}{9}\frac{DF}{m_0}
-\frac{8}{3}\frac{F^2}{m_0}+\frac{16}{9}\left(b_D-b_F\right)-\frac{32}{3}b_1
+\frac{32}{9}\left(b_2+b_3\right)+\frac{8}{3}m_0b_5\nonumber\\&-&\frac{8}{9}m_0
\left(b_6+b_7\right)       ~,\nonumber\\
\tilde{\epsilon}_{9,\Xi^0}&=& \left(8+48D^2-96DF+48F^2\right)\left(b_D-b_F\right)    ~,\nonumber\\
\tilde{\epsilon}_{10,\Xi^0}&=&  \left(\frac{40}{3}D^2+16DF+24F^2\right)b_D+\left(8D^2+48DF-24F^2\right)b_F   ~,\nonumber\\
\tilde{\epsilon}_{11,\Xi^0}&=&    \frac{8}{9}\frac{D^2}{m_0}+\frac{16}{9}\frac{DF}{m_0}
-\frac{8}{3}\frac{F^2}{m_0}+\frac{40}{9}\left(b_D-b_F\right)-\frac{32}{3}b_1
+\frac{32}{9}\left(b_2+b_3\right)+\frac{8}{3}m_0b_5
\nonumber\\&-&\frac{8}{9}m_0\left(b_6+b_7\right)     ~.
\eeqa
Note that from the dimension two (four) Lagrangian, the LECs $b_{4,8} \, (d_{4,6,7})$
do not contribute to the strong isospin breaking  corrections. There are no contributions
$\sim b_0$ because this operator is only sensitive to the sum of the quark masses and
also not affected by the $\Lambda-\Sigma^0$ mixing.
While these expressions appear very voluminous and contain a fair amount of LECs, 
we show in the next paragraph how these can be constrained by reducing to the SU(2)
case and some phenomenological results. 

%%%%%%%% matching for two flavors %%%%%%%%% %%%%%%%%%%%%%%%%%%%%%%%%%%%%%%%%%%%%%%%%%
\section{Matching to the two-flavor case}
\label{sec:match}
%%%%%%%%%%%%%%%%%%%%%%%%%%%%%%%%%%%%%%%%%%%%%%%%%%%%%%%%%%%%%%%%%%%%%%%%%%%%%%%%%%%%%
\subsection{Matching equations}
\label{sec:matcheq}
The quark mass expansions of the baryon masses given in the preceding section contain
a sizeable amount of parameters (LECs). This does not pose a problem if one has sufficiently
many lattice data at various quark masses. However, the fitting procedure to find the
real minimum in this parameter space will be somewhat tedious and thus it is important
to find further constraints on these parameters. This can be achieved for certain 
combinations of the LECs by matching to the SU(2) result for the nucleon mass. In the
isospin limit $m_u = m_d$, we have $m_p = m_n = m_N$, and the quark mass expansion
of $m_N$ is given by \cite{SFM,KM1,BL}
\beqa\label{mNsu2}
m_N &=& \tilde{m}_0 - 4 c_1 \tilde{M}_\pi^2 - \frac{3 g_A^2}{32\pi F_\pi^2} \tilde{M}_\pi^3
+ \frac{3}{F_\pi^2} \left( 2c_1 -\frac{c_2}{4}-c_3 - \frac{g_A^2}{4\tilde{m}_0} \right) 
\,\tilde{M}_\pi^4 \, \bar\mu
\nonumber\\
&& \qquad\qquad + \left(-4 \bar e_1 + \frac{3}{8}\frac{1}{(4\pi F_\pi)^2}c_2 - \frac{3}{4}
\frac{g_A^2}{(4\pi F_\pi)^2}\frac{1}{\tilde m_0}\right) \, \tilde{M}_\pi^4 + {\cal O}(\tilde{M}_\pi^5)~.
\eeqa
Here, $\tilde m_0$ is the nucleon mass in the SU(2) chiral limit with $m_u = m_d =0$
and $m_s$ fixed at its physical value and $\tilde{M}_\pi$ is the SU(2) lowest order pion mass distinguished from its SU(3) counterpart $\bar{M}_\pi$ in Eqs.~(\ref{mesmass2}) by corrections in $m_s$. Strictly speaking, the same is true for the above $F_\pi$, yet, to the order we are working here, the difference is of no relevance. The $c_i$ are scale-independent LECs from the
second order pion-nucleon Lagrangian \cite{BKMrev} and $e_1$ is the sole scale-dependent
combination of LECs that contributes to the nucleon mass \cite{SFM,FMMS}. In
the notation of \cite{FMMS}, we have $e_1= 4e_{38}+ e_{115}/2+e_{116}/2$.
Note that again we work at the
scale given by the SU(2) nucleon mass in the chiral limit and thus use the LEC $\bar e_1$. These
particular LECs are constrained from the various analyses of pion-nucleon scattering
and pionic hydrogen/deuterium in chiral perturbation theory 
and the reaction $\pi N \to \pi\pi N$, see the next paragraph
for a detailed discussion of this topic.
Their numerical values can also be understood in 
terms of resonance saturation, see \cite{BKMlec}; in particular they incorporate the
important contribution from the $\Delta (1232)$ resonance. Mapping the quark mass
expansion of $m_N$ onto Eq.~(\ref{mNsu2}), we obtain the following matching conditions:

\bigskip
\medskip \noindent \underline{Chiral limit mass:}
\beqa\label{matchmass}
\tilde{m}_0&=&\left(1+\left(-\frac{17}{12}b_5+\frac{17}{36}b_6-\frac{89}{108}b_7-\frac{13}{9}b_8
-\frac{1}{4}b_9\right)\frac{\hat{M}_K^4}{(4 \pi F_\pi)^2}\right.\nonumber\\&+&\left.\left(\frac{3}{2}b_5
-\frac{1}{2}b_6+\frac{3}{2}b_7+2b_8+\frac{1}{2}b_9\right)\frac{\hat{M}_K^4}{(4 \pi F_\pi)^2}
\ln\left(\frac{\hat{M}_K^2}{m_0^2}\right)\right.\nonumber\\
&+&\left.\left(\frac{4}{3}b_5-\frac{4}{9}b_6
+\frac{4}{27}b_7+\frac{8}{9}b_8\right)\frac{\hat{M}_K^4}{(4 \pi F_\pi)^2}
\ln\left(\frac{4 \hat{M}_K^2}{3 m_0^2}\right)\right)m_0\nonumber\\
%\eeqa
%\beqa
&+& 4\left(-b_0-b_D+b_F\right)\hat{M}_K^2
+\left(-\left(\frac{5}{12}+\frac{1}{9 \sqrt{3}}\right)D^2+\left(\frac{1}{2}
+\frac{2}{3 \sqrt{3}}\right)DF\right.\nonumber\\&-&\left.\left(\frac{3}{4}
+\frac{1}{\sqrt{3}}\right)F^2\right)\frac{\hat{M}_K^3}{4 \pi F_\pi^2}
+\left(\left(\frac{52}{9}D^2-\frac{40}{3}DF+4F^2\right)\frac{b_D}{(4 \pi F_\pi)^2}\right.\nonumber\\
&+&\left.\left(-\frac{20}{3}D^2+8    DF-12F^2\right)\frac{b_F}{(4 \pi F_\pi)^2}
+16\left(-d_1+d_2-d_3+d_5-d_6-d_7\right)\right)\hat{M}_K^4\nonumber\\
&+&\left(8b_0
+\left(6+\frac{26}{3}D^2-20DF+6F^2\right)b_D+\left(-2-10D^2+12DF-18F^2\right)b_F\right.\nonumber\\
&-&\left.6b_1+2b_2-6b_3-8b_4\right)\frac{\hat{M}_K^4}{(4 \pi F_\pi)^2}
\ln\left(\frac{\hat{M}_K^2}{m_0^2}\right)\nonumber\\
&+&\left(\frac{32}{9}\left(b_0+b_D-b_F\right)
-\frac{16}{3}b_1+\frac{16}{9}b_2-\frac{16}{27}b_3-\frac{32}{9}b_4\right)\frac{\hat{M}_K^4}{(4 \pi F_\pi)^2}
\ln\left(\frac{4 \hat{M}_K^2}{3 m_0^2}\right)
\nonumber%\\
\eeqa
\beqa
&+&\left(\left(-\frac{53}{54}D^2+\frac{17}{9}DF-\frac{17}{6}F^2\right)\frac{\hat{M}_K^4}{(4 \pi F_\pi)^2}\right.
\nonumber\\
&+&\left.\left(-\frac{5}{6}D^2+DF-\frac{3}{2}F^2\right)\frac{\hat{M}_K^4}{(4 \pi F_\pi)^2}
\ln\left(\frac{ \hat{M}_K^2}{ m_0^2}\right)\right.\nonumber\\
&+&\left.\left(-\frac{4}{27}D^2+\frac{8}{9}DF-\frac{4}{3}F^2\right)\frac{\hat{M}_K^4}{(4 \pi F_\pi)^2}
\ln\left(\frac{4 \hat{M}_K^2}{3 m_0^2}\right)%\right.\nonumber\\
%\eeqa
%\beqa
%&+&\left.\left(\left(-\frac{13}{18}D^2+\frac{5}{3}DF-\frac{1}{2}F^2\right)b_D
%+\left(\frac{5}{6}D^2-DF+\frac{3}{2}F^2\right)b_F\right)\frac{\hat{M}_K^5}{4 \pi F_\pi^2}
\right)\frac{1}{m_0}
%\nonumber\\
%\eeqa
%\beqa
%&+&\left(\left(\left(\frac{5}{96}+\frac{1}{54 \sqrt{3}}\right)D^2
%-\left(\frac{1}{16}+\frac{1}{9 \sqrt{3}}\right)DF+\left(\frac{3}{32}
%+\frac{1}{6 \sqrt{3}}\right)F^2\right)\frac{\hat{M}_K^5}{4 \pi F_\pi^2}\right.\nonumber\\
%&+&\left.\left(\left(-\frac{26}{9}D^2+\frac{20}{3}DF-2F^2\right)b_D
%+\left(\frac{10}{3}D^2-4DF+6F^2\right)b_F\right)\frac{\hat{M}_K^6}{(4 \pi F_\pi)^2}\right)\frac{1}{m_0^2}
%\nonumber\\
%&+&\left(\left(\frac{167}{972}D^2-\frac{59}{162}DF+\frac{59}{108}F^2\right)\frac{\hat{M}_K^6}{(4 \pi F_\pi)^2}
%+\left(\left(-\frac{13}{144}D^2+\frac{5}{24}DF-\frac{1}{16}F^2\right)b_D\right.\right.\nonumber\\
%&+&\left.\left.\left(\frac{5}{48}D^2-\frac{1}{8}DF+\frac{3}{16}F^2\right)b_F\right)
%\frac{\hat{M}_K^7}{4 \pi F_\pi^2}\right)\frac{1}{m_0^3}
%\nonumber\\
+{\cal O}\left(\hat{M}_K^5\right)~,
\eeqa

\medskip
\noindent \underline{Axial-vector coupling:}
\beq\label{matchaxial}
g_A = D+F+{\cal O}\left(\hat{M}_K^2\right)~,
\eeq

\medskip
\noindent \underline{Dimension two LECs:}
\beqa\label{matchlec1}
c_1&=&\left(\left(-\frac{3}{8}b_5+\frac{1}{8}b_6-\frac{3}{8}b_7-\frac{1}{2}b_8-\frac{1}{8}b_9\right)\frac{\hat{M}_K^2}{(4 \pi F_\pi)^2}
\ln\left(\frac{ \hat{M}_K^2}{ m_0^2}\right)\right.\nonumber\\&+&\left.\left(-\frac{1}{6}b_5+\frac{1}{18}b_6-\frac{1}{54}b_7-\frac{1}{9}b_8\right)\frac{\hat{M}_K^2}{(4 \pi F_\pi)^2}
\ln\left(\frac{4 \hat{M}_K^2}{3 m_0^2}\right)\right)m_0\nonumber\\&+&b_0+\frac{1}{2}\left(b_D+b_F\right)+\left(\left(\frac{5}{64}+\frac{1}{96\sqrt{3}}\right)D^2
-\left(\frac{3}{32}+\frac{1}{16 \sqrt{3}}\right)DF\right.\nonumber\\&+&\left.\left(\frac{9}{64}
+\frac{3}{32\sqrt{3}}\right)F^2\right)\frac{\hat{M}_K}{4\pi F_\pi^2}+\left(\left(16 L_4-32L_6
-\frac{11}{9}\frac{1}{(4 \pi)^2}\right)\frac{b_0}{F_\pi^2}\right.\nonumber\\
&+&\left.\left(8L_4-16 L_6+\left(-\frac{35}{36}-\frac{13}{12}D^2+\frac{5}{2}DF
-\frac{3}{4}F^2\right)\frac{1}{(4\pi)^2}\right)\frac{b_D}{F_\pi^2}\right.\nonumber\\
&+&\left.\left(8L_4-16 L_6+\left(\frac{17}{36}+\frac{5}{4}D^2-\frac{3}{2}DF
+\frac{9}{4}F^2\right)\frac{1}{(4\pi)^2}\right)\frac{b_F}{F_\pi^2}\right.\nonumber\\
%\eeqa
%\beqa
&+&\left.\left(\frac{13}{12}b_1-\frac{13}{36}b_2+\frac{85}{108}b_3+\frac{11}{9}b_4\right)\frac{1}{(4\pi F_\pi)^2}
+4\left(-d_1+d_3\right)-2d_5+8d_6\right)\hat{M}_K^2\nonumber\\&+&\left(-2b_0-\frac{3}{2}b_D
+\frac{1}{2}b_F+\frac{3}{2}b_1-\frac{1}{2}b_2+\frac{3}{2}b_3+2b_4\right)\frac{\hat{M}_K^2}{(4 \pi F_\pi)^2}
\ln\left(\frac{ \hat{M}_K^2}{ m_0^2}\right)\nonumber\\&+&\left(\frac{2}{9}\left(-b_0-b_D+b_F\right)
+\frac{2}{3}b_1-\frac{2}{9}b_2+\frac{2}{27}b_3+\frac{4}{9}b_4\right)\frac{\hat{M}_K^2}{(4 \pi F_\pi)^2}
\ln\left(\frac{4 \hat{M}_K^2}{3 m_0^2}\right)\nonumber\\
&+&\left(\left(\frac{49}{144}D^2-\frac{13}{24}DF+\frac{13}{16}F^2\right)\frac{\hat{M}_K^2}{(4 \pi F_\pi)^2}\right.\nonumber\\&+&\left.\left(\frac{5}{24}D^2-\frac{1}{4}DF+\frac{3}{8}F^2\right)\frac{\hat{M}_K^2}{(4 \pi F_\pi)^2}
\ln\left(\frac{ \hat{M}_K^2}{ m_0^2}\right)\right.\nonumber\\&+&\left.\left(\frac{1}{54}D^2-\frac{1}{9}DF+\frac{1}{6}F^2\right)\frac{\hat{M}_K^2}{(4 \pi F_\pi)^2}
\ln\left(\frac{4 \hat{M}_K^2}{3 m_0^2}\right)\right)\frac{1}{m_0}
+{\cal O}  \left(\hat{M}_K^3\right)~,
\eeqa
\beqa\label{matchlec2}
\frac{1}{4}c_2+c_3&=&-\left(\frac{1}{4}\left(b_5+b_6+b_7\right)+\frac{1}{2}b_8\right)m_0+b_1+b_2+b_3+2b_4+{\cal O}  \left(\hat{M}_K\right)~,
\eeqa
\noindent \underline{Dimension four LEC with recoil correction:}
\beqa\label{matchlec4}
-4\bar{e}_1&+&\frac{3}{8}\frac{1}{(4 \pi F_\pi)^2}c_2=\nonumber\\& &\left(\left(\frac{1}{12}b_5-\frac{19}{36}b_6+\frac{1}{108}b_7-\frac{7}{36}b_8+\frac{1}{8}b_9\right)\frac{1}{(4\pi F_\pi)^2}\right.\nonumber\\&+&\left.\left(\frac{3}{8}b_5-\frac{1}{8}b_6+\frac{3}{8}b_7+\frac{1}{2}b_8+\frac{1}{8}b_9\right)\frac{1}{(4 \pi F_\pi)^2}
\ln\left(\frac{ \hat{M}_K^2}{ m_0^2}\right)\right.\nonumber\\&+&\left.\left(\frac{1}{12}b_5-\frac{1}{36}b_6+\frac{1}{108}b_7+\frac{1}{18}b_8\right)\frac{1}{(4 \pi F_\pi)^2}
\ln\left(\frac{4 \hat{M}_K^2}{3 m_0^2}\right)\right)m_0\nonumber\\&+&\left(-\left(\frac{5}{128}+\frac{1}{384\sqrt{3}}\right)D^2+\left(\frac{3}{64}+\frac{1}{64 \sqrt{3}}\right)DF
-\left(\frac{9}{128}+\frac{3}{128 \sqrt{3}}\right)F^2\right)\frac{1}{\hat{M}_K 4 \pi F_\pi^2}\nonumber\\
&+&\left(\frac{10}{3}b_0+\left(\frac{89}{36}-\frac{91}{36}D^2+\frac{35}{6}DF-\frac{7}{4}F^2\right)b_D
+\left(-\frac{3}{4}+\frac{35}{12}D^2-\frac{7}{2}DF+\frac{21}{4}F^2\right)b_F\right.\nonumber\\
&-&\left.\frac{11}{4}b_1+\frac{11}{12}b_2-\frac{83}{36}b_3-\frac{10}{3}b_4\right)\frac{1}{(4 \pi F_\pi)^2}
-4\left(d_1+d_2+d_3\right)-8 d_5-16d_6-8d_7\nonumber\\&+&\left(2b_0+\left(\frac{3}{2}-\frac{13}{6}D^2+5DF
-\frac{3}{2}F^2\right)b_D+\left(-\frac{1}{2}+\frac{5}{2}D^2-3DF+\frac{9}{2}F^2\right)b_F\right.\nonumber\\
&-&\left.\frac{3}{2}b_1+\frac{1}{2}b_2-\frac{3}{2}b_3-2b_4\right)\frac{1}{(4 \pi F_\pi)^2}
\ln\left(\frac{ \hat{M}_K^2}{ m_0^2}\right)\nonumber\\&+&\left(\frac{2}{9}b_0+\frac{1}{9}\left(b_D+b_F\right)
-\frac{1}{3}b_1+\frac{1}{9}b_2-\frac{1}{27}b_3-\frac{2}{9}b_4\right)\frac{1}{(4 \pi F_\pi)^2}
\ln\left(\frac{4 \hat{M}_K^2}{3 m_0^2}\right)\nonumber\\&+&\left(\left(-\frac{235}{432}D^2+\frac{55}{72}DF-\frac{55}{48}F^2\right)\frac{1}{(4\pi F_\pi)^2}\right.\nonumber\\&+&\left.\left(-\frac{5}{24}D^2+\frac{1}{4}DF-\frac{3}{8}F^2\right)\frac{1}{(4 \pi F_\pi)^2}
\ln\left(\frac{ \hat{M}_K^2}{ m_0^2}\right)\right.\nonumber\\&+&\left.\left(-\frac{1}{108}D^2+\frac{1}{18}DF-\frac{1}{12}F^2\right)\frac{1}{(4 \pi F_\pi)^2}
\ln\left(\frac{4 \hat{M}_K^2}{3 m_0^2}\right)\right)\frac{1}{m_0}+{\cal O}  \left(\hat{M}_K\right)~,
\eeqa
where $\hat M_K^2 = m_s B$ is the leading strange quark contribution to the kaon mass.
The first of these relations Eq.~(\ref{matchmass}) connects the non--chiral
part of the nucleon mass (generated by the gluon condensate) in the SU(2) limit
with the one in the SU(3) case. Obviously, the only difference are terms
proportional to  powers of the strange quark mass (modulo
logarithms). The difference between $m_0$ and $\tilde m_0$ is what is usually
called the strangeness contribution to the nucleon
mass. Eq.~(\ref{matchaxial}) is the usual reduction of the two axial--vector
couplings in SU(3) to the single coupling $g_A$ for the two--flavor case. To
the order we are working, this relation is unaffected by quark mass
corrections. A similar statement holds for the relation between the dynamic dimension
two LECs~\footnote{This name stems from the fact that they essentially
contribute to pion/kaon-nucleon scattering.} displayed in
Eq.~(\ref{matchlec2}).  Quite differently, the relation between the leading
order symmetry breaking LEC in SU(2), called $c_1$, and its three SU(3) 
counterparts $b_{0,D,F}$, has sizeable quark mass corrections, simply because
the corresponding operators contribute already to the baryon masses at first
non-trivial order. In particular, one has corrections from the third order
loop contributions that generate the terms $\sim \hat M_K \sim \sqrt{m_s}$.
Finally, note that in the last relation the expansion of the loop functions
$\bar I_{K,\eta}$ in powers of $M_\pi^2$ (using $\bar M_K^2 = \hat M_K^2 + \bar M_\pi^2/2$)
generates the contributions $\sim 1/\hat M_K \sim 1 / \sqrt{m_s}$.

\medskip\noindent In the isospin breaking sector, we consider only the
leading dimension two operator that is parameterized by the LEC $c_5$ in 
SU(2) \cite{BKMlec}. From matching the leading SU(3) terms  $\sim (m_u-m_d)$ 
the following relation is obtained:
\beqa
c_5&=&\left(\left(\frac{1}{6}b_5+\frac{1}{18}\left(b_6-b_7\right)\right)\frac{\hat{M}_K^2}{(4 \pi F_\pi)^2}-\frac{1}{4}\left(b_5+b_6+b_7+b_9\right)
\frac{\hat{M}_K^2}{(4 \pi F_\pi)^2}\ln\left(\frac{ \hat{M}_K^2}{ m_0^2}\right)
\right.\nonumber\\
&+&\left.\left(-\frac{1}{3}b_5+\frac{1}{9}\left(-b_6+b_7\right)\right)
\frac{\hat{M}_K^2}{(4 \pi F_\pi)^2}\ln\left(\frac{4 \hat{M}_K^2}{3 m_0^2}\right)\right)m_0\nonumber\\&+&b_D+b_F+\left(-\left(\frac{1}{32}+\frac{1}{12 \sqrt{3}}\right)D^2
+\left(\frac{3}{16}+\frac{1}{6\sqrt{3}}\right)DF+\left(\frac{3}{32}
+\frac{1}{4\sqrt{3}}\right)F^2\right)\frac{\hat{M}_K}{4\pi F_\pi^2}\nonumber\\
&+&\left(\left(16 L_4-32L_6+\left(-\frac{1}{2}-\frac{17}{18}D^2-DF
+\frac{3}{2}F^2\right)\frac{1}{(4\pi)^2}\right)\frac{b_D}{F_\pi^2}\right.\nonumber\\
&+&\left.\left(16 L_4-32L_6+\left(-\frac{1}{2}\left(1+D^2\right)+3DF
+\frac{3}{2}F^2\right)\frac{1}{(4\pi)^2}\right)\frac{b_F}{F_\pi^2}\right.\nonumber\\
&+&\left.\frac{1}{2}\left(b_1+b_2+b_3\right)\frac{1}{(4\pi F_\pi)^2}
+8\left(-d_1+d_3\right)+4d_5\right)\hat{M}_K^2\nonumber\\
&+&\left(-\left(1+\frac{8}{3}D^2\right)b_D-b_F+b_1+b_2+b_3\right)
\frac{\hat{M}_K^2}{(4 \pi F_\pi)^2}\ln\left(\frac{ \hat{M}_K^2}{ m_0^2}\right)
\nonumber\\
&+&\left(\frac{4}{3}b_1+\frac{4}{9}\left(b_2-b_3\right)\right)
\frac{\hat{M}_K^2}{(4 \pi F_\pi)^2}\ln\left(\frac{4 \hat{M}_K^2}{3 m_0^2}\right)\nonumber\\&+&\left(\left(-\frac{17}{72}D^2+\frac{35}{36}DF+\frac{17}{24}F^2\right)\frac{\hat{M}_K^2}{(4 \pi F_\pi)^2}\right.\nonumber\\&+&\left.\left(-\frac{1}{12}D^2+\frac{1}{2}DF+\frac{1}{4}F^2\right)
\frac{\hat{M}_K^2}{(4 \pi F_\pi)^2}\ln\left(\frac{ \hat{M}_K^2}{ m_0^2}\right)
\right.\nonumber\\
&+&\left.\left(-\frac{1}{9}D^2+\frac{2}{9}DF+\frac{1}{3}F^2\right)
\frac{\hat{M}_K^2}{(4 \pi F_\pi)^2}\ln\left(\frac{4 \hat{M}_K^2}{3 m_0^2}\right)\right)\frac{1}{m_0}
+{\cal O}  \left(\hat{M}_K^3\right)~.
\eeqa

\subsection{Bounds on the low-energy constants}
\label{sec:bounds}

In this paragraph we collect what is known from various sources on the LECs appearing
in the baryon masses. The SU(2) parameters have indeed been determined from the analysis
of pion-nucleon and nucleon-nucleon scattering. Combining the results from 
Refs.~\cite{BKMlec,FMS,BuM,RTS}, we find
\beq
c_1 = -0.9^{+0.5}_{-0.2}~,~~ c_2 = 3.3 \pm 0.2~,~~ c_3 = -5.0^{+1.6}_{-1.0}~,~~
c_4 =  3.5^{+0.5}_{-0.2}~,
\eeq
where all numbers are given in GeV$^{-1}$. Furthermore, 
%$e_1 (\lambda =1~{\rm GeV})$ 
naturalness gives  $-1 \lesssim e_1 \lesssim 1\,$GeV$^{-3}$ (for $\lambda = 1\,$GeV), 
which is consistent with the  loose determination in \cite{FMiso}. 
Consequently, the left-hand-side (lhs) of Eq.~(\ref{matchlec2}) can be conservatively bounded
by
\beq\label{A}
\frac{1}{4}c_2 + c_3 = -5.2 \ldots -2.5 ~{\rm GeV}^{-1}~,
\eeq
and similarly for the lhs of Eq.~(\ref{matchlec4})
\beq\label{B}
3~{\rm GeV}^{-3} \leq
-4\bar{e}_1 + \frac{3}{8}\frac{1}{(4 \pi F_\pi)^2}c_2 \leq 5~{\rm GeV}^{-3}~.
\eeq
These constraints are also consistent with recent lattice interpolations of
SU(2) results, one has for Eqs.(\ref{A}, \ref{B}) $-3.1 \, (-3.9 \ldots -2.6)$ and
$5 \, (3.2 \ldots 5.7)$ in \cite{BHMcutoff} (\cite{PHW}), respectively. We now turn
to estimating the SU(3) LECs $b_i$ and $d_i$. The symmetry breaking LECs $b_{0,D,F}$
can be determined from a third order fit to the baryon masses. We collect here the
results from the comprehensive study in \cite{BKMmass},
\beq
-0.79 \leq b_0 \leq -0.70~,~~ 0.01 \leq b_D \leq 0.07~,~~ -0.61 \leq b_F \leq -0.48~,
\eeq
with all numbers given in GeV$^{-1}$. The uncertainties on these numbers are certainly
underestimated by looking only at the baryon masses, see the discussion in \cite{BM}. 
The resonance saturation estimates presented in that paper also give somewhat different
values, but it should be noted that this method is not expected to work well for
the LECs related to symmetry breaking. For orientation, one can use Eq.~(68) of \cite{BM}
for getting a rough idea of the size of the dimension four LECs $d_i$, provided
one performs the relabeling $d_7^{\rm BM} \to d_6$ and  $d_8^{\rm BM} \to d_7$.
For the dynamical dimension two LECs $b_i$ $(i=1, \ldots,9)$, resonance saturation
should work better. We map here the results of \cite{BM} onto our notation
so that one can get
some estimates about the size of some combinations of the dynamical dimension
two LECs (we refrain here from redoing the calculations of that paper since
these numbers only serve as a rough guide). We find the following relations
\beqa\label{biconv}
b_1^{\rm BM} &\hat{=}& b_1 - \frac{m_0}{4}b_5 - \frac{m_0}{16}b_9 ~, 
\nonumber\\
b_2^{\rm BM} &\hat{=}& b_2 - \frac{m_0}{4}b_6~,\nonumber\\
b_3^{\rm BM} &\hat{=}& b_3 - \frac{m_0}{4}b_7 - \frac{3 m_0}{16}b_9 ~, 
\nonumber\\
b_8^{\rm BM} &\hat{=}& b_4 - \frac{m_0}{4}b_8 + \frac{m_0}{8}b_9 ~, 
\eeqa
and the pertinent resonance saturation estimates for the $b_i^{\rm BM}$ can
be found in Eq.~(68) of  \cite{BM}. Last, we consider the leading
isospin breaking LEC $c_5$. It was determined in \cite{BKMlec} 
$c_5 =-0.09\pm 0.01$~GeV$^{-1}$. Note that in \cite{weinmass} a somewhat larger
value was obtained based on a leading order SU(3) estimate for the strong proton--neutron
mass difference.

%%%%%%%%%%%%%%%%%%%%%%%%%%%%%%%%%%%%%%%%%%%%%%%%%%%%%%%%%%%%%%%%%%%%%%%%%%%%%%%%%%%%%%%%%%%%%
\section{Sigma terms}
\label{sec:sigma}
%%%%%%%%%%%%%%%%%%%%%%%%%%%%%%%%%%%%%%%%%%%%%%%%%%%%%%%%%%%%%%%%%%%%%%%%%%%%%%%%%%%%%%%%%%%%%

Directly related to the baryon masses are the so-called sigma terms that measure
the contribution of the QCD quark mass term to any baryon mass,
\beq\label{sigmadef}
\sigma_{qB} (0) = \langle B| m_q \bar q q |B\rangle~.
\eeq
Here $q$ denotes any of the light quark flavors u, d or s. These scalar-isoscalar
operators can be directly read off from the quark mass expansion of the baryon masses
given in the preceding section. We refrain from displaying the lengthy formulas here.
Instead, we briefly discuss the so-called pion-nucleon sigma term. It measures
the strength of the light quark condensate in the proton (and can  be determined from
the analytically continued isoscalar pion-nucleon scattering amplitude $\bar D^+ (\nu,t)$),
\beq\label{sigmapin}
\sigma_{\pi N}(0) = \hat m \, \langle p | \bar u u  + \bar d d|p\rangle
= \hat{m} \, \frac{\partial m_N}{\partial \hat m}~,
\eeq
where the last equality follows from the Feynman-Hellmann theorem \cite{FH}.
More precisely, what is given in Eq.~(\ref{sigmapin}) is the pion--nucleon
sigma term in the isospin limit, its generalization to include isospin violation
is obtained by the substitution $\hat m (\bar u u +\bar d d) \to m_u \bar u u 
+ m_d \bar d d$. We will only consider the isospin symmetric case here and
refer the reader to Refs.\cite{MS,MM} for a discussion of the effects generated
by strong and electromagnetic isospin violation. In terms of the so-called 
strangeness fraction $y$, the sigma term can be expressed as
\beq
\sigma_{\pi N}(0) = \frac{\hat \sigma}{1-y }~,
\eeq
with
\beqa
\hat \sigma &=& \hat m \, \langle p | \bar u u  + \bar d d - 2\bar s s|p\rangle ~,\nonumber\\
y &=& \frac{2 \langle p |\bar s s|p\rangle}{\langle p | \bar u u  + \bar d d|p\rangle}~. 
\eeqa
These two quantities can also be read off from the quark mass expansion of the nucleon mass via
\beqa
\hat \sigma &=& \hat{m} \, \frac{\partial m_N}{\partial \hat m} - 
2 \hat{m} \, \frac{\partial m_N}{\partial m_s}~, \nonumber\\
y &=& 2 \,\frac{\partial m_N}{\partial m_s} \, 
\left(\frac{\partial m_N}{\partial \hat m}\right)^{-1}~,
\eeqa
which allows for a direct determination from the lattice result for $m_N = m_N (m_u,
m_d,m_s)$.  Chiral perturbation
theory calculations to order $m_q^2$ give $\hat \sigma = 35\pm 5\,$MeV \cite{Juerg}, 
 $\hat \sigma = 36\pm 7\,$MeV \cite{BM}, and $\hat \sigma = 33\pm 3\,$MeV \cite{BB}. For a recent 
analysis of the sigma term in view of SU(2) lattice results, see \cite{PHW}. 
For earlier lattice studies of this interesting quantity, see \cite{sigma1}-\cite{sigma5}.
Similarly, one
can also analyze the two kaon-nucleon sigma terms, $\sigma_{KN}^{(1)} (0) = (\hat m + m_s)
\langle p|\bar u u+ \bar s s|p\rangle/2$ and $\sigma_{KN}^{(2)} (0) = (\hat m + m_s)
\langle p|-\bar u u+ 2\bar d d + \bar s s|p\rangle/2$, in the isospin limit. 
The strange quark contribution to the nucleon mass can be directly expressed as a linear
combination of the pion-nucleon and the two kaon-nucleon sigma terms  via
\beq
m_s \,\langle p| \bar s s|p\rangle = \left(\frac{1}{2} - \frac{\bar M_\pi^2}{4\bar M_K^2}
\right) \, \left[ 3\sigma_{KN}^{(1)} (0) +\sigma_{KN}^{(2)} (0)\right] + \left(\frac{1}{2}-
\frac {\bar M_K^2}{\bar M_\pi^2}\right)\, \sigma_{\pi N}(0)~.
\eeq
Similar analyses can be performed for the other baryons, but only in the case of
the nucleon sigma terms one can hope to compare lattice results to the ones of
phenomenological investigations.

%%%%%%%%%%%%%%%%%%%%%%%%%%%%%%%%%%%%%%%%%%%%%%%%%%%%%%%%%%%%%%%%%%%%%%%%%%%%%%%%%%%%%%%%%%%%%
\section{Summary and conclusions}
\label{sec:sum}
%%%%%%%%%%%%%%%%%%%%%%%%%%%%%%%%%%%%%%%%%%%%%%%%%%%%%%%%%%%%%%%%%%%%%%%%%%%%%%%%%%%%%%%%%%%%%

We summarize the main results of our work:
\begin{itemize}
\item[1)]We have calculated the octet baryon masses to fourth order in the
chiral expansion within a Lorentz-invariant formulation of baryon chiral
perturbation theory. In contrast to earlier works (with the exception of 
Ref.~\cite{Juerg} which utilizes an UV cut-off) we have systematically 
included strong isospin breaking, 
$m_u \neq m_d$. This amounts to considering all terms quadratic in the quark masses.
\item[2)]To disentangle the dependence of the baryon masses on the three light
quark masses, one has to consider the dimension two and four Lagrangians displayed
in Eq.~(\ref{leff2}) and Eq.~(\ref{leff4}), respectively. At dimension two, one has
three symmetry breaking LECs $(b_0,b_D,b_F)$ and nine dynamical LECs, that
enter at fourth order in the tadpole diagrams. At dimension four, we have seven LECs,
which appear in various combinations in the masses.
\item[3)]We have derived chiral extrapolation functions for the octet ground state
baryons, including all isospin breaking terms linear in the mixing angle $\ve$,
see Section~\ref{sec:extra}.
These constitute the main result of this paper and might be used to analyze unquenched
three-flavor simulations at varying quark masses above their physical values.
The equations collected in this section can be obtained as a Mathematica notebook
from the authors upon request. The corresponding meson mass representations of the
baryon masses that are not truncated at order $\ve$ are given in App.~\ref{sec:meson}.
\item[4)]We have performed the matching to the two-flavor case to obtain constraints
on various combinations of dimension two and four SU(3) 
low-energy constants, see Section~\ref{sec:match}. We have 
also reviewed the determination of LECs in SU(2) and the symmetry breaking dimension
two SU(3) LECs and also given resonance saturation
estimates for the SU(3) LECs, based on the earlier work in \cite{BM}.
\item[5)]The various sigma terms as defined in Eq.~(\ref{sigmadef}) can be obtained
by differentiation of the mass formulas with respect to the quark masses. The 
corresponding extrapolation functions can be obtained from the authors upon request. 
%\item[6)]
%\item[7)]
\end{itemize}
 
\bigskip

\section*{Acknowledgements}
We thank Bastian Kubis for some useful comments in the initial stages of this investigation.

\appendix

\newpage

%%%%%%%%%%%%%%%%%%%%%%%%%%%%%%%%%%%%%%%%%%%%%%%%%%%%%%%%%%%%%%%%%%%%%%%%%%%%%%%%%%%%%%%%%%%%
\section{Meson mass representation}
\def\theequation{\Alph{section}.\arabic{equation}}
\setcounter{equation}{0}
\label{sec:meson}
%%%%%%%%%%%%%%%%%%%%%%%%%%%%%%%%%%%%%%%%%%%%%%%%%%%%%%%%%%%%%%%%%%%%%%%%%%%%%%%%%%%%%%%%%%%%%%
Here, we give the explicit representations of the baryon masses in terms of the
Goldstone boson masses. The local terms generated by the dimension two and four
insertions from the effective Lagrangian are expressed in terms of the charged pion
and kaon masses throughout. The loop contributions at third and fourth order are
expressed in terms of the loop functions $\bar I_P$, $\bar\mu_P$, and $\bar{I}^{12}_P$,
compare Eq.~(\ref{Ibar}), Eq.~(\ref{mubar}), and  Eq.~(\ref{I12bar}), respectively.
Here, $P$ stands for any of the charged and neutral pseudoscalar mesons.  
To arrive at the quark mass expansion given in Section\ref{sec:extra}, one has to 
expand these loop functions as given there and further expand the various mixing 
functions to linear order in $\ve$. Note that only the lowest order meson mass
relations are used to convert to the quark masses, that means in the quark mass representation we employ the chiral
value of $B$ throughout (while the meson mass representation is formulated in physical masses to the order we are working here).

\medskip\noindent
The various contributions to the baryon masses can be compactly written as
\beqa\label{masses234}
m_B^{(2)} &=& \gamma_B^\pi \, M_{\pi^+}^2 + \gamma_B^K \, M_{K^+}^2~, \\
m_B^{(3)} &=& \delta_{1,B} \, \bar{I}_{\pi^+} + \delta_{2,B} \, \bar{I}_{\pi^0} +
              \delta_{3,B} \, \bar{I}_{K^+} + \delta_{4,B} \, \bar{I}_{K^0} +
              \delta_{5,B} \, \bar{I}_{\eta}~,  \\   
m_B^{(4)} &=& \epsilon_{1,B} \, \bar{\mu}_{\pi^+} + \epsilon_{2,B} \, \bar{\mu}_{\pi^0} +
              \epsilon_{3,B} \, \bar{\mu}_{K^+} + \epsilon_{4,B} \, \bar{\mu}_{K^0} +
              \epsilon_{5,B} \, \bar{\mu}_{\eta}  \nonumber\\
          &+& \epsilon_{6,B} \, \bar{I}^{12}_{\pi^+} + \epsilon_{7,B} \, \bar{I}^{12}_{\pi^0} +
              \epsilon_{8,B} \, \bar{I}^{12}_{K^+} + \epsilon_{9,B} \, \bar{I}^{12}_{K^0} +
              \epsilon_{10,B} \, \bar{I}^{12}_{\eta}  \nonumber\\
          &+& \epsilon_{11,B}^\pi \, M_{\pi^+}^4 + \epsilon_{11,B}^K \, M_{K^+}^4 +
              \epsilon_{11,B}^{\pi K} \, M_{\pi^+}^2  M_{K^+}^2~, 
\eeqa
with 
\beqa
\delta_{i,B}   &=& \frac{m_0}{F_\pi^2}\,\left(
                   \delta_{i,B}^\pi \, M_{\pi^+}^2 +  \delta_{i,B}^K  \,  
                   M_{K^+}^2 \right)~,
                   \quad\qquad\qquad\,\, (i = 1,\ldots,5)~, \\
\epsilon_{i,B} &=& \frac{1}{F_\pi^2}\,\left( \epsilon_{i,B}^\pi \,  M_{\pi^+}^4 + \epsilon_{i,B}^K \,  M_{K^+}^4 +
                   \epsilon_{i,B}^{\pi K} \, M_{\pi^+}^2  M_{K^+}^2\right)~, 
                   \quad (i = 1,\ldots,5) ~,  \\
\epsilon_{j,B} &=& \frac{m_0^2}{F_\pi^2}\,\left( \epsilon_{j,B}^\pi \,  M_{\pi^+}^4 + \epsilon_{j,B}^K \,  M_{K^+}^4 +
                   \epsilon_{j,B}^{\pi K} \, M_{\pi^+}^2  M_{K^+}^2\right)~, 
                   \quad (j = 6,\ldots,10) ~.    
\eeqa
For the $\Lambda$ and $\Sigma^0$ one has additional fourth order terms due to the
mixing which take the form
\beqa
\delta m_{\Sigma^0}^{(4)} &=& \frac{1}{\eta_7 \left(M_{\pi^+}^2- M_{K^+}^2\right)}\,
\sum_{i=1}^6\left(\eta_i^\pi\, M_{\pi^+}^4 + \eta_i^K \,  M_{K^+}^4 +  \eta_i^{\pi K }\, 
 M_{\pi^+}^2  M_{K^+}^2\right)\,\frac{m_0^2}{F_\pi^4} \, \bar{A}_i~, 
\nonumber \\
\delta m_{\Lambda }^{(4)} &=& - \delta m_{\Sigma^0}^{(4)}~,\label{etafactors}
%\delta m_{\Lambda }^{(4)} &=& -\frac{1}{\eta_7 \left(M_{\pi^+}^2- M_{K^+}^2\right)}\,
%\sum_{i=1}^6\left(\eta_i^\pi\, M_{\pi^+}^4 + \eta_i^K \,  M_{K^+}^4 +  \eta_i^{\pi K }\, 
% M_{\pi^+}^2  M_{K^+}^2\right)\,\frac{m_0^2}{F_\pi^2} \, \bar{\Gamma}_i~, \label{etafactors}
\eeqa
with $\bar A_1 = \bar{I}^2_{\pi^+}$, $\bar A_2 = \bar{I}_{\pi^+}\bar{I}_{K^+}$,
$\bar A_3 = \bar{I}_{\pi^+}\bar{I}_{K^0}$, $\bar A_4 = \bar{I}^2_{K^+}$,
$\bar A_5 = \bar{I}_{K^+}\bar{I}_{K^0}$, and
$\bar A_6 = \bar{I}_{K^0}^2\,$.

\noindent
In what follows, we only display the non-vanishing coefficients $\gamma_B$, $\delta_B$,
and $\epsilon_B$.

\medskip\noindent
The second order coefficients $\gamma_B$ are: 
\beqa
\gamma_{\Sigma^+}^\pi&=&\left(-6+4I 
\right)b_0-4 b_D+8\left(1-I\right)
b_F ~,\nonumber\\
%\eeqa
%\beqa
\gamma_{\Sigma^+}^K&=&-4Ib_0+8
\left(-1+I\right)b_F~,\nonumber\\
%\eeqa
%\begin{eqnarray}
\gamma_{\Sigma^-}^\pi&=&\left(-6+4I 
\right)b_0-4 b_D+8\left(-1+I\right)
b_F ~,\nonumber\\
%\eeqa
%\beqa
\gamma_{\Sigma^-}^K&=&-4Ib_0+8
\left(1-I\right)b_F~,\nonumber\\
%\end{eqnarray}
%\begin{eqnarray}
\gamma_{\Sigma^0}^\pi&=&\left(-6+4I 
\right)b_0+\left(-4+\frac{8}{3}I
-\frac{8}{3}\cos(2 \ve)+\frac{1}{\sqrt{3}}\left(8-\frac{32}{3}
I\right)\sin(2 \ve)\right)
b_D~,\nonumber\\
%\eeqa
%\beqa
\gamma_{\Sigma^0}^K&=&-4Ib_0+\left(-\frac{8}{3}I+\frac{8}{3}
\cos(2 \ve)+\frac{1}{\sqrt{3}}\left(-8+\frac{32}{3}
I \right)\sin(2 \ve)\right)b_D~,\nonumber\\
%\end{eqnarray}
%\begin{eqnarray}
\gamma_{p}^\pi&=&\left(-6+4I \right)
b_0+\left(4-8I \right)b_F~,\nonumber\\
%\eeqa
%\beqa
\gamma_{p}^K&=&-4I b_0-4 b_D
+\left(-4+8I \right)b_F~,\nonumber\\
%\end{eqnarray}
%\begin{eqnarray}
\gamma_{\Xi^-}^\pi&=&\left(-6+4I 
\right)b_0+\left(-4+8I \right)b_F~,\nonumber\\
%\eeqa
%\beqa
\gamma_{\Xi^-}^K&=&-4I b_0-4 b_D+
\left(4-8I \right)b_F~,\nonumber\\
%\end{eqnarray}
%\begin{eqnarray}
\gamma_{n}^\pi&=&\left(-6+4I \right)
b_0+8\left(-1+I \right)b_D-4b_F~,\nonumber\\
%\eeqa
%\beqa
\gamma_{n}^K&=&-4I b_0+
\left(4-8I \right)b_D+4b_F~,\nonumber\\
%\end{eqnarray}
%\begin{eqnarray}
\gamma_{\Xi^0}^\pi&=&\left(-6+4I 
\right)b_0+8\left(-1+I \right)
b_D+4b_F~,\nonumber\\
%\eeqa
%\beqa
\gamma_{\Xi^0}^K&=&-4I b_0+
\left(4-8I \right)b_D-4 b_F~,\nonumber\\
%\end{eqnarray}
%\begin{eqnarray}
\gamma_{\Lambda}^\pi&=&\left(-6+4I 
\right)b_0+\left(-4+\frac{8}{3}I +
\frac{8}{3}\cos(2\ve)+\frac{1}{\sqrt{3}}\left(-8+\frac{32}{3}
I \right)\sin(2\ve)\right)b_D~,\nonumber\\
%\eeqa
%\beqa
\gamma_{\Lambda}^K&=&-4I b_0+\left(-\frac{8}{3}I 
-\frac{8}{3}\cos(2\ve)+\frac{1}{\sqrt{3}}
\left(8-\frac{32}{3}I \right)
\sin(2\ve)\right)b_D~,
\end{eqnarray}
with
\beq
I = \frac{\sqrt{3}}{\sqrt{3}-\tan(2\ve)} \simeq 1 + \frac{2\ve}{\sqrt{3}}~,
\eeq
and furthermore $I=1$ in the isospin limit $m_u=m_d$.

\medskip\noindent
 The third order coefficients $\delta_B$ are: 
\beq
\begin{tabular}{ll}
$\delta_{1,\Sigma^+}^\pi = \frac{2}{3}D^2+2F^2$~,&
$\delta_{1,\Sigma^-}^\pi =\frac{2}{3}D^2+2F^2$~,\\
$\delta_{1,\Sigma^0}^\pi = \frac{2}{3}\left(1-\cos(2\ve)\right)D^2
+2\left(1+\cos(2\ve)\right)F^2$~, & 
$\delta_{1,p}^\pi = D^2+2DF+F^2$~,\\
$\delta_{1,\Xi^-}^\pi = D^2-2DF+F^2$~,&\\
$\delta_{1,n}^\pi = D^2+2DF+F^2$~,\\
$\delta_{1,\Xi^0}^\pi = D^2-2DF+F^2$~,&\\
$\delta_{1,\Lambda}^\pi =\frac{2}{3}\left(1+\cos(2\ve)\right)D^2
+2\left(1-\cos(2\ve)\right)F^2$~,&\\
\end{tabular}
\eeq
\beqa
\delta_{2,\Sigma^+}^\pi&=&\left(\frac{1}{3}-\frac{4}{9}I+\frac{1}{9}\cos(2\ve)+\frac{1}{\sqrt{3}}
\left(-\frac{2}{3}+\frac{10}{9}I 
\right)\sin(2\ve)\right)D^2+\left(-\frac{4}{3}+\frac{4}{3}
I\right.\nonumber\\& +&\left.\frac{1}{\sqrt{3}}
\left(2-\frac{4}{3}I \right)
\sin(2\ve)\right)D F+\left(1+\cos(2\ve)+
\frac{2}{\sqrt{3}}\left(-1+I \right)
\sin(2\ve)\right)F^2~,\nonumber\\
%\eeqa
%\beqa
\delta_{2,\Sigma^+}^K&=&\left(\frac{4}{9}I-\frac{4}{9}\cos(2\ve)+\frac{1}{\sqrt{3}}
\left(\frac{2}{3}-\frac{10}{9}I 
\right)\sin(2\ve)\right)D^2+\frac{4}{3}\left(1-I+\frac{ \sin(2\ve)       }{\sqrt{3}}I\right)D F\nonumber\\&+&\frac{2}{\sqrt{3}}\left(1-I\right)\sin(2\ve)F^2~,\nonumber
\eeqa
\beqa
\delta_{2,\Sigma^-}^\pi&=&\left(\frac{1}{3}-\frac{4}{9}I+\frac{1}{9}\cos(2\ve)+\frac{1}{\sqrt{3}}
\left(-\frac{2}{3}+\frac{10}{9}I 
\right)\sin(2\ve)\right)D^2+\left(\frac{4}{3}-\frac{4}{3}
I \right.\nonumber\\&+&\left.\frac{1}{\sqrt{3}}
\left(-2+\frac{4}{3}I\right)\sin(2\ve)\right)DF+\left(1+\cos(2\ve)+\frac{2}{\sqrt{3}}\left(-1+I\right)\sin(2\ve)\right)F^2~,\nonumber\\
%\eeqa
%\beqa
\delta_{2,\Sigma^-}^K&=&\left(\frac{4}{9}I-\frac{4}{9}\cos(2\ve)+\frac{1}{\sqrt{3}}
\left(\frac{2}{3}-\frac{10}{9}I 
\right)\sin(2\ve)\right)D^2+\frac{4}{3}\left(-1+I-\frac{ \sin(2\ve)}{\sqrt{3}} I
\right)DF\nonumber\\&+&\frac{2}{\sqrt{3}}\left(1-I 
\right)\sin(2\ve)F^2~,\nonumber
\eeqa
\beqa
\delta_{2,\Sigma^0}^\pi&=&\left(\frac{2}{3}-\frac{4}{9}I+\frac{4}{9}\cos(2\ve)+\frac{1}{\sqrt{3}}
\left(-\frac{4}{3}+\frac{16}{9}I 
\right)\sin(2\ve)\right)D^2~,\nonumber\\
%\eeqa
%\beqa
\delta_{2,\Sigma^0}^K&=&\left(\frac{4}{9}I-\frac{4}{9}\cos(2\ve)+\frac{1}{\sqrt{3}}
\left(\frac{4}{3}-\frac{16}{9}I 
\right)\sin(2\ve)\right)D^2~,\nonumber
\eeqa
\beqa
\delta_{2,p}^\pi&=&\left(\frac{2}{3}-\frac{4}{9}I+\frac{5}{18}\cos(2\ve)+\frac{1}{\sqrt{3}}
\left(-\frac{7}{6}+\frac{10}{9}I 
\right)\sin(2\ve)\right)D^2+\left(-\frac{2}{3}+\frac{4}{3}
I +\frac{1}{3}\cos(2\ve)\right.\nonumber\\
&+&\left.\frac{1}{\sqrt{3}}\left(1-\frac{4}{3}I\right)\sin(2\ve)\right)DF
+\left(\frac{1}{2}\cos(2\ve)+\frac{1}{\sqrt{3}}\left(-\frac{1}{2}
+2I \right)\sin(2\ve)\right)F^2~,\nonumber\\
%\eeqa
%\beqa
\delta_{2,p}^K&=&\left(-\frac{1}{3}+\frac{4}{9}I-\frac{1}{9}\cos(2\ve)+\frac{1}{\sqrt{3}}\left(\frac{2}
{3}-\frac{10}{9}I \right)
\sin(2\ve)\right)D^2+\left(\frac{2}{3}-\frac{4}{3}
I +\frac{2}{3}\cos(2\ve)\right.\nonumber\\
&+&\left.\frac{4}{3}\frac{ \sin(2\ve)}{\sqrt{3}} I\right)DF+\left(1-\cos(2\ve)+\frac{2}{\sqrt{3}}
\left(1-I \right)\sin(2\ve)
\right)F^2~,\nonumber
\eeqa
\beqa
\delta_{2,\Xi^-}^\pi&=&\left(\frac{2}{3}-\frac{4}{9}I+\frac{5}{18}\cos(2\ve)+\frac{1}{\sqrt{3}}
\left(-\frac{7}{6}+\frac{10}{9}I 
\right)\sin(2\ve)\right)D^2+\left(\frac{2}{3}-\frac{4}{3}
I -\frac{1}{3}\cos(2\ve)\right.\nonumber\\
&+&\left.\frac{1}{\sqrt{3}}\left(-1+\frac{4}{3}I\right)\sin(2\ve)\right)DF+\left(\frac{1}{2}\cos(2\ve)+\frac{1}{\sqrt{3}}\left(-\frac{1}{2}
+2I \right)\sin(2\ve)\right)F^2~,\nonumber\\
\delta_{2,\Xi^-}^K&=&\left(-\frac{1}{3}+\frac{4}{9}I -\frac{1}{9}\cos(2\ve)+\frac{1}{\sqrt{3}}\left(\frac{2}{3}-\frac{10}{9}I \right)\sin(2\ve)\right)D^2+\left(-\frac{2}{3}+\frac{4}{3}I -\frac{2}{3}\cos(2\ve)\right.\nonumber\\&-&\left.\frac{4}{3}\frac{\sin(2\ve) }{\sqrt{3}}I\right)DF+\left(1-\cos(2\ve)+\frac{2}{\sqrt{3}}\left(1-I \right)\sin(2\ve)\right)F^2~,\nonumber
\eeqa
\beqa
\delta_{2,n}^\pi&=&\left(\frac{2}{9}I +\frac{5}{18}\cos(2\ve)+\frac{1}{\sqrt{3}}\left(-\frac{1}{6}+\frac{4}{9}I \right)\sin(2\ve)\right)D^2+\left(\frac{2}{3}+\frac{1}{3}\cos(2\ve)-\frac{\sin(2\ve)}{\sqrt{3}}\right)DF\nonumber\\&+&\left(2-2I +\frac{1}{2}\cos(2\ve)+\frac{1}{\sqrt{3}}\left(-\frac{7}{2}+4I \right)\sin(2\ve)\right)F^2~,\nonumber\\
%\eeqa
%\beqa
\delta_{2,n}^K&=&\left(\frac{1}{3}-\frac{2}{9}I -\frac{1}{9}\cos(2\ve)+\frac{1}{\sqrt{3}}\left(\frac{2}{3}-\frac{4}{9}I \right)\sin(2\ve)\right)D^2+\frac{2}{3}\left(-1+\cos(2\ve)\right)DF\nonumber\\&+&\left(-1+2I -\cos(2\ve)+\frac{1}{\sqrt{3}}\left(2-4I \right)\sin(2\ve)\right)F^2~,\nonumber
\eeqa
\beqa
\delta_{2,\Xi^0}^\pi&=&\left(\frac{2}{9}I +\frac{5}{18}\cos(2\ve)+\frac{1}{\sqrt{3}}\left(-\frac{1}{6}+\frac{4}{9}I \right)\sin(2\ve)\right)D^2+\left(-\frac{2}{3}-\frac{1}{3}\cos(2\ve)+\frac{\sin(2\ve)}{\sqrt{3}}\right)DF\nonumber\\&+&\left(2-2I +\frac{1}{2}\cos(2\ve)+\frac{1}{\sqrt{3}}\left(-\frac{7}{2}+4I \right)\sin(2\ve)\right)F^2~,\nonumber\\
%\eeqa
%\beqa
\delta_{2,\Xi^0}^K&=&\left(\frac{1}{3}-\frac{2}{9}I -\frac{1}{9}\cos(2\ve)+\frac{1}{\sqrt{3}}\left(\frac{2}{3}-\frac{4}{9}I \right)\sin(2\ve)\right)D^2+\frac{2}{3}\left(1-\cos(2\ve)\right)DF\nonumber\\&+&\left(-1+2I -\cos(2\ve)+\frac{1}{\sqrt{3}}\left(2-4I \right)\sin(2\ve)\right)F^2~,\nonumber
\eeqa
\beqa
\delta_{2,\Lambda}^\pi&=&\left(\frac{2}{3}-\frac{4}{9}I +\frac{4}{9}\cos(2\ve)+\frac{1}{\sqrt{3}}\left(-\frac{4}{3}+\frac{16}{9}I \right)\sin(2\ve)\right)D^2~,\nonumber\\
%\eeqa
%\beqa
\delta_{2,\Lambda}^K&=&\left(\frac{4}{9}I -\frac{4}{9}\cos(2\ve)+\frac{1}{\sqrt{3}}\left(\frac{4}{3}-\frac{16}{9}I \right)\sin(2\ve)\right)D^2~,
\eeqa
%\beqa
%\delta_{3,\Sigma^+}^\pi&=&0~,
%\eeqa
\beq
\begin{tabular}{ll}
$\delta_{3,\Sigma^+}^K=D^2+2D F+F^2$~,&
%\eeqa
%\beqa
%\delta_{3,\Sigma^-}^\pi&=&0~,
%\eeqa
%\beqa
$\delta_{3,\Sigma^-}^K=D^2-2DF+F^2$~,\\\end{tabular}\nonumber\eeq
%\eeqa
%\beqa
%\delta_{3,\Sigma^0}^\pi&=&0~,
%\eeqa
%\beqa
\beq
\delta_{3,\Sigma^0}^K=\left(\frac{2}{3}+\frac{1}{3}\cos(2\ve)-\frac{\sin(2\ve)}{\sqrt{3}}\right)D^2+\left(2-\cos(2\ve)+\sqrt{3}\sin(2\ve)\right)F^2~,
\nonumber\eeq
%\eeqa
%\beqa
%\delta_{3,p}^\pi&=&0~,
%\eeqa
%\beqa
\beq
\begin{tabular}{ll}
$\delta_{3,p}^K=\frac{2}{3}D^2+2F^2$~,&
%\eeqa
%\beqa
%\delta_{3,\Xi^-}^\pi&=&0~,\eeqa
%\beqa
$\delta_{3,\Xi^-}^K=\frac{2}{3}D^2+2F^2$~,\\
%\eeqa
%\beqa
%\delta_{3,n}^\pi&=&0~,
%\eeqa
%\beqa
$\delta_{3,n}^K=D^2-2DF+F^2$~,&
%\eeqa
%\beqa
%\delta_{3,\Xi^0}^\pi&=&0~,
%\eeqa
%\beqa
$\delta_{3,\Xi^0}^K=D^2+2DF+F^2$~,\\
\end{tabular}\eeq
%\eeqa
%\beqa
%\delta_{3,\Lambda}^\pi&=&0~,
%\eeqa
%\beqa
\beq
\delta_{3,\Lambda}^K=\left(\frac{2}{3}-\frac{1}{3}\cos(2\ve)+\frac{\sin(2\ve)}{\sqrt{3}}\right)D^2+\left(2+\cos(2\ve)-\sqrt{3}\sin(2\ve)\right)F^2~,\nonumber
\eeq
\beq\begin{tabular}{ll}
$\delta_{4,\Sigma^+}^\pi=2 \left(1-I \right)\left(D^2-2D F+F^2\right)$~,&
%\eeqa
%\beqa
$\delta_{4,\Sigma^+}^K=\left(-1+2I \right)\left(D^2-2 D F+F^2\right)$~,\\
%\eeqa
%\beqa
$\delta_{4,\Sigma^-}^\pi=2\left(1-I \right)\left(D^2+2DF+F^2\right)$~,&
%\eeqa
%\beqa
$\delta_{4,\Sigma^-}^K=\left(-1+2I \right)\left(D^2+2DF+F^2\right)$~,\\\end{tabular}\nonumber
\eeq
\beqa
\delta_{4,\Sigma^0}^\pi&=&\left(\frac{4}{3}-\frac{4}{3}I +\frac{1}{\sqrt{3}}\left(2-\frac{8}{3}I \right)\sin(2\ve)\right)D^2+\left(4-4I +\frac{1}{\sqrt{3}}\left(-6+8I \right)\sin(2\ve)\right)F^2~,
\nonumber\\
%\beqa
\delta_{4,\Sigma^0}^K&=&\left(-\frac{2}{3}+\frac{4}{3}I +\frac{1}{3}\cos(2\ve)+\frac{1}{\sqrt{3}}\left(-1+\frac{8}{3}I \right)\sin(2\ve)\right)D^2\nonumber\\&+&\left(-2+4I -\cos(2\ve)+\frac{1}{\sqrt{3}}\left(3-8I \right)\sin(2\ve)\right)F^2~,\nonumber
\eeqa
\beq\begin{tabular}{ll}
$\delta_{4,p}^\pi=2\left(1-I \right)\left(D^2-2DF+F^2\right)$~,&
%\eeqa
%\beqa
$\delta_{4,p}^K=\left(-1+2I \right)\left(D^2-2DF+F^2\right)$~,\\
%\eeqa
%\beqa
$\delta_{4,\Xi^-}^\pi=2\left(1-I \right)\left(D^2+2DF+F^2\right)$~,&
%\eeqa
%\beqa
$\delta_{4,\Xi^-}^K=\left(-1+2I \right)\left(D^2+2DF+F^2\right)$~,\\
%\eeqa
%\beqa
$\delta_{4,n}^\pi=\frac{4}{3}\left(1-I \right)D^2+4\left(1-I \right)F^2$~,&
%\eeqa
%\beqa
$\delta_{4,n}^K=\left(-\frac{2}{3}+\frac{4}{3}I \right)D^2+\left(-2+4I \right)F^2$~,\\
%\beqa
$\delta_{4,\Xi^0}^\pi=\frac{4}{3}\left(1-I \right)D^2+4\left(1-I \right)F^2$~,&
%\eeqa
%\beqa
$\delta_{4,\Xi^0}^K=\left(-\frac{2}{3}+\frac{4}{3}I \right)D^2+\left(-2+4I \right)F^2$~,\\
\end{tabular}
\eeq
\beqa
\delta_{4,\Lambda}^\pi&=&\left(\frac{4}{3}-\frac{4}{3}I +\frac{1}{\sqrt{3}}\left(-2+\frac{8}{3}I \right)\sin(2\ve)\right)D^2+\left(4-4I +\frac{1}{\sqrt{3}}\left(6-8I \right)\sin(2\ve)\right)F^2~,\nonumber\\
%\eeqa
%\beqa
\delta_{4,\Lambda}^K&=&\left(-\frac{2}{3}+\frac{4}{3}I -\frac{1}{3}\cos(2\ve)+\frac{1}{\sqrt{3}}\left(1-\frac{8}{3}I \right)\sin(2\ve)\right)D^2\nonumber\\&+&\renewcommand{\beqa}{\begin{eqnarray}}\left(-2+4I +\cos(2\ve)+\frac{1}{\sqrt{3}}\left(-3+8I \right)\sin(2\ve)\right)F^2~,\nonumber
\eeqa
\beqa
\delta_{5,\Sigma^+}^\pi&=&\left(\frac{1}{3}-\frac{4}{9}I -\frac{1}{9}\cos(2\ve)+\frac{1}{\sqrt{3}}\left(\frac{2}{3}-\frac{10}{9}I \right)\sin(2\ve)\right)D^2+\left(-\frac{4}{3}+\frac{4}{3}I\right.\nonumber\\& +&\left.\frac{1}{\sqrt{3}}\left(-2+\frac{4}{3}I \right)\sin(2\ve)\right)D F+\left(1-\cos(2\ve)+\frac{2}{\sqrt{3}}\left(1-I \right)\sin(2\ve)\right)F^2~,\nonumber\\
%\eeqa
%\beqa
\delta_{5,\Sigma^+}^K&=&\left(\frac{4}{9}I +\frac{4}{9}\cos(2\ve)+\frac{1}{\sqrt{3}}\left(-\frac{2}{3}+\frac{10}{9}I \right)\sin(2\ve)\right)D^2+\frac{4}{3}\left(1-I -\frac{ \sin(2\ve)}{\sqrt{3}}I\right)DF\nonumber\\&+&\frac{2}{\sqrt{3}}\left(-1+I \right)\sin(2\ve)F^2~,\nonumber
\eeqa
\beqa
\delta_{5,\Sigma^-}^\pi&=&\left(\frac{1}{3}-\frac{4}{9}I -\frac{1}{9}\cos(2\ve)+\frac{1}{\sqrt{3}}\left(\frac{2}{3}-\frac{10}{9}I \right)\sin(2\ve)\right)D^2+\left(\frac{4}{3}-\frac{4}{3}I\right.\nonumber\\& +&\left.\frac{1}{\sqrt{3}}\left(2-\frac{4}{3}I \right)\sin(2\ve)\right)DF+\left(1-\cos(2\ve)+\frac{2}{\sqrt{3}}\left(1-I \right)\sin(2\ve)\right)F^2~,\nonumber\\
%\eeqa
%\beqa
\delta_{5,\Sigma^-}^K&=&\left(\frac{4}{9}I +\frac{4}{9}\cos(2\ve)+\frac{1}{\sqrt{3}}\left(-\frac{2}{3}+\frac{10}{9}I \right)\sin(2\ve)\right)D^2+\frac{4}{3}\left(-1+I +\frac{ \sin(2\ve)}{\sqrt{3}}I\right)DF\nonumber\\&+&\frac{2}{\sqrt{3}}\left(-1+I \right)\sin(2\ve)F^2~,\nonumber
\eeqa
\beqa
\delta_{5,\Sigma^0}^\pi&=&\left(\frac{2}{3}-\frac{4}{9}I -\frac{4}{9}\cos(2\ve)+\frac{1}{\sqrt{3}}\left(\frac{4}{3}-\frac{16}{9}I \right)\sin(2\ve)\right)D^2~,\nonumber\\
%\eeqa
%\beqa
\delta_{5,\Sigma^0}^K&=&\left(\frac{4}{9}I +\frac{4}{9}\cos(2\ve)+\frac{1}{\sqrt{3}}\left(-\frac{4}{3}+\frac{16}{9}I \right)\sin(2\ve)\right)D^2~,\nonumber
\eeqa
\beqa
\delta_{5,p}^\pi&=&\left(\frac{2}{3}-\frac{4}{9}I -\frac{5}{18}\cos(2\ve)+\frac{1}{\sqrt{3}}\left(\frac{7}{6}-\frac{10}{9}I \right)\sin(2\ve)\right)D^2+\left(-\frac{2}{3}+\frac{4}{3}I-\frac{1}{3}\cos(2\ve)\right.\nonumber\\&+&\left.\frac{1}{\sqrt{3}}\left(-1+\frac{4}{3}I \right)\sin(2\ve)\right)DF+\left(-\frac{1}{2}\cos(2\ve)+\frac{1}{\sqrt{3}}\left(\frac{1}{2}-2I \right)\sin(2\ve)\right)F^2~,\nonumber\\
%\eeqa
%\beqa
\delta_{5,p}^K&=&\left(-\frac{1}{3}+\frac{4}{9}I +\frac{1}{9}\cos(2\ve)+\frac{1}{\sqrt{3}}\left(-\frac{2}{3}+\frac{10}{9}I \right)\sin(2\ve)\right)D^2+\left(\frac{2}{3}-\frac{4}{3}I -\frac{2}{3}\cos(2\ve)\right.\nonumber\\&-&\left.\frac{4}{3}\frac{ \sin(2\ve)}{\sqrt{3}}I\right)DF+\left(1+\cos(2\ve)+\frac{2}{\sqrt{3}}\left(-1+I \right)\sin(2\ve)\right)F^2~,\nonumber
\eeqa
\beqa
\delta_{5,\Xi^-}^\pi&=&\left(\frac{2}{3}-\frac{4}{9}I -\frac{5}{18}\cos(2\ve)+\frac{1}{\sqrt{3}}\left(\frac{7}{6}-\frac{10}{9}I \right)\sin(2\ve)\right)D^2+\left(\frac{2}{3}-\frac{4}{3}I +\frac{1}{3}\cos(2\ve)\right.\nonumber\\&+&\left.\frac{1}{\sqrt{3}}\left(1-\frac{4}{3}I \right)\sin(2\ve)\right)DF\nonumber+\left(-\frac{1}{2}\cos(2\ve)+\frac{1}{\sqrt{3}}\left(\frac{1}{2}-2I \right)\sin(2\ve)\right)F^2~,\nonumber\\
%\eeqa
%\beqa
\delta_{5,\Xi^-}^K&=&\left(-\frac{1}{3}+\frac{4}{9}I +\frac{1}{9}\cos(2\ve)+\frac{1}{\sqrt{3}}\left(-\frac{2}{3}+\frac{10}{9}I \right)\sin(2\ve)\right)D^2+\left(-\frac{2}{3}+\frac{4}{3}I +\frac{2}{3}\cos(2\ve)\right.\nonumber\\&+&\left.\frac{4}{3}\frac{ \sin(2\ve)}{\sqrt{3}}I\right)DF+\left(1+\cos(2\ve)+\frac{2}{\sqrt{3}}\left(-1+I \right)\sin(2\ve)\right)F^2~,\nonumber
\eeqa
\beqa
\delta_{5,n}^\pi&=&\left(\frac{2}{9}I -\frac{5}{18}\cos(2\ve)+\frac{1}{\sqrt{3}}\left(\frac{1}{6}-\frac{4}{9}I \right)\sin(2\ve)\right)D^2+\left(\frac{2}{3}-\frac{1}{3}\cos(2\ve)+\frac{\sin(2\ve)}{\sqrt{3}}\right)DF\nonumber\\&+&\left(2-2I -\frac{1}{2}\cos(2\ve)+\frac{1}{\sqrt{3}}\left(\frac{7}{2}-4I \right)\sin(2\ve)\right)F^2~,\nonumber\\
%\eeqa
%\beqa
\delta_{5,n}^K&=&\left(\frac{1}{3}-\frac{2}{9}I +\frac{1}{9}\cos(2\ve)+\frac{1}{\sqrt{3}}\left(-\frac{2}{3}+\frac{4}{9}I \right)\sin(2\ve)\right)D^2-\frac{2}{3}\left(1+\cos(2\ve)\right)DF\nonumber\\&+&\left(-1+2I +\cos(2\ve)+\frac{1}{\sqrt{3}}\left(-2+4I \right)\sin(2\ve)\right)F^2~,\nonumber
\eeqa
\beqa
\delta_{5,\Xi^0}^\pi&=&\left(\frac{2}{9}I -\frac{5}{18}\cos(2\ve)+\frac{1}{\sqrt{3}}\left(\frac{1}{6}-\frac{4}{9}I \right)\sin(2\ve)\right)D^2+\left(-\frac{2}{3}+\frac{1}{3}\cos(2\ve)-\frac{\sin(2\ve)}{\sqrt{3}}\right)DF\nonumber\\&+&\left(2-2I -\frac{1}{2}\cos(2\ve)+\frac{1}{\sqrt{3}}\left(\frac{7}{2}-4I \right)\sin(2\ve)\right)F^2~,\nonumber\\
%\eeqa
%\beqa
\delta_{5,\Xi^0}^K&=&\left(\frac{1}{3}-\frac{2}{9}I +\frac{1}{9}\cos(2\ve)+\frac{1}{\sqrt{3}}\left(-\frac{2}{3}+\frac{4}{9}I \right)\sin(2\ve)\right)D^2+\frac{2}{3}\left(1+\cos(2\ve)\right)DF\nonumber\\&+&\left(-1+2I +\cos(2\ve)+\frac{1}{\sqrt{3}}\left(-2+4I \right)\sin(2\ve)\right)F^2~,\nonumber
\eeqa
\beqa
\delta_{5,\Lambda}^\pi&=&\left(\frac{2}{3}-\frac{4}{9}I -\frac{4}{9}\cos(2\ve)+\frac{1}{\sqrt{3}}\left(\frac{4}{3}-\frac{16}{9}I \right)\sin(2\ve)\right)D^2~,\nonumber\\
%\eeqa
%\beqa
\delta_{5,\Lambda}^K&=&\left(\frac{4}{9}I +\frac{4}{9}\cos(2\ve)+\frac{1}{\sqrt{3}}\left(-\frac{4}{3}+\frac{16}{9}I \right)\sin(2 \ve)\right)D^2~.
\end{eqnarray}

\medskip\noindent
Finally, the fourth order coefficients $\epsilon_B$ are (since these formulas are very long, 
we only give here the expressions for the proton and the neutron. The others, including the coefficients $\eta_i,\,i=1,\ldots7$, in Eq.~(\ref{etafactors}) of the non-diagonal contributions to the $\Sigma^0$ and $\Lambda$ masses, can be obtained
from the authors upon request):
\beqa
\epsilon_{1,p}^\pi&=&4b_0+\left(2-4D^2-8DF-4F^2\right)\left(b_D+b_F\right)-2\left(b_1+b_2+b_3\right)-4b_4+\frac{m_0}{2}\left(b_5+b_6+b_7\right)\nonumber\\&+&m_0b_8+\left(4D^2+8DF+4F^2\right)\left(b_D+b_F\right)I~,\nonumber\\
%\eeqa
%\beqa
%\epsilon_{1,p}^K&=&
%\eeqa
%\beqa
\epsilon_{1,p}^{\pi K}&=&\left(4D^2+8DF+4F^2\right)\left(b_D+b_F\right)\left(1-I\right)~,\nonumber
\eeqa
\beqa
\epsilon_{1,n}^\pi&=&4b_0+\left(2+4D^2+8DF+4F^2\right)\left(b_D+b_F\right)-2\left(b_1+b_2+b_3\right)-4b_4+\frac{m_0}{2}\left(b_5+b_6+b_7\right)\nonumber\\&+&m_0b_8-\left(4D^2+8DF+4F^2\right)\left(b_D+b_F\right)I~,\nonumber\\
%\eeqa
%\beqa
%\epsilon_{1,n}^K&=&
%\eeqa
%\beqa
\epsilon_{1,n}^{\pi K}&=&\left(4D^2+8DF+4F^2\right)\left(b_D+b_F\right)\left(-1+I\right)~,
\eeqa

\beqa
\epsilon_{2,p}^\pi&=&\frac{17}{3}b_0+\frac{2}{3}b_D-\frac{4}{3}b_F-\frac{2}{3}b_1+\frac{4}{3}b_2-\frac{26}{9}b_3-\frac{14}{3}b_4+\frac{m_0}{6}b_5-\frac{m_0}{3}b_6+\frac{13}{18}m_0b_7+\frac{7}{6}m_0b_8\nonumber\\&+&\left(4b_0+\frac{8}{3}\left(b_D-b_1\right)+\frac{16}{9}b_2-\frac{56}{27}b_3-\frac{40}{9}b_4+\frac{2}{3}m_0b_5-\frac{4}{9}m_0b_6+\frac{14}{27}m_0b_7+\frac{10}{9}m_0b_8\right)I^2\nonumber\\&+&\left(-8b_0+\frac{8}{3}\left(-b_D+b_F+b_1\right)-\frac{32}{9}b_2+\frac{40}{9}b_3+8b_4-\frac{2}{3}m_0b_5+\frac{8}{9}m_0b_6-\frac{10}{9}m_0b_7\right.\nonumber\\&-&\left.2m_0b_8\right)I+\left(\frac{4}{3}b_0+\frac{1}{3}b_D+\frac{5}{3}b_F-\frac{1}{3}b_1-\frac{5}{9}b_2-\frac{13}{27}b_3-\frac{8}{9}b_4+\frac{m_0}{12}b_5+\frac{5}{36}m_0b_6\right.\nonumber\\&+&\left.\frac{13}{108}m_0b_7+\frac{2}{9}m_0b_8\right)\cos(2\ve)+\frac{1}{\sqrt{3}}\left(-10b_0-b_D+3b_F+b_1-\frac{7}{3}b_2+5b_3+8b_4\right.\nonumber\\&-&\left.\frac{m_0}{4}b_5+\frac{7}{12}m_0b_6-\frac{5}{4}m_0b_7-2m_0b_8+\left(-8b_0-\frac{8}{3}b_D+8b_F+\frac{8}{3}b_1-\frac{40}{9}b_2+\frac{104}{27}b_3\right.\right.\nonumber\\&+&\left.\left.\frac{64}{9}b_4-\frac{2}{3}m_0b_5+\frac{10}{9}m_0b_6-\frac{26}{27}m_0b_7-\frac{16}{9}m_0b_8\right)I^2+\left(16b_0+4b_D-\frac{20}{3}b_F-4b_1\right.\right.\nonumber\\&+&\left.\left.\frac{52}{9}b_2-\frac{220}{27}b_3-\frac{128}{9}b_4+m_0b_5-\frac{13}{9}m_0b_6+\frac{55}{27}m_0b_7+\frac{32}{9}m_0b_8\right)I\right)\sin(2\ve)~,\nonumber
\eeqa

\beqa
\epsilon_{2,p}^K&=&\frac{8}{3}\left(b_0+b_D+b_F-b_1\right)-\frac{8}{9}b_3-\frac{8}{3}b_4+\frac{2}{3}m_0b_5+\frac{2}{9}m_0b_7+\frac{2}{3}m_0b_8+\left(4b_0\right.\nonumber\\&+&\left.\frac{8}{3}\left(b_D-b_1\right)+\frac{16}{9}b_2-\frac{56}{27}b_3-\frac{40}{9}b_4+\frac{2}{3}m_0b_5-\frac{4}{9}m_0b_6+\frac{14}{27}m_0b_7+\frac{10}{9}m_0b_8\right)I^2\nonumber\\&+&\left(-4b_0-\frac{8}{3}b_D-\frac{16}{3}b_F+\frac{8}{3}b_1-\frac{8}{9}b_2+\frac{8}{3}b_3+\frac{16}{3}b_4-\frac{2}{3}m_0b_5+\frac{2}{9}m_0b_6-\frac{2}{3}m_0b_7\right.\nonumber\\&-&\left.\frac{4}{3}m_0b_8\right)I+\left(\frac{8}{3}\left(-b_0-b_D+b_F+b_1\right)-\frac{8}{9}b_2+\frac{8}{27}b_3+\frac{16}{9}b_4-\frac{2}{3}m_0b_5+\frac{2}{9}m_0b_6\right.\nonumber\\&-&\left.\frac{2}{27}m_0b_7-\frac{4}{9}m_0b_8\right)\cos(2\ve)+\frac{1}{\sqrt{3}}\left(4\left(b_D+b_F-b_1\right)+\frac{4}{3}\left(-b_2+b_3\right)+m_0b_5\right.\nonumber\\&+&\left.\frac{m_0}{3}\left(b_6-b_7\right)+\left(-8b_0-\frac{8}{3}b_D+8b_F+\frac{8}{3}b_1-\frac{40}{9}b_2+\frac{104}{27}b_3+\frac{64}{9}b_4-\frac{2}{3}m_0b_5\right.\right.\nonumber\\&+&\left.\left.\frac{10}{9}m_0b_6-\frac{26}{27}m_0b_7-\frac{16}{9}m_0b_8\right)I^2+\left(4\left(b_0-b_D\right)-\frac{20}{3}b_F+4b_1+\frac{28}{9}b_2-\frac{100}{27}b_3\right.\right.\nonumber\\&-&\left.\left.\frac{32}{9}b_4-m_0b_5-\frac{7}{9}m_0b_6+\frac{25}{27}m_0b_7+\frac{8}{9}m_0b_8\right)I\right)\sin(2\ve)~,\nonumber
\eeqa

\vfill\eject

\beqa
\epsilon_{2,p}^{\pi K}&=&-\frac{16}{3}b_0+\frac{4}{3}\left(-b_D-b_F+b_1-b_2\right)+\frac{28}{9}b_3+\frac{16}{3}b_4+\frac{m_0}{3}\left(-b_5+b_6\right)-\frac{7}{9}m_0b_7\nonumber\\&-&\frac{4}{3}m_0b_8+\left(-8b_0+\frac{16}{3}\left(-b_D+b_1\right)-\frac{32}{9}b_2+\frac{112}{27}b_3+\frac{80}{9}b_4-\frac{4}{3}m_0b_5+\frac{8}{9}m_0b_6\right.\nonumber\\&-&\left.\frac{28}{27}m_0b_7-\frac{20}{9}m_0b_8\right)I^2+\left(12b_0+\frac{16}{3}b_D+\frac{8}{3}b_F-\frac{16}{3}b_1+\frac{40}{9}b_2-\frac{64}{9}b_3-\frac{40}{3}b_4\right.\nonumber\\&+&\left.\frac{4}{3}m_0b_5-\frac{10}{9}m_0b_6+\frac{16}{9}m_0b_7+\frac{10}{3}m_0b_8\right)I+\left(\frac{4}{3}\left(b_0+b_D-b_F-b_1\right)+\frac{4}{9}b_2\right.\nonumber\\&-&\left.\frac{4}{27}b_3-\frac{8}{9}b_4+\frac{m_0}{3}b_5-\frac{m_0}{9}b_6+\frac{m_0}{27}b_7+\frac{2}{9}m_0b_8\right)\cos(2\ve)\nonumber\\&+&\frac{1}{\sqrt{3}}\left(10b_0-4b_F+\frac{8}{3}b_2-\frac{16}{3}b_3-8b_4-\frac{2}{3}m_0b_6+\frac{4}{3}m_0b_7+2m_0b_8\right.\nonumber\\&+&\left.\left(16b_0+\frac{16}{3}b_D-16b_F-\frac{16}{3}b_1+\frac{80}{9}b_2-\frac{208}{27}b_3-\frac{128}{9}b_4+\frac{4}{3}m_0b_5-\frac{20}{9}m_0b_6\right.\right.\nonumber\\&+&\left.\left.\frac{52}{27}m_0b_7+\frac{32}{9}m_0b_8\right)I^2+\left(-20b_0+\frac{40}{3}b_F-\frac{80}{9}b_2+\frac{320}{27}b_3+\frac{160}{9}b_4\right.\right.\nonumber\\&+&\left.\left.\frac{20}{9}m_0b_6-\frac{80}{27}m_0b_7-\frac{40}{9}m_0b_8\right)I\right)\sin(2\ve)~,\nonumber
\eeqa

\beqa
\epsilon_{2,n}^\pi&=&\frac{17}{3}b_0+6b_D+4b_F-\frac{26}{3}b_1-\frac{4}{3}b_2-\frac{2}{9}b_3-\frac{14}{3}b_4+\frac{13}{6}m_0b_5+\frac{m_0}{3}b_6+\frac{m_0}{18}b_7\nonumber\\&+&\frac{7}{6}m_0b_8+\left(4b_0+\frac{8}{3}b_D-8b_1-\frac{8}{27}b_3-\frac{40}{9}b_4+2m_0b_5+\frac{2}{27}m_0b_7+\frac{10}{9}m_0b_8\right)I^2\nonumber\\&+&\left(-8\left(b_0+b_D\right)-\frac{8}{3}b_F+16b_1+\frac{8}{9}b_2+8b_4-4m_0b_5-\frac{2}{9}m_0b_6-2m_0b_8\right)I\nonumber\\&+&\left(\frac{4}{3}b_0+\frac{1}{3}b_D+\frac{5}{3}b_F-\frac{1}{3}b_1-\frac{5}{9}b_2-\frac{13}{27}b_3-\frac{8}{9}b_4+\frac{m_0}{12}b_5+\frac{5}{36}m_0b_6+\frac{13}{108}m_0b_7\right.\nonumber\\&+&\left.\frac{2}{9}m_0b_8\right)\cos(2\ve)+\frac{1}{\sqrt{3}}\left(-10b_0-11b_D-7b_F+15b_1+\frac{7}{3}b_2+\frac{1}{3}b_3+8b_4\right.\nonumber\\&-&\left.\frac{15}{4}m_0b_5-\frac{7}{12}m_0b_6-\frac{m_0}{12}b_7-2m_0b_8+\left(-8b_0-\frac{32}{3}b_D+16b_1-\frac{16}{27}b_3\right.\right.\nonumber\\&+&\left.\left.\frac{64}{9}b_4-4m_0b_5+\frac{4}{27}m_0b_7-\frac{16}{9}m_0b_8\right)I^2+\left(16b_0+\frac{52}{3}b_D+\frac{20}{3}b_F-28b_1\right.\right.\nonumber\\&-&\left.\left.\frac{20}{9}b_2-\frac{4}{27}b_3-\frac{128}{9}b_4+7m_0b_5+\frac{5}{9}m_0b_6+\frac{m_0}{27}b_7+\frac{32}{9}m_0b_8\right)I\right)\sin(2\ve)~,\nonumber
\eeqa

\vfill\eject

\beqa
\epsilon_{2,n}^K&=&\frac{8}{3}\left(b_0-b_1\right)-\frac{8}{9}b_3-\frac{8}{3}b_4+\frac{2}{3}m_0 b_5+\frac{2}{9}m_0 b_7+\frac{2}{3}m_0 b_8\nonumber\\&+&\left(4 b_0+\frac{8}{3} b_D-8b_1-\frac{8}{27}b_3-\frac{40}{9}b_4+2m_0b_5+\frac{2}{27}m_0b_7+\frac{10}{9}m_0b_8\right)I^2\nonumber\\&+&\left(-4b_0-\frac{8}{3}b_F+8b_1+\frac{8}{9}\left(b_2+b_3\right)+\frac{16}{3}b_4-2m_0b_5-\frac{2}{9}m_0\left(b_6+b_7\right)-\frac{4}{3}m_0b_8\right)I\nonumber\\&+&\left(\frac{8}{3}\left(-b_0-b_D+b_F+b_1\right)-\frac{8}{9}b_2+\frac{8}{27}b_3+\frac{16}{9}b_4-\frac{2}{3}m_0b_5+\frac{2}{9}m_0b_6-\frac{2}{27}m_0b_7\right.\nonumber\\&-&\left.\frac{4}{9}m_0b_8\right)\cos(2\ve)+\frac{1}{\sqrt{3}}\left(4\left(-b_D-b_F+b_1\right)+\frac{4}{3}\left(b_2-b_3\right)-m_0b_5+\frac{m_0}{3}\left(-b_6+b_7\right)\right.\nonumber\\&+&\left.\left(-8b_0-\frac{32}{3}b_D+16b_1-\frac{16}{27}b_3+\frac{64}{9}b_4-4m_0b_5+\frac{4}{27}m_0b_7-\frac{16}{9}m_0b_8\right)I^2\right.\nonumber\\&+&\left.\left(4b_0+\frac{28}{3}b_D+\frac{20}{3}b_F-12b_1-\frac{20}{9}b_2+\frac{44}{27}b_3-\frac{32}{9}b_4+3m_0b_5+\frac{5}{9}m_0b_6\right.\right.\nonumber\\&-&\left.\left.\frac{11}{27}m_0b_7+\frac{8}{9}m_0b_8\right)I\right)\sin(2\ve)~,\nonumber
\eeqa

\beqa
\epsilon_{2,n}^{\pi K}&=&-\frac{16}{3}b_0-4\left(b_D+b_F\right)+\frac{28}{3}b_1+\frac{4}{3}b_2+\frac{4}{9}b_3+\frac{16}{3}b_4-\frac{7}{3}m_0 b_5-\frac{m_0}{3}b_6-\frac{m_0}{9} b_7\nonumber\\&-&\frac{4}{3}m_0 b_8+\left(-8 b_0-\frac{16}{3} b_D+16b_1+\frac{16}{27}b_3+\frac{80}{9}b_4-4m_0 b_5-\frac{4}{27}m_0 b_7-\frac{20}{9}m_0 b_8\right)I^2\nonumber\\&+&\left(12b_0+8b_D+\frac{16}{3}b_F-24b_1-\frac{16}{9}b_2-\frac{8}{9}b_3-\frac{40}{3}b_4+6m_0b_5+\frac{4}{9}m_0b_6+\frac{2}{9}m_0b_7\right.\nonumber\\&+&\left.\frac{10}{3}m_0b_8\right)I+\left(\frac{4}{3}\left(b_0+b_D-b_F-b_1\right)+\frac{4}{9}b_2-\frac{4}{27}b_3-\frac{8}{9}b_4+\frac{m_0}{3}b_5-\frac{m_0}{9}b_6\right.\nonumber\\&+&\left.\frac{m_0}{27}b_7+\frac{2}{9}m_0b_8\right)\cos(2\ve)+\frac{1}{\sqrt{3}}\left(10b_0+12b_D+8b_F-16b_1-\frac{8}{3}b_2-8b_4\right.\nonumber\\&+&\left.4m_0b_5+\frac{2}{3}m_0b_6+2m_0b_8+\left(16b_0+\frac{64}{3}b_D-32b_1+\frac{32}{27}b_3-\frac{128}{9}b_4+8m_0b_5\right.\right.\nonumber\\&-&\left.\left.\frac{8}{27}m_0b_7+\frac{32}{9}m_0b_8\right)I^2+\left(-20b_0-\frac{80}{3}b_D-\frac{40}{3}b_F+40b_1+\frac{40}{9}b_2-\frac{40}{27}b_3\right.\right.\nonumber\\&+&\left.\left.\frac{160}{9}b_4-10m_0b_5-\frac{10}{9}m_0b_6+\frac{10}{27}m_0b_7-\frac{40}{9}m_0b_8\right)I\right)\sin(2\ve)~,
\eeqa

\beqa
\epsilon_{3,p}^K&=&4b_0+\left(4+\frac{8}{3}D^2+\frac{8}{3}DF\right)b_D+\left(\frac{4}{3}D^2+4F^2\right)b_F-4\left(b_1+b_3+b_4\right)\nonumber\\&+&m_0\left(b_5+b_7+b_8\right)+\frac{m_0}{2}b_9-\left(\left(\frac{16}{9}D^2+\frac{16}{3}DF\right)b_D+\left(\frac{8}{3}D^2+8F^2\right)b_F\right)I~,\nonumber\\
%\eeqa
%\beqa
\epsilon_{3,p}^{\pi K}&=&-\frac{8}{3}\left(D^2+DF\right)b_D-\left(\frac{4}{3}D^2+4F^2\right)b_F\nonumber\\&+&\left(\left(\frac{16}{9}D^2+\frac{16}{3}DF\right)b_D+\left(\frac{8}{3}D^2+8F^2\right)b_F\right)I~,\nonumber
\eeqa

\beqa
\epsilon_{3,n}^K&=&4b_0+\left(2-2D^2+4DF-2F^2\right)\left(b_D-b_F\right)+2\left(-b_1+b_2-b_3\right)-4b_4\nonumber\\&+&\frac{m_0}{2}\left(b_5-b_6+b_7\right)+m_0b_8+\left(4D^2-8DF+4F^2\right)\left(b_D-b_F\right)I~,\nonumber\\
%\eeqa
%\beqa
\epsilon_{3,n}^{\pi K}&=&\left(2D^2-4DF+2F^2\right)\left(b_D-b_F\right)+\left(-4D^2+8DF-4F^2\right)\left(b_D-b_F\right)I~,
\eeqa

\beqa
\epsilon_{4,p}^\pi&=&16b_0+\left(8-4D^2+8DF-4F^2\right)\left(b_D-b_F\right)+8\left(-b_1+b_2-b_3\right)-16b_4\nonumber\\&+&2m_0\left( b_5-b_6+b_7\right)+4m_0 b_8+\left(16 b_0+8\left( b_D-b_F-b_1+b_2-b_3\right)-16b_4\right.\nonumber\\&+&\left.2m_0 \left(b_5- b_6+b_7\right)+4m_0 b_8\right)I^2+\left(-32b_0+\left(-16+4D^2-8DF+4F^2\right)\left(b_D-b_F\right)\right.\nonumber\\&+&\left.16\left(b_1-b_2+b_3\right)+32b_4+4m_0\left(-b_5+b_6-b_7\right)-8m_0b_8\right)I~,\nonumber\\
%\eeqa
%\beqa
\epsilon_{4,p}^K&=&4b_0+\left(2-2D^2+4DF-2F^2\right)\left(b_D-b_F\right)+2\left(-b_1+b_2-b_3\right)-4b_4\nonumber\\&+&\frac{m_0}{2}\left( b_5-b_6+b_7\right)+m_0 b_8+\left(16 b_0+8\left( b_D-b_F-b_1+b_2-b_3\right)-16b_4\right.\nonumber\\&+&\left.2m_0\left( b_5- b_6+b_7\right)+4m_0 b_8\right)I^2+\left(-16b_0+\left(-8+4D^2-8DF+4F^2\right)\left(b_D-b_F\right)\right.\nonumber\\&+&\left.8\left(b_1-b_2+b_3\right)+16b_4+2m_0\left(-b_5+b_6-b_7\right)-4m_0b_8\right)I~,\nonumber\\
%\eeqa
%\beqa
\epsilon_{4,p}^{\pi K}&=&-16b_0+\left(-8+6D^2-12DF+6F^2\right)\left(b_D-b_F\right)+8\left(b_1-b_2+b_3\right)+16b_4\nonumber\\&+&2m_0\left(- b_5+b_6-b_7\right)-4m_0 b_8+\left(-32 b_0+16\left(- b_D+b_F+b_1-b_2+b_3\right)+32b_4\right.\nonumber\\&+&\left.4m_0 \left(-b_5+ b_6-b_7\right)-8m_0 b_8\right)I^2+\left(48b_0+\left(24-8D^2+16DF-8F^2\right)\left(b_D-b_F\right)\right.\nonumber\\&+&\left.24\left(-b_1+b_2-b_3\right)-48b_4+6m_0\left(b_5-b_6+b_7\right)+12m_0b_8\right)I~,\nonumber
\eeqa

\beqa
\epsilon_{4,n}^\pi&=&16b_0+\left(16+\frac{16}{3}\left(D^2+DF\right)\right)b_D+\left(\frac{8}{3}D^2+8F^2\right)b_F-16\left(b_1+b_3+b_4\right)\nonumber\\&+&4m_0\left( b_5+b_7+b_8\right)+2m_0 b_9+\left(16 b_0+\left(16+\frac{64}{9}D^2\right)b_D-16\left(b_1+b_3+b_4\right)\right.\nonumber\\&+&\left.4m_0 \left(b_5+ b_7+b_8\right)+2m_0 b_9\right)I^2+\left(-32b_0-\left(32+\frac{112}{9}D^2+\frac{16}{3}DF\right)b_D\right.\nonumber\\&-&\left.\left(\frac{8}{3}D^2+8F^2\right)b_F+32\left(b_1+b_3+b_4\right)-8m_0\left(b_5+b_7+b_8\right)-4m_0b_9\right)I~,\nonumber
\eeqa
\beqa
\epsilon_{4,n}^K&=&4b_0+\left(4+\frac{8}{3}\left(D^2+DF\right)\right)b_D+\left(\frac{4}{3}D^2+4F^2\right)b_F-4\left(b_1+b_3+b_4\right)\nonumber\\&+&m_0\left( b_5+b_7+b_8\right)+\frac{m_0}{2} b_9+\left(16 b_0+\left(16+\frac{64}{9}D^2\right) b_D-16\left(b_1+b_3+b_4\right)\right.\nonumber\\&+&\left.4m_0\left( b_5+ b_7+b_8\right)+2m_0 b_9\right)I^2+\left(-16b_0-\left(16+\frac{80}{9}D^2+\frac{16}{3}DF\right)b_D\right.\nonumber\\&-&\left.\left(\frac{8}{3}D^2+8F^2\right)b_F+16\left(b_1+b_3+b_4\right)-4m_0\left(b_5+b_7+b_8\right)-2m_0b_9\right)I~,\nonumber
\eeqa
\beqa
\epsilon_{4,n}^{\pi K}&=&-16b_0-\left(16+8\left(D^2+DF\right)\right)b_D-\left(4D^2+12F^2\right)b_F+16\left(b_1+b_3+b_4\right)\nonumber\\&-&4m_0\left( b_5+b_7+b_8\right)-2m_0 b_9+\left(-32 b_0-\left(32+\frac{128}{9}D^2\right) b_D+32\left(b_1+b_3+b_4\right)\right.\nonumber\\&-&\left.8m_0\left( b_5+ b_7+b_8\right)-4m_0 b_9\right)I^2+\left(48b_0+\left(48+\frac{64}{3}D^2+\frac{32}{3}DF\right)b_D\right.\\&+&\left.\left(\frac{16}{3}D^2+16F^2\right)b_F-48\left(b_1+b_3+b_4\right)+12m_0\left(b_5+b_7+b_8\right)+6m_0b_9\right)I~,\nonumber
\eeqa

\beqa
\epsilon_{5,p}^\pi&=&\frac{17}{3}b_0+\frac{2}{3}b_D-\frac{4}{3}b_F-\frac{2}{3}b_1+\frac{4}{3}b_2-\frac{26}{9}b_3-\frac{14}{3}b_4+\frac{m_0}{6} b_5-\frac{m_0}{3} b_6+\frac{13}{18}m_0 b_7+\frac{7}{6}m_0 b_8\nonumber\\&+&\left(4 b_0+\frac{8}{3}\left( b_D- b_1\right)+\frac{16}{9}b_2-\frac{56}{27}b_3-\frac{40}{9}b_4+\frac{2}{3}m_0 b_5-\frac{4}{9}m_0 b_6+\frac{14}{27}m_0 b_7+\frac{10}{9}m_0 b_8\right)I^2\nonumber\\&+&\left(-8b_0+\frac{8}{3}\left(-b_D+b_F+b_1\right)-\frac{32}{9}b_2+\frac{40}{9}b_3+8b_4-\frac{2}{3}m_0b_5+\frac{8}{9}m_0b_6-\frac{10}{9}m_0b_7\right.\nonumber\\&-&\left.2m_0b_8\right)I+\left(-\frac{4}{3}b_0-\frac{1}{3}b_D-\frac{5}{3}b_F+\frac{1}{3}b_1+\frac{5}{9}b_2+\frac{13}{27}b_3+\frac{8}{9}b_4-\frac{m_0}{12}b_5-\frac{5}{36}m_0b_6\right.\nonumber\\&-&\left.\frac{13}{108}m_0b_7-\frac{2}{9}m_0b_8\right)\cos(2\ve)+\frac{1}{\sqrt{3}}\left(10b_0+b_D-3b_F-b_1+\frac{7}{3}b_2-5b_3-8b_4\right.\nonumber\\&+&\left.\frac{m_0}{4}b_5-\frac{7}{12}m_0b_6+\frac{5}{4}m_0b_7+2m_0b_8+\left(8b_0+\frac{8}{3}b_D-8b_F-\frac{8}{3}b_1+\frac{40}{9}b_2-\frac{104}{27}b_3\right.\right.\nonumber\\&-&\left.\left.\frac{64}{9}b_4+\frac{2}{3}m_0b_5-\frac{10}{9}m_0b_6+\frac{26}{27}m_0b_7+\frac{16}{9}m_0b_8\right)I^2+\left(-16b_0-4b_D+\frac{20}{3}b_F\right.\right.\nonumber\\&+&\left.\left.4b_1-\frac{52}{9}b_2+\frac{220}{27}b_3+\frac{128}{9}b_4-m_0b_5+\frac{13}{9}m_0b_6-\frac{55}{27}m_0b_7-\frac{32}{9}m_0b_8\right)I\right)\sin(2\ve)~,\nonumber
\eeqa
\beqa
\epsilon_{5,p}^K&=&\frac{8}{3}\left(b_0+b_D+b_F-b_1\right)-\frac{8}{9}b_3-\frac{8}{3}b_4+\frac{2}{3}m_0 b_5+\frac{2}{9}m_0 b_7+\frac{2}{3}m_0 b_8+\left(4 b_0\right.\nonumber\\&+&\left.\frac{8}{3}\left( b_D-b_1\right)+\frac{16}{9}b_2-\frac{56}{27}b_3-\frac{40}{9}b_4+\frac{2}{3}m_0 b_5-\frac{4}{9}m_0 b_6+\frac{14}{27}m_0 b_7+\frac{10}{9}m_0 b_8\right)I^2\nonumber\\&+&\left(-4b_0-\frac{8}{3}b_D-\frac{16}{3}b_F+\frac{8}{3}b_1-\frac{8}{9}b_2+\frac{8}{3}b_3+\frac{16}{3}b_4-\frac{2}{3}m_0b_5+\frac{2}{9}m_0b_6-\frac{2}{3}m_0b_7\right.\nonumber\\&-&\left.\frac{4}{3}m_0b_8\right)I+\left(\frac{8}{3}\left(b_0+b_D-b_F-b_1\right)+\frac{8}{9}b_2-\frac{8}{27}b_3-\frac{16}{9}b_4+\frac{2}{3}m_0b_5-\frac{2}{9}m_0b_6\right.\nonumber\\&+&\left.\frac{2}{27}m_0b_7+\frac{4}{9}m_0b_8\right)\cos(2\ve)+\frac{1}{\sqrt{3}}\left(4\left(-b_D-b_F+b_1\right)+\frac{4}{3}\left(b_2-b_3\right)-m_0b_5\right.\nonumber\\&+&\left.\frac{m_0}{3}\left(-b_6+b_7\right)+\left(8b_0+\frac{8}{3}b_D-8b_F-\frac{8}{3}b_1+\frac{40}{9}b_2-\frac{104}{27}b_3-\frac{64}{9}b_4+\frac{2}{3}m_0b_5\right.\right.\nonumber\\&-&\left.\left.\frac{10}{9}m_0b_6+\frac{26}{27}m_0b_7+\frac{16}{9}m_0b_8\right)I^2+\left(4\left(-b_0+b_D\right)+\frac{20}{3}b_F-4b_1-\frac{28}{9}b_2+\frac{100}{27}b_3\right.\right.\nonumber\\&+&\left.\left.\frac{32}{9}b_4+m_0b_5+\frac{7}{9}m_0b_6-\frac{25}{27}m_0b_7-\frac{8}{9}m_0b_8\right)I\right)\sin(2\ve)~,\nonumber
\eeqa
\beqa
\epsilon_{5,p}^{\pi K}&=&-\frac{16}{3}b_0+\frac{4}{3}\left(-b_D-b_F+b_1-b_2\right)+\frac{28}{9}b_3+\frac{16}{3}b_4+\frac{m_0}{3}\left(- b_5+b_6\right)-\frac{7}{9}m_0 b_7\nonumber\\&-&\frac{4}{3}m_0 b_8+\left(-8 b_0+\frac{16}{3}\left(- b_D+b_1\right)-\frac{32}{9}b_2+\frac{112}{27}b_3+\frac{80}{9}b_4-\frac{4}{3}m_0 b_5+\frac{8}{9}m_0 b_6\right.\nonumber\\&-&\left.\frac{28}{27}m_0 b_7-\frac{20}{9}m_0 b_8\right)I^2+\left(12b_0+\frac{16}{3}b_D+\frac{8}{3}b_F-\frac{16}{3}b_1+\frac{40}{9}b_2-\frac{64}{9}b_3-\frac{40}{3}b_4\right.\nonumber\\&+&\left.\frac{4}{3}m_0b_5-\frac{10}{9}m_0b_6+\frac{16}{9}m_0b_7+\frac{10}{3}m_0b_8\right)I+\left(\frac{4}{3}\left(-b_0-b_D+b_F+b_1\right)-\frac{4}{9}b_2\right.\nonumber\\&+&\left.\frac{4}{27}b_3+\frac{8}{9}b_4-\frac{m_0}{3}b_5+\frac{m_0}{9}b_6-\frac{m_0}{27}b_7-\frac{2}{9}m_0b_8\right)\cos(2\ve)\nonumber\\&+&\frac{1}{\sqrt{3}}\left(-10b_0+4b_F-\frac{8}{3}b_2+\frac{16}{3}b_3+8b_4+\frac{2}{3}m_0b_6-\frac{4}{3}m_0b_7-2m_0b_8\right.\nonumber\\&+&\left.\left(-16b_0-\frac{16}{3}b_D+16b_F+\frac{16}{3}b_1-\frac{80}{9}b_2+\frac{208}{27}b_3+\frac{128}{9}b_4-\frac{4}{3}m_0b_5+\frac{20}{9}m_0b_6\right.\right.\nonumber\\&-&\left.\left.\frac{52}{27}m_0b_7-\frac{32}{9}m_0b_8\right)I^2+\left(20b_0-\frac{40}{3}b_F+\frac{80}{9}b_2-\frac{320}{27}b_3-\frac{160}{9}b_4-\frac{20}{9}m_0b_6\right.\right.\nonumber\\&+&\left.\left.\frac{80}{27}m_0b_7+\frac{40}{9}m_0b_8\right)I\right)\sin(2\ve)~,\nonumber
\eeqa

\beqa
\epsilon_{5,n}^\pi&=&\frac{17}{3}b_0+6b_D+4b_F-\frac{26}{3}b_1-\frac{4}{3}b_2-\frac{2}{9}b_3-\frac{14}{3}b_4+\frac{13}{6}m_0b_5+\frac{m_0}{3}b_6+\frac{m_0}{18}b_7\nonumber\\&+&\frac{7}{6}m_0b_8+\left(4b_0+\frac{8}{3}b_D-8b_1-\frac{8}{27}b_3-\frac{40}{9}b_4+2m_0b_5+\frac{2}{27}m_0b_7+\frac{10}{9}m_0b_8\right)I^2\nonumber\\&+&\left(-8\left(b_0+b_D\right)-\frac{8}{3}b_F+16b_1+\frac{8}{9}b_2+8b_4-4m_0b_5-\frac{2}{9}m_0b_6-2m_0b_8\right)I\nonumber\\&+&\left(-\frac{4}{3}b_0-\frac{1}{3}b_D-\frac{5}{3}b_F+\frac{1}{3}b_1+\frac{5}{9}b_2+\frac{13}{27}b_3+\frac{8}{9}b_4-\frac{m_0}{12}b_5-\frac{5}{36}m_0b_6-\frac{13}{108}m_0b_7\right.\nonumber\\&-&\left.\frac{2}{9}m_0b_8\right)\cos(2\ve)+\frac{1}{\sqrt{3}}\left(10b_0+11b_D+7b_F-15b_1-\frac{7}{3}b_2-\frac{1}{3}b_3-8b_4\right.\nonumber\\&+&\left.\frac{15}{4}m_0b_5+\frac{7}{12}m_0b_6+\frac{m_0}{12}b_7+2m_0b_8+\left(8b_0+\frac{32}{3}b_D-16b_1+\frac{16}{27}b_3\right.\right.\nonumber\\&-&\left.\left.\frac{64}{9}b_4+4m_0b_5-\frac{4}{27}m_0b_7+\frac{16}{9}m_0b_8\right)I^2+\left(-16b_0-\frac{52}{3}b_D-\frac{20}{3}b_F+28b_1\right.\right.\nonumber\\&+&\left.\left.\frac{20}{9}b_2+\frac{4}{27}b_3+\frac{128}{9}b_4-7m_0b_5-\frac{5}{9}m_0b_6-\frac{m_0}{27}b_7-\frac{32}{9}m_0b_8\right)I\right)\sin(2\ve)~,\nonumber
\eeqa

\vfill\eject

\beqa
\epsilon_{5,n}^K&=&\frac{8}{3}\left(b_0-b_1\right)-\frac{8}{9}b_3-\frac{8}{3}b_4+\frac{2}{3}m_0 b_5+\frac{2}{9}m_0 b_7+\frac{2}{3}m_0 b_8+\left(4 b_0+\frac{8}{3} b_D-8b_1-\frac{8}{27}b_3\right.\nonumber\\&-&\left.\frac{40}{9}b_4+2m_0 b_5+\frac{2}{27}m_0 b_7+\frac{10}{9}m_0 b_8\right)I^2+\left(-4b_0-\frac{8}{3}b_F+8b_1+\frac{8}{9}\left(b_2+b_3\right)\right.\nonumber\\&+&\left.\frac{16}{3}b_4-2m_0b_5-\frac{2}{9}m_0\left(b_6+b_7\right)-\frac{4}{3}m_0b_8\right)I+\left(\frac{8}{3}\left(b_0+b_D-b_F-b_1\right)+\frac{8}{9}b_2\right.\nonumber\\&-&\left.\frac{8}{27}b_3-\frac{16}{9}b_4+\frac{2}{3}m_0b_5-\frac{2}{9}m_0b_6+\frac{2}{27}m_0b_7+\frac{4}{9}m_0b_8\right)\cos(2\ve)\nonumber\\&+&\frac{1}{\sqrt{3}}\left(4\left(b_D+b_F-b_1\right)+\frac{4}{3}\left(-b_2+b_3\right)+m_0b_5+\frac{m_0}{3}\left(b_6-b_7\right)\right.\nonumber\\&+&\left.\left(8b_0+\frac{32}{3}b_D-16b_1+\frac{16}{27}b_3-\frac{64}{9}b_4+4m_0b_5-\frac{4}{27}m_0b_7+\frac{16}{9}m_0b_8\right)I^2\right.\nonumber\\&+&\left.\left(-4b_0-\frac{28}{3}b_D-\frac{20}{3}b_F+12b_1+\frac{20}{9}b_2-\frac{44}{27}b_3+\frac{32}{9}b_4-3m_0b_5-\frac{5}{9}m_0b_6\right.\right.\nonumber\\&+&\left.\left.\frac{11}{27}m_0b_7-\frac{8}{9}m_0b_8\right)I\right)\sin(2\ve)~,\nonumber
\eeqa
\beqa
\epsilon_{5,n}^{\pi K}&=&-\frac{16}{3}b_0-4\left(b_D+b_F\right)+\frac{28}{3}b_1+\frac{4}{3}b_2+\frac{4}{9}b_3+\frac{16}{3}b_4-\frac{7}{3}m_0 b_5-\frac{m_0}{3}b_6-\frac{m_0}{9} b_7\nonumber\\&-&\frac{4}{3}m_0 b_8+\left(-8 b_0-\frac{16}{3} b_D+16b_1+\frac{16}{27}b_3+\frac{80}{9}b_4-4m_0 b_5-\frac{4}{27}m_0 b_7-\frac{20}{9}m_0 b_8\right)I^2\nonumber\\&+&\left(12b_0+8b_D+\frac{16}{3}b_F-24b_1-\frac{16}{9}b_2-\frac{8}{9}b_3-\frac{40}{3}b_4+6m_0b_5+\frac{4}{9}m_0b_6+\frac{2}{9}m_0b_7\right.\nonumber\\&+&\left.\frac{10}{3}m_0b_8\right)I+\left(\frac{4}{3}\left(-b_0-b_D+b_F+b_1\right)-\frac{4}{9}b_2+\frac{4}{27}b_3+\frac{8}{9}b_4-\frac{m_0}{3}b_5+\frac{m_0}{9}b_6\right.\nonumber\\&-&\left.\frac{m_0}{27}b_7-\frac{2}{9}m_0b_8\right)\cos(2\ve)+\frac{1}{\sqrt{3}}\left(-10b_0-12b_D-8b_F+16b_1+\frac{8}{3}b_2+8b_4\right.\nonumber\\&-&\left.4m_0b_5-\frac{2}{3}m_0b_6-2m_0b_8+\left(-16b_0-\frac{64}{3}b_D+32b_1-\frac{32}{27}b_3+\frac{128}{9}b_4-8m_0b_5\right.\right.\nonumber\\&+&\left.\left.\frac{8}{27}m_0b_7-\frac{32}{9}m_0b_8\right)I^2+\left(20b_0+\frac{80}{3}b_D+\frac{40}{3}b_F-40b_1-\frac{40}{9}b_2+\frac{40}{27}b_3\right.\right.\nonumber\\&-&\left.\left.\frac{160}{9}b_4+10m_0b_5+\frac{10}{9}m_0b_6-\frac{10}{27}m_0b_7+\frac{40}{9}m_0b_8\right)I\right)\sin(2\ve)~,
\eeqa

\beqa
\epsilon_{6,p}^\pi&=&\left(16D^2+32DF+16F^2\right)\left(b_D+b_F\right)\left(1-I\right)~,\nonumber\\
%\eeqa
%\beqa
%\epsilon_{6,p}^K&=&
%\eeqa
%\beqa
\epsilon_{6,p}^{\pi K}&=&\left(16D^2+32DF+16F^2\right)\left(b_D+b_F\right)\left(-1+I\right)~,\nonumber
\eeqa

\beqa
\epsilon_{6,n}^\pi&=&\left(16D^2+32DF+16F^2\right)\left(b_D+b_F\right)\left(-1+I\right)~,\nonumber\\
%\eeqa
%\beqa
%\epsilon_{6,n}^K&=&
%\eeqa
%\beqa
\epsilon_{6,n}^{\pi K}&=&\left(16D^2+32DF+16F^2\right)\left(b_D+b_F\right)\left(1-I\right)~,
\eeqa

\beqa
\epsilon_{8,p}^K&=&-\frac{32}{3}\left(D^2+DF\right)b_D-\left(\frac{16}{3}D^2+16F^2\right)b_F\nonumber\\&+&\left(\left(\frac{64}{9}D^2+\frac{64}{3}DF\right)b_D+\left(\frac{32}{3}D^2+32F^2\right)b_F\right)I~,\nonumber\\
%\eeqa
%\beqa
\epsilon_{8,p}^{\pi K}&=&\frac{32}{3}\left(D^2+DF\right)b_D+\left(\frac{16}{3}D^2+16F^2\right)b_F\nonumber\\&-&\left(\left(\frac{64}{9}D^2+\frac{64}{3}DF\right)b_D+\left(\frac{32}{3}D^2+32F^2\right)b_F\right)I~,
\eeqa

\beqa
\epsilon_{8,n}^K&=&\left(8D^2-16DF+8F^2\right)\left(b_D-b_F\right)+\left(-16D^2+32DF-16F^2\right)\left(b_D-b_F\right)I~,\nonumber\\
%\eeqa
%\beqa
\epsilon_{8,n}^{\pi K}&=&\left(-8D^2+16DF-8F^2\right)\left(b_D-b_F\right)+\left(16D^2-32DF+16F^2\right)\left(b_D-b_F\right)I~,\nonumber
\eeqa

\beqa
\epsilon_{9,p}^\pi&=&\left(16D^2-32DF+16F^2\right)\left(b_D-b_F\right)\left(1-I\right)~,\nonumber\\
%\eeqa
%\beqa
\epsilon_{9,p}^K&=&\left(8D^2-16DF+8F^2\right)\left(b_D-b_F\right)+\left(-16D^2+32DF-16F^2\right)\left(b_D-b_F\right)I~,\nonumber\\
%\eeqa
%\beqa
\epsilon_{9,p}^{\pi K}&=&\left(-24D^2+48DF-24F^2\right)\left(b_D-b_F\right)+\left(32D^2-64DF+32F^2\right)\left(b_D-b_F\right)I~,\nonumber
\eeqa

\beqa
\epsilon_{9,n}^\pi&=&-\frac{64}{3}\left(D^2+DF\right)b_D-\left(\frac{32}{3}D^2+32F^2\right)b_F-\frac{256}{9}D^2b_DI^2\nonumber\\&+&\left(\left(\frac{448}{9}D^2+\frac{64}{3}DF\right)b_D+\left(\frac{32}{3}D^2+32F^2\right)b_F\right)I~,\nonumber\\
%\eeqa
%\beqa
\epsilon_{9,n}^K&=&-\frac{32}{3}\left(D^2+DF\right)b_D-\left(\frac{16}{3}D^2+16F^2\right)b_F-\frac{256}{9}D^2b_DI^2\nonumber\\&+&\left(\left(\frac{320}{9}D^2+\frac{64}{3}DF\right)b_D+\left(\frac{32}{3}D^2+32F^2\right)b_F\right)I~,\nonumber\\
%\eeqa
%\beqa
\epsilon_{9,n}^{\pi K}&=&32\left(D^2+DF\right)b_D+\left(16D^2+48F^2\right)b_F+\frac{512}{9}D^2b_DI^2\nonumber\\&-&\left(\left(\frac{256}{3}D^2+\frac{128}{3}DF\right)b_D+\left(\frac{64}{3}D^2+64F^2\right)b_F\right)I~,
\eeqa

\vfill\eject

\beqa
\epsilon_{11,p}^\pi&=&\left(-144L_4-48L_5+288L_6+96L_8\right)\frac{b_0}{F_\pi^2}+\left(96L_4+32L_5-192L_6-64L_8\right)\frac{b_F}{F_\pi^2}\nonumber\\&+&\frac{m_0}{(4 \pi)^2 F_\pi^2}\left(-\frac{17}{12}b_5+\frac{13}{12}b_6-\frac{71}{36}b_7-\frac{11}{3}b_8\right)-16d_1+24d_5-36d_6-44d_7\nonumber\\&+&\left(\left(-64L_4+128L_6\right)\frac{b_0}{F_\pi^2}+\left(128L_4-256L_6\right)\frac{b_F}{F_\pi^2}\right.\nonumber\\&+&\left.\frac{m_0}{(4 \pi)^2 F_\pi^2}\left(-\frac{5}{3}b_5+\frac{13}{9}b_6-\frac{41}{27}b_7-\frac{28}{9}b_8\right)-64d_1+32d_5-16d_6-48d_7\right)I^2\nonumber\\&+&\left(\left(192L_4+32L_5-384L_6-64L_8\right)\frac{b_0}{F_\pi^2}+\left(-256L_4-64L_5+512L_6+128L_8\right)\frac{b_F}{F_\pi^2}\right.\nonumber\\&+&\left.\frac{m_0}{(4 \pi)^2 F_\pi^2}\left(\frac{8}{3}b_5-\frac{26}{9}b_6+\frac{28}{9}b_7+6b_8\right)+64\left(d_1-d_5\right)+48d_6+80d_7\right)I~,\nonumber
\eeqa
\beqa
\epsilon_{11,p}^K&=&\left(-32L_5+64L_8\right)\left(\frac{b_D}{F_\pi^2}+\frac{b_F}{F_\pi^2}\right)+\frac{m_0}{(4 \pi)^2 F_\pi^2}\left(-\frac{17}{12}b_5+\frac{1}{4}b_6-\frac{35}{36}b_7-\frac{5}{3}b_8-\frac{1}{4}b_9\right)\nonumber\\&-&16\left(d_1+d_2+d_3\right)-32d_7+\left(\left(-64L_4+128L_6\right)\frac{b_0}{F_\pi^2}+\left(128L_4-256L_6\right)\frac{b_F}{F_\pi^2}\right.\nonumber\\&+&\left.\frac{m_0}{(4 \pi)^2 F_\pi^2}\left(-\frac{5}{3}b_5+\frac{13}{9}b_6-\frac{41}{27}b_7-\frac{28}{9}b_8\right)-64d_1+32d_5-16d_6-48d_7\right)I^2\nonumber\\&+&\left(\left(-32L_5+64L_8\right)\frac{b_0}{F_\pi^2}+\left(-64L_4+128L_6\right)\frac{b_D}{F_\pi^2}\right.\nonumber\\&+&\left.\left(64\left(-L_4+L_5\right)+128\left(L_6-L_8\right)\right)\frac{b_F}{F_\pi^2}+\frac{m_0}{(4 \pi)^2 F_\pi^2}\left(\frac{5}{3}b_5-\frac{11}{9}b_6+\frac{5}{3}b_7+\frac{10}{3}b_8\right)\right.\nonumber\\&+&\left.64d_1+32d_2-16d_5+64d_7\right)I~,\nonumber
\eeqa
\beqa
\epsilon_{11,p}^{\pi K}&=&\left(-96L_4+192L_6\right)\left(\frac{b_D}{F_\pi^2}+\frac{b_F}{F_\pi^2}\right)+\frac{m_0}{(4 \pi)^2 F_\pi^2}\left(\frac{4}{3}\left(b_5-b_6\right)+\frac{16}{9}b_7+\frac{10}{3}b_8\right)\nonumber\\&+&32d_1+16d_2-24d_5+64d_7+\left(\left(128L_4-256L_6\right)\frac{b_0}{F_\pi^2}+\left(-256L_4+512L_6\right)\frac{b_F}{F_\pi^2}\right.\nonumber\\&+&\left.\frac{m_0}{(4 \pi)^2 F_\pi^2}\left(\frac{10}{3}b_5-\frac{26}{9}b_6+\frac{82}{27}b_7+\frac{56}{9}b_8\right)+128d_1-64d_5+32d_6+96d_7\right)I^2\nonumber\\&+&\left(\left(-192L_4+384L_6\right)\frac{b_0}{F_\pi^2}+\left(64L_4-128L_6\right)\frac{b_D}{F_\pi^2}+\left(320L_4-640L_6\right)\frac{b_F}{F_\pi^2}\right.\nonumber\\&+&\left.\frac{m_0}{(4 \pi)^2 F_\pi^2}\left(-\frac{13}{3}b_5+\frac{37}{9}b_6-\frac{43}{9}b_7-\frac{28}{3}b_8\right)-128d_1-32d_2+80d_5-48d_6\right.\nonumber\\&-&\left.144d_7\right)I~,\nonumber
\eeqa

\beqa
\epsilon_{11,n}^\pi&=&\left(-144L_4-48L_5+288L_6+96L_8\right)\frac{b_0}{F_\pi^2}+\left(-192L_4-64L_5+384L_6+128L_8\right)\frac{b_D}{F_\pi^2}\nonumber\\&+&\left(-96L_4-32L_5+192L_6+64L_8\right)\frac{b_F}{F_\pi^2}-\frac{m_0}{(4 \pi)^2 F_\pi^2}\left(\frac{53}{12}b_5+\frac{7}{12}b_6+\frac{83}{36}b_7+\frac{11}{3}b_8\right.\nonumber\\&+&\left.b_{9}\right)-16d_1-32d_2-64d_3-24d_5-36d_6-44d_7+\left(\left(-64L_4+128L_6\right)\frac{b_0}{F_\pi^2}\right.\nonumber\\&+&\left.\left(-128L_4+256L_6\right)\frac{b_D}{F_\pi^2}-\frac{m_0}{(4 \pi)^2 F_\pi^2}\left(4b_5+\frac{56}{27}b_7+\frac{28}{9}b_8+b_9\right)-64d_3-16d_6\right.\nonumber\\&-&\left.48d_7\right)I^2+\left(\left(192L_4+32L_5-384L_6-64L_8\right)\frac{b_0}{F_\pi^2}+\left(320L_4+64L_5-640L_6\right.\right.\nonumber\\&-&\left.\left.128L_8\right)\frac{b_D}{F_\pi^2}+\left(64L_4-128L_6\right)\frac{b_F}{F_\pi^2}+\frac{m_0}{(4 \pi)^2 F_\pi^2}\left(8b_5+\frac{2}{9}b_6+4b_7+6b_8+2b_9\right)\right.\nonumber\\&+&\left.32d_2+128d_3+16d_5+48d_6+80d_7\right)I~,\nonumber
\eeqa
\beqa
\epsilon_{11,n}^K&=&\left(32L_5-64L_8\right)\left(\frac{b_D}{F_\pi^2}+\frac{b_F}{F_\pi^2}\right)+\frac{m_0}{(4 \pi)^2 F_\pi^2}\left(-\frac{17}{12}b_5+\frac{1}{4}b_6-\frac{35}{36}b_7-\frac{5}{3}b_8-\frac{1}{4}b_9\right)\nonumber\\&-&16\left(d_1+d_2+d_3\right)-32d_7+\left(\left(-64L_4+128L_6\right)\frac{b_0}{F_\pi^2}+\left(-128L_4+256L_6\right)\frac{b_D}{F_\pi^2}\right.\nonumber\\&-&\left.\frac{m_0}{(4 \pi)^2 F_\pi^2}\left(4b_5+\frac{56}{27}b_7+\frac{28}{9}b_8+b_9\right)-64d_3-16d_6-48d_7\right)I^2\nonumber\\&+&\left(\left(-32L_5+64L_8\right)\frac{b_0}{F_\pi^2}+\left(64\left(L_4-L_5\right)+128\left(-L_6+L_8\right)\right)\frac{b_D}{F_\pi^2}\right.\nonumber\\&+&\left.\left(64L_4-128L_6\right)\frac{b_F}{F_\pi^2}+\frac{m_0}{(4 \pi)^2 F_\pi^2}\left(4b_5+\frac{2}{9}b_6+\frac{20}{9}b_7+\frac{10}{3}b_8+b_9\right)\right.\nonumber\\&+&\left.32d_2+64d_3+16d_5+64d_7\right)I~,\nonumber
\eeqa
\beqa
\epsilon_{11,n}^{\pi K}&=&\left(96L_4-192L_6\right)\left(\frac{b_D}{F_\pi^2}+\frac{b_F}{F_\pi^2}\right)+\frac{m_0}{(4 \pi)^2 F_\pi^2}\left(\frac{13}{3}b_5+\frac{1}{3}b_6+\frac{19}{9}b_7+\frac{10}{3}b_8+b_9\right)\nonumber\\&+&32d_1+48d_2+64d_3+24d_5+64d_7+\left(\left(128L_4-256L_6\right)\frac{b_0}{F_\pi^2}\right.\nonumber\\&+&\left.\left(256L_4-512L_6\right)\frac{b_D}{F_\pi^2}+\frac{m_0}{(4 \pi)^2 F_\pi^2}\left(8b_5+\frac{112}{27}b_7+\frac{56}{9}b_8+2b_9\right)+128d_3\right.\nonumber\\&+&\left.32d_6+96d_7\right)I^2+\left(\left(-192L_4+384L_6\right)\frac{b_0}{F_\pi^2}+\left(-384L_4+768L_6\right)\frac{b_D}{F_\pi^2}\right.\nonumber\\&+&\left.\left(-128L_4+256L_6\right)\frac{b_F}{F_\pi^2}-\frac{m_0}{(4 \pi)^2 F_\pi^2}\left(12b_5+\frac{4}{9}b_6+\frac{56}{9}b_7+\frac{28}{3}b_8+3b_9\right)\right.\nonumber\\&-&\left.64d_2-192d_3-32d_5-48d_6-144d_7\right)I~.
\eeqa

For completeness, we also give the fourth order corrections to the meson masses in the
form they were used to arrive at the quark mass representations of the baryon masses.
More precisely, in the second order baryon mass corrections $m_B^{(2)}=
\sum_P \gamma_B^P M_P^2$ we
have contributions from the meson masses $M_P^2 = \bar{M}_P^2 + M_P^{(4)}$, where 
$M_P^{(4)}$ is the fourth-order correction. We remark that only the charged
pion and kaon mass appear in the terms $m_B^{(2)}$
due to our conventions.  The generic form of these corrections is
\beqa
F_\pi^2\, M_P^{(4)} &=&
\left(\alpha_{1,P}^\pi\, M_{\pi^+}^4 +\alpha_{1,P}^K\, M_{K^+}^4
      + \alpha_{1,P}^{\pi K}\, M_{\pi^+}^2\, M_{K^+}^2\right)\, \bar{\mu}_{\pi^+} \nonumber\\
&+& 
\left(\alpha_{2,P}^\pi\, M_{\pi^+}^4 +\alpha_{2,P}^K\, M_{K^+}^4
      + \alpha_{2,P}^{\pi K}\, M_{\pi^+}^2\, M_{K^+}^2\right)\, \bar{\mu}_{\pi^0} \nonumber\\
&+& 
\left(\alpha_{3,P}^\pi\, M_{\pi^+}^4 +\alpha_{3,P}^K\, M_{K^+}^4
      + \alpha_{3,P}^{\pi K}\, M_{\pi^+}^2\, M_{K^+}^2\right)\, \bar{\mu}_{K^+} \nonumber\\
&+& 
\left(\alpha_{4,P}^\pi\, M_{\pi^+}^4 +\alpha_{4,P}^K\, M_{K^+}^4
      + \alpha_{4,P}^{\pi K}\, M_{\pi^+}^2\, M_{K^+}^2\right)\, \bar{\mu}_{K^0} \nonumber\\
&+& 
\left(\alpha_{5,P}^\pi\, M_{\pi^+}^4 +\alpha_{5,P}^K\, M_{K^+}^4
      + \alpha_{5,P}^{\pi K}\, M_{\pi^+}^2\, M_{K^+}^2\right)\, \bar{\mu}_{\eta} \nonumber\\
&+& 
    \,\, \alpha_{6,P}^\pi\,  M_{\pi^+}^4 +\alpha_{6,P}^K\, M_{K^+}^4
     + \alpha_{6,P}^{\pi K}\, M_{\pi^+}^2\, M_{K^+}^2~.
\eeqa
The non-vanishing coefficients  for $P = \pi^+, K^+$ are
\beqa
\alpha_{2,\pi^+}^\pi&=&\frac{1}{6}+\frac{1}{9}I+\frac{2}{9}\cos(2\ve)
+\frac{1}{\sqrt{3}}\left(-\frac{1}{3}+\frac{2}{9}I\right)\sin(2\ve)~,
\nonumber\\
%\eeqa
%\beqa
\alpha_{2,\pi^+}^{\pi K}&=&-\frac{1}{9}I+\frac{1}{9}\cos(2\ve)
+\frac{1}{\sqrt{3}}\left(\frac{1}{3}-\frac{2}{9}I\right)\sin(2\ve)~,
\nonumber\\
%\eeqa
%\beqa
\alpha_{5,\pi^+}^\pi&=&\frac{1}{6}+\frac{1}{9}I-\frac{2}{9}\cos(2\ve)
+\frac{1}{\sqrt{3}}\left(\frac{1}{3}-\frac{2}{9}I\right)\sin(2\ve)~,
\nonumber\\
%\eeqa
%\beqa
\alpha_{5,\pi^+}^{\pi K}&=&-\frac{1}{9}I-\frac{1}{9}\cos(2\ve)
+\frac{1}{\sqrt{3}}\left(-\frac{1}{3}+\frac{2}{9}I\right)\sin(2\ve)~,
\nonumber\\
%\eeqa
%\beqa
\alpha_{6,\pi^+}^\pi&=&\left(-24+16I\right)L_4-8L_5+\left(48-32I\right)L_6+16L_8~,
\nonumber\\
%\eeqa
%\beqa
\alpha_{6,\pi^+}^{\pi K}&=&-16IL_4+32IL_6~,
\eeqa

\vspace{-0.6cm}

\beqa
\alpha_{2,K^+}^K&=&\frac{1}{3}-\frac{1}{9}I-\frac{2}{9}\cos(2\ve)
+\frac{1}{\sqrt{3}}\left(\frac{1}{3}-\frac{2}{9}I\right)\sin(2\ve)~,
\nonumber\\
%\eeqa
%\beqa
\alpha_{2,K^+}^{\pi K}&=&-\frac{1}{6}+\frac{1}{9}I+\frac{1}{18}\cos(2\ve)
+\frac{1}{\sqrt{3}}\left(\frac{1}{6}+\frac{2}{9}I\right)\sin(2\ve)~,
\nonumber\\
%\eeqa
%\beqa
\alpha_{5,K^+}^K&=&\frac{1}{3}-\frac{1}{9}I+\frac{2}{9}\cos(2\ve)
+\frac{1}{\sqrt{3}}\left(-\frac{1}{3}+\frac{2}{9}I\right)\sin(2\ve)~,
\nonumber\\
%\eeqa
%\beqa
\alpha_{5,K^+}^{\pi K}&=&-\frac{1}{6}+\frac{1}{9}I-\frac{1}{18}\cos(2\ve)
-\frac{1}{\sqrt{3}}\left(\frac{1}{6}+\frac{2}{9}I\right)\sin(2\ve)~,
\nonumber\\
%\eeqa
%\beqa
\alpha_{6,K^+}^K&=&-16IL_4-8L_5+32IL_6+16L_8~,
\nonumber\\
%\eeqa
%\beqa
\alpha_{6,K^+}^{\pi K}&=&\left(-24+16I\right)L_4+\left(48-32I\right)L_6~.
\eeqa
We note that in the isospin limit $m_u= m_d$, these formulas coincide with the
ones in the classic paper \cite{GL85}. We stress again that the $L_i$ are the
renormalized values taken at the scale $\lambda = m_0$ and similarly for the
chiral logarithms hidden in the $\bar \mu_P$. These LECs are usually given at
the scale $\lambda=M_\rho$, with $M_\rho =770\,$MeV the $\rho$-meson mass. The
corresponding $\beta$-functions to evolve the $L_i$ from  $\lambda=M_\rho$ to
$\lambda = m_0$ are given in \cite{GL85}. 

%%%%%%%%%%%%%%%%%%%%%%%%%%%%%%%%%%%%%%%%%%%%%%%%%%%%%%%%%%%%%%%%%%%%%%%%%%%%%%%%%%%%%%%%%%%%
\section{Analysis of tapdole graphs}
\def\theequation{\Alph{section}.\arabic{equation}}
\setcounter{equation}{0}
\label{app:tadpole}
%%%%%%%%%%%%%%%%%%%%%%%%%%%%%%%%%%%%%%%%%%%%%%%%%%%%%%%%%%%%%%%%%%%%%%%%%%%%%%%%%%%%%%%%%%%%%
Here, we give a detailed analysis of the tadpole diagrams. We had already 
enumerated in Eq.~(\ref{leff2}) the terms in the dimension two Lagrangian
that lead to a non-trivial quark mass dependence of the baryon masses.
Apart from the terms listed there,  one also has terms with two covariant
derivatives of the form (we stick to the labeling of \cite{BM})
\beqa
{\mathcal L}^{(2,2)}
&=&\frac{b_{15}}{2}\Big({\rm Tr}\left\{\overline 
B\left[u_{\mu},\left[u_{\nu},\left[D^{\mu},\left[D^{\nu},B\right]\right]
\right]\right]\right\}+{\rm Tr}\left\{\overline B\left[D_{\nu},\left[D_{\mu},
\left[u^{\nu},\left[u^{\mu},B\right]\right]\right]\right]\right\}\Big)
\nonumber\\&+&\frac{b_{16}}{4}\Big({\rm Tr}\left\{\overline B\left[u_{\mu},
\left\{u_{\nu},\left[D^{\mu},\left[D^{\nu},B\right]\right]\right\}\right]
\right\}+{\rm Tr}\left\{\overline B\left[D_{\nu},\left[D_{\mu},\left\{u^{\nu},
\left[u^{\mu},B\right]\right\}\right]\right]\right\}\nonumber\\
& & +{\rm Tr}\left\{\overline B\left\{u_{\mu},\left[u_{\nu},\left[D^{\mu},
\left[D^{\nu},B\right]\right]\right]\right\}\right\}+
{\rm Tr}\left\{\overline B\left[D_{\nu},\left[D_{\mu},\left[u^{\nu},
\left\{u^{\mu},B\right\}\right]\right]\right]\right\}\Big)\nonumber\\
&+&\frac{b_{17}}{2}\Big({\rm Tr}\left\{\overline B\left\{u_{\mu},
\left\{u_{\nu},\left[D^{\mu},\left[D^{\nu},B\right]\right]\right\}
\right\}\right\}+{\rm Tr}\left\{\overline B\left[D_{\nu},\left[D_{\mu},
\left\{u^{\nu},\left\{u^{\mu},B\right\}\right\}\right]\right]\right\}\Big)
\nonumber\\&+&\frac{b_{18}}{2}\Big({\rm Tr}\left\{\overline B u_{\mu}\right\}
{\rm Tr}\left\{u_{\nu}\left[D^{\mu},\left[D^{\nu},B\right]\right]\right\}
+{\rm Tr}\left\{\overline B u_{\nu}\right\}{\rm Tr}\left\{\left[D_{\mu},
\left[D^{\nu},u^{\mu}B\right]\right]\right\}\nonumber\\& &+{\rm Tr}\left
\{\overline B\left[D_{\mu},\left[D_{\nu},u^{\nu}\right]\right]\right\}
{\rm Tr}\left\{u^{\mu}B\right\}+{\rm Tr}\left\{\overline B\left[D_{\mu},
u_{\nu}\right]\right\}{\rm Tr}\left\{\left[D^{\nu},u^{\mu}B\right]\right\}
\nonumber\\ & &+{\rm Tr}\left\{\overline B\left[D_{\nu},u^{\nu}\right]\right\}
{\rm Tr}\left\{\left[D_{\mu},u^{\mu}B\right]\right\}\Big)\nonumber\\
&+&b_{19}\Big({\rm Tr}\left\{\overline B\left[D_{\mu},\left[D_{\nu},B\right]
\right]\right\}{\rm Tr}\left\{u^{\mu}u^{\nu}\right\}+{\rm Tr}\left\{\overline 
B\left[D_{\nu},B\right]\right\}{\rm Tr}\left\{\left[D_{\mu},u^{\mu}u^{\nu}
\right]\right\}\Big)~.
\eeqa 
\FIGURE[l]{
%\begin{figure}[htb]
%\centerline{
\psfig{file=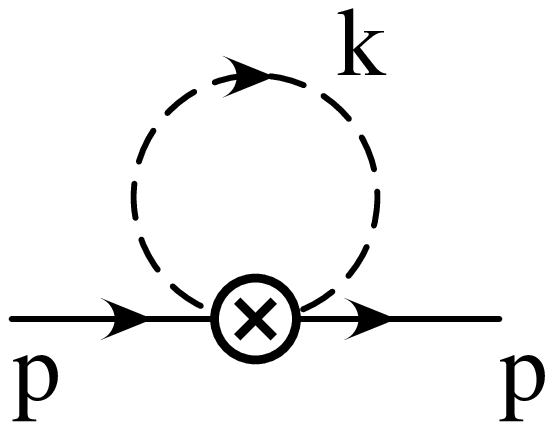,width=1.in}
%}
\caption{Tadpole graph.}
\label{fig:tadpole}
%\end{figure}
}
\noindent
Therefore, the tadpole diagram shown in Fig.~\ref{fig:tadpole} with the
four-momenta as given there can lead to the following structures
(we only display the non-vanishing ones)
\beq
{\cal A}_{\rm tadpole} \sim \int_I \frac{d^4k}{(2\pi)^4} \,
\left\{ {\bf 1}, k_\mu k^\mu, k_\mu k_\nu
p^\mu p^\nu, k_\mu k_\nu p^\mu \gamma^\mu \right\}
\, \frac{i}{k^2 - \bar{M}_P^2}~,
\eeq
where the factors $\sim k_\mu$ are generated from the chiral building blocks
$u_\mu$ whereas the factors of $p_\mu$ stem from the covariant derivatives
$D_\mu$ acting on the baryon fields. The terms $\sim {\bf 1}$ stem from the
explicit symmetry breaking $\sim \chi_+$ encoded in the terms proportional to
$b_{0,D,F}$. Upon integration over the meson momentum in the loop
one obtains
\beqa\label{tadstrcutures}
{\mathcal A}^{(1)}  &=&  \int_I \frac{d^4k}{(2\pi)^4} \,
\frac{i}{k^2 - \bar{M}_P^2} = \bar\Delta_P~,
\nonumber\\
{\mathcal A}^{(2)} &=&  \int_I \frac{d^4k}{(2\pi)^4} \,
  \frac{i \,k^2}{k^2 - \bar{M}_P^2} = \bar{M}^2_P \,\bar\Delta_P~,
\nonumber\\
{\mathcal A}^{(3)} &=&  \int_I \frac{d^4k}{(2\pi)^4} \,
p^\mu p^\nu  \frac{i \,k_\mu k_\nu}{k^2 - \bar{M}_P^2} =
p^2 \, \left( \frac{1}{4}\bar{M}^2_P \,\bar\Delta_P
-\frac{1}{8}\frac{\bar{M}_P^4}{(4\pi)^2}\right)~,
\nonumber\\
{\mathcal A}^{(4)} &=& \int_I \frac{d^4k}{(2\pi)^4} \,
\gamma^\mu p^\nu  \frac{i \,k_\mu k_\nu}{k^2 - \bar{M}_P^2} =
p\!\!\!/ \, \left( \frac{1}{4}\bar{M}^2_P \,\bar\Delta_P
-\frac{1}{8}\frac{\bar{M}_P^4}{(4\pi)^2}\right)~,
%\nonumber\\
%{\mathcal A}^{(5)} \sigma^{\mu\nu} \frac{i \,k_\mu k_\nu}{k^2 - \bar{M}_P^2} &=&
%0~,
\eeqa
with $\bar\Delta_P$ given in Eq.~(\ref{D2}). One sees that the terms with one or two 
derivatives generate quark mass dependent structures $\sim M_P^4$ that can not be
absorbed in the LECs of the terms $\sim u_\mu u^\mu \sim k_\mu k^\mu$. However,
these additional terms are only of importance if one is interested in varying the
quark masses (the procedure of retaining only the dimension two terms $\sim b_{1,2,3,4}$
in \cite{BM} for the analysis of the chiral expansion of the baryon masses was thus
correct). Furthermore, Eq.~(\ref{tadstrcutures}) also shows that the quark mass 
dependence of the terms with one or two covariant derivatives acting on the baryon
fields is the same since the diagrams have to be evaluated on the mass shell, 
$p\!\!\!/ = m_0,\, p^2 = m_0^2$.

%%%%%%%%%%%%%%%%%%%%%%%%%%%%%%%%%%%%%%%%%%%%%%%%%%%%%%%%%%%%%%%%%%%%%%%%%%%%%%%%%%%%%%%%%%%%
\section{Renormalization and \boldmath{$\beta$}-functions}
\def\theequation{\Alph{section}.\arabic{equation}}
\setcounter{equation}{0}
\label{app:renorm}
%%%%%%%%%%%%%%%%%%%%%%%%%%%%%%%%%%%%%%%%%%%%%%%%%%%%%%%%%%%%%%%%%%%%%%%%%%%%%%%%%%%%%%%%%%%%% 
Here, we briefly discuss the renormalization of the fourth order loop graphs. We
use standard dimensional regularization, the UV infinities are mapped onto simple
poles $\sim 1/(d-4)$, where $d$ is the number of space-time dimensions. This leads
to the following representation of the fourth order LECs $d_i$ in Eq.~(\ref{leff4})
\beq
d_i = d_i^r (m_0) + \frac{\Gamma_i}{F_\pi^2} \, \bar L~,
\eeq
with $\bar L$ defined in Eq.~(\ref{mubar}) (remember that we work at the
scale $\lambda = m_0$). The corresponding $\beta$-functions are given by
\beqa
\Gamma_1 &=& -\frac{1}{6}\, b_1 + \frac{1}{18}\, b_3 + \frac{9}{4}\,D\,F\,b_F
+ \frac{1}{4} \left( \frac{7}{9} + \frac{23}{6}\,D^2 +\frac{9}{2}\,F^2\right)\, b_D
\nonumber\\
&& + \frac{1}{24\,m_0} \left(\frac{1}{3}\,D^2 - F^2\right) + \frac{m_0}{24}\left(
    b_5 - \frac{1}{3}b_7 \right)~, \nonumber\\
\Gamma_2 &=& \frac{1}{4}\, b_2 + \frac{5}{2}\,D\,F\,b_D +  \frac{1}{4} 
 \left( \frac{1}{3} + 5\,D^2 + 9\,F^2\right)\, b_F + \frac{1}{8\,m_0}\, D\,F
-\frac{m_0}{16}\,b_6~,\nonumber \\
\Gamma_3 &=& -\frac{3}{4}\, b_1 - \frac{1}{12}\, b_3 + \frac{9}{4}\,D\,F\,b_F
+ \frac{1}{2} \left( 1 + \frac{13}{4}\,D^2 +\frac{9}{4}\,F^2\right)\, b_D
\nonumber\\
&& + \frac{1}{16\,m_0} \left(D^2 - 3\, F^2\right) + \frac{m_0}{16}\left(
    3b_5 + \frac{1}{3}b_7 + b_9\right)~, \nonumber\\
\Gamma_4 &=& \frac{3}{2}\, b_1 - \frac{1}{18}\,b_3 - 
 \left( \frac{11}{18} + 4\,D^2 \right)\, b_D + \frac{1}{8\,m_0}\, \left( 3\,F^2
- D^2\right) -\frac{m_0}{4}\,\left(\frac{3}{2}b_5 - \frac{1}{18}b_7+\frac{1}{3}b_9
\right)~,\nonumber \\
\Gamma_5 &=& -\frac{1}{18} \,\left( 13\,b_2 -11\,b_F \right) - \frac{13}{36}\,
\left(\frac{1}{m_0} D\,F - \frac{1}{2}\, m_0 \, b_6 \right) ~, \nonumber \\
\Gamma_6 &=& \frac{11}{18}\, b_0 - \frac{1}{4}\,b_1 - \frac{35}{108}\,b_3 - \frac{11}{18}\,b_4 
+\frac{1}{3} \left( \frac{11}{12} - D^2 \right)\, b_D - \frac{1}{16\,m_0}\, \left(\frac{35}{27}\,D^2
+ F^2\right)\nonumber \\ 
&& +\frac{m_0}{8}\,\left(\frac{1}{2}\, b_5 + \frac{35}{54}\, b_7+\frac{11}{9}\, b_8
\right)~,\nonumber \\
\Gamma_7 &=& \frac{5}{6}\, b_0 - \frac{1}{4}\,b_1 - \frac{17}{36}\,b_3 - \frac{5}{6}\,b_4 
- 3\,D\,F\,b_F +\frac{1}{2} \left( \frac{7}{18} - \frac{7}{3}\, D^2 - 3\, F^2 \right)\, b_D \nonumber \\ 
&&- \frac{1}{16\,m_0}\, \left(\frac{17}{9}\,D^2 + F^2\right)
+\frac{m_0}{8}\,\left(\frac{1}{2}\,b_5 + \frac{17}{18}\,b_7+\frac{5}{3}\,b_8 
\right)~.
\eeqa
These  agree with the $\beta$-functions in \cite{BM} (if one makes use of Eqs.~(\ref{biconv}))
up to kinetic energy insertions and a few typographical errors in that paper.

\newpage
%%%%%%%%%%%%%%%%%%%%%%%%%%%%%%%%%%%%%

%%%%%%%%%%%%%%%%%%%%%%%%%%%%%%%%%%%%%
\end{document}